\documentclass[journal,draftclsnofoot,onecolumn,12pt,twoside]{IEEEtran}
\usepackage[usenames, dvipsnames]{color}
\definecolor{mycolor}{RGB}{0, 0, 0}
\usepackage{algorithm} %ctan.org\pkg\algorithms
\usepackage{algpseudocode}
\usepackage[neverdecrease]{paralist}
\usepackage{graphicx}
\usepackage{amssymb}
\usepackage{amsfonts}
\usepackage[cmex10]{amsmath}
\usepackage{dsfont}
\usepackage{float}
\usepackage{url}
\usepackage{bm}
\usepackage{hhline}
\usepackage{enumerate}
\usepackage{pgf,tikz}
\usepackage{mathrsfs}
\usepackage[framed,numbered,autolinebreaks,useliterate]{mcode}
\usetikzlibrary{arrows}

\IEEEoverridecommandlockouts

\newtheorem{Proposition}{Proposition}
\newtheorem{Definition}{Definition}
\newtheorem{Example}{Example}
\newtheorem{theorem}{Theorem}

\newtheorem{lemma}[theorem]{Lemma}

\newtheorem{rem}{Remark}[section]

 \usepackage[colorlinks=true]{hyperref}
\hypersetup{urlcolor=blue, citecolor=blue,linkcolor=purple}

\makeatletter
\let\@@pmod\pmod
\DeclareRobustCommand{\pmod}{\@ifstar\@pmods\@@pmod}
\def\@pmods#1{\mkern4mu({\operator@font mod}\mkern 6mu#1)}
\makeatother

\usepackage{tikz}
\usetikzlibrary{shapes,arrows}
\usepackage{verbatim}
\usetikzlibrary{arrows}

\newcommand\Circle[1]{%
  \tikz[baseline=(char.base)]\node[circle,draw,inner sep=2pt] (char) {#1};}

\tikzstyle{startstop} = [rectangle, rounded corners, minimum width=1.5cm, minimum height=0.65cm,text centered, draw=black, fill=blue!3]
\tikzstyle{io} = [trapezium, trapezium left angle=70, trapezium right angle=110, minimum width=3cm, minimum height=1cm, text centered, draw=black, fill=blue!30]
\tikzstyle{process} = [rectangle, minimum width=3cm, minimum height=1cm, text centered, draw=black, fill=orange!30]
\tikzstyle{decision} = [diamond, minimum width=3cm, minimum height=1cm, text centered, draw=black, fill=green!30]
\tikzstyle{arrow} = [thick,->,>=stealth]

\ifCLASSINFOpdf
\else
\fi
\begin{document}
\title{Construction and Encoding of QC-LDPC Codes Using Group Rings}
%--------------------------------------------------------------------------------------------main
%\author{Dariush Kiani ,~\IEEEmembership{Member,~IEEE,}
%        Hassan Khodaiemehr,~\IEEEmembership{Fellow,~OSA,}
 %       and~Jane~Doe,~\IEEEmembership{Life~Fellow,~IEEE}% <-this % stops a space
%\thanks{M. Shell is with the Department
%of Electrical and Computer Engineering, Georgia Institute of Technology, Atlanta,
%\thanks{Manuscript received April 19, 2005; revised January 11, 2007.}}
%-------------------------------------------------------------------------------------------

\author{Hassan Khodaiemehr and %~\IEEEmembership{Member,~IEEE,}
Dariush Kiani%,~\IEEEmembership{Member,~IEEE,}
\thanks{H. Khodaiemehr and D. Kiani  are with the Department
of Mathematics and Computer Science, Amirkabir University of Technology (Tehran Polytechnic), Tehran, Iran (emails: h.khodaiemehr@aut.ac.ir and d.kiani@aut.ac.ir).

D. Kiani is also with the  School of Mathematics, Institute for Research in Fundamental Sciences (IPM), P.O. Box 19395-5746, Tehran, Iran.
}}% <-this % stops a space
\maketitle

\begin{abstract}
%\boldmath
Quasi-cyclic (QC) low-density parity-check (LDPC) codes which are known as QC-LDPC codes, have many applications due to their simple encoding implementation by means of cyclic shift registers. In this paper, we construct QC-LDPC codes from group rings. A group ring is a free module (at the same time a ring) constructed in a natural way from any given ring and any given group. We present a structure based on the elements of a group ring for constructing QC-LDPC codes. Some of the previously addressed methods for constructing QC-LDPC codes based on finite fields are special cases of the proposed construction method. The constructed QC-LDPC codes perform very well over the additive white Gaussian noise (AWGN) channel with iterative decoding in terms of bit-error probability and block-error probability. Simulation results demonstrate that the proposed codes have competitive performance in comparison with the similar existing LDPC codes.
Finally, we propose a new encoding method for the proposed group ring based QC-LDPC codes that can be implemented faster than the current encoding methods. The encoding complexity of the proposed method is analyzed mathematically, and indicates a significate reduction in the required number of operations, even when compared to the available efficient encoding methods that have linear time and space complexities.
\end{abstract}


% \fi
\IEEEpeerreviewmaketitle

\section{Introduction}

%\IEEEPARstart{T}{his} demo file is intended to serve as a ``starter file''
%for IEEE journal papers produced under \LaTeX\ using
%IEEEtran.cls version 1.7 and later.

\IEEEPARstart{T}{he} subfield of algebraic coding dominated the field of channel coding theory for the first couple of decades. In the $1960$s, the objective of  most of the research  in channel coding was the development of  algebraic block codes, particularly cyclic codes. The main focus  of algebraic coding theory is the design of $[n,k,d]$-codes that maximize the minimum distance $d$ for a given $(n,k)$ over a finite field $\mathbb{F}_q$.  The algebraic coding paradigm used the structure of finite fields to design efficient encoding and error-correction procedures for linear block
codes operating on a hard-decision channel. Cyclic codes are codes that are invariant under cyclic shifts of $n$-tuple codewords. They were first investigated by  Prange in $1957$~\cite{00} and became the primary focus of research after the publication of Peterson's pioneering text in $1961$~\cite{01}. Cyclic codes have a nice algebraic theory and attractive simple encoding and decoding procedures based on cyclic shift-register implementations. Hamming, Golay, and shortened Reed-Muller (RM) codes can be put into cyclic form. The main achievement in this field was the invention of BCH and Reed-Solomon (RS) codes in $1959$ and $1960$~\cite{02,03,04}.

Even though, binary algebraic block codes can be used on the additive white Gaussian noise (AWGN) channel, they have not proved to be the way to approach  channel capacity on this channel.
Approaching the Shannon limit on the AWGN channel for an error-correction paradigm requires the operation of the decoder on the vector of \emph{soft decisions}
%, which is the real-valued received sequence,
and minimizing the Euclidean distance, rather than working on \emph{hard decisions}
%, which are obtained by quantizing the received sequence into  two levels,
and minimizing the Hamming distance. It can be shown that using hard decisions generally costs $2$ to $3$dB loss in the decoding performance. Thus, in order to approach the Shannon limit on the AWGN channel, the error-correction paradigm of the algebraic coding must be modified to accommodate soft decisions.

\emph{Probabilistic coding} is  an alternative line of development, that was more directly inspired by Shannon's
probabilistic approach to coding. Whereas the algebraic coding theory aims to find specific codes that maximize the
minimum distance $d$ for a given $(n,k)$, probabilistic coding is more concerned with finding the codes that optimize the average performance as a function of coding and decoding complexities. Gallager's doctoral thesis \cite{05} on low-density parity-check (LDPC) codes was motivated by the problem of finding a class of \emph{random-like} codes that could be decoded near capacity with quasi-optimal performance and feasible complexity. Gallager's LDPC codes and his iterative  \emph{a posteriori probability} (APP) decoding algorithm, which seems to have been the first appearance of the now-ubiquitous \emph{sum-product algorithm} (SPA) or \emph{belief propagation}, were long before their time.  Gallager's LDPC codes were forgotten for more than $30$ years due to their high complexity for the technology of that time.
In $1996$, MacKay \cite{06} showed empirically that near-Shannon-limit performance could be obtained with long LDPC codes and iterative decoding.

However, even though LDPC codes have a good error performance in the AWGN channel, their encoding complexity was a drawback for their implementation until the recent two decades and the invention of the \emph{quasi-cyclic} LDPC (QC-LDPC) codes. It has been shown that QC-LDPC codes can perform as well as other types of LDPC codes in most of the applications \cite{5,6,7,8,9,10,11,12}. They have also been applied in the design of efficient multidimensional signal constellations for the AWGN channels \cite{mypaper1,mypaper2}.
QC-LDPC codes have advantages over other types of LDPC codes in hardware implementation of encoding \cite{4} and decoding \cite{6}, \cite{12}. Thus, most of the LDPC codes adopted as standard codes for
various next-generation communication and storage systems, are quasi-cyclic.
In \cite{4}, the authors proposed a method for encoding QC-LDPC codes with linear time and space complexities in the length of the code.
These features have made the design of QC-LDPC codes an attractive research area and lots of methods, including algebraic methods, are proposed for constructing QC-LDPC codes. Among these methods, the methods based on finite fields are the most related ones to this paper~\cite{12,3,QC_cyclic}.
% It is shown that the time and space complexities of encoding  for QC-LDPC code is linearly proportional
%to the number of parity bits (the number of rows of parity-check matrix) of the code for serial encoding, and to
%the length of the code \cite{4}.

%Different construction methods of QC-LDPC codes are available in the literature and  the construction of QC-LDPC codes from finite fields  can be found in \cite{12,3,QC_cyclic}.

Looking at the evolution of channel coding theory, which can be found in the brilliant survey of Costello and Forney  \cite{07}, reveals a trade off between the complexity of encoding in the transmitter and error performance of decoding in the receiver. Indeed, if we want to have a low complexity encoding, the structure of  the code must be close to the structure of the algebraic codes (like cyclic codes) and if we want to have a code with good error performance in the AWGN channel, the structure of the code should be close to the structure of the capacity-approaching codes (like LDPC codes). In this paper, we propose a family of codes that have both advantages at the same time. This family of codes can be obtained by group ring encodings, which are called  \emph{group ring based codes} through this paper. The codes from group ring encodings, which are presented in \cite{0,1}, are submodules (sometimes ideals) in a group ring.
%Obtaining the generator and the parity-check matrices of the group ring based codes are presented in \cite{1}.
An isomorphism between a group ring and a certain subring of the matrices can be used to obtain the parity-check and the generator matrices of the codes directly from the  elements of the group ring \cite{1}. The properties of the group ring based codes, such as the self-duality or owning a low density parity-check matrix, have simple algebraic descriptions. Examples of  LDPC and self-dual group ring based codes are presented in \cite{1}.

%We apply the construction method of QC-LDPC codes from finite fields \cite{3} together with the existing connection between the elements of a group ring and the ring of the matrices \cite{2} enables us to
In this paper, we present a new method for constructing and encoding QC-LDPC codes from group rings. We exploit the available implementations of the fast Fourier transform (FFT) over group rings \cite{group_ring_FFT} to increase the efficiency of the proposed encoding method, compared to the  existent encoding methods for QC-LDPC codes \cite{4}.

The rest of this paper is organized as follows: Section~\ref{sec2} gives  a brief introduction to  group rings and their matrix representation. In Section~\ref{sec3}, we present the design criteria for constructing the parity-check matrix of the QC-LDPC codes based on finite fields.  In Section~\ref{sec4}, we present the design criteria for constructing the parity-check matrix of the QC-LDPC codes from an element of a group algebra. In Section~\ref{sec5}, we present a new method for constructing QC-LDPC codes based on group rings with a cyclic underlying group. In Section~\ref{Abelian groups}, we present the construction of QC-LDPC codes based on  group rings with a non-cyclic Abelian underlying group. A new encoding method for the proposed  QC-LDPC codes  based on group ring is presented in Section~\ref{sec5_2}.
%We also present a new  encoding procedure of the proposed structures.
The implementation and  complexity analysis of the proposed encoding method is given in Section~\ref{implementation}.
Section~\ref{simulation} is concerned with  numerical and  simulation results.  Section~\ref{conclusion} contains  concluding remarks.

\textbf{Notation}: Matrices and vectors are denoted by bold upper
and lower case letters. We denote the groups and rings  by upper case letters. The $i^{th}$ element of a vector $\mathbf{a}$ is denoted
by $a_i$ and the  entry $(i,j)$ of a matrix $\mathbf{A}$ is denoted by
$A_{i,j}$; $[\,\,]^t$ denotes the transposition for vectors and matrices.
\section{Preliminaries on Group Rings}\label{sec2}
Let $G$ be a multiplicative group and let $R$ be a ring. The \emph{group ring} of $G$ over $R$, which is denoted by $R[G]$ (or simply $RG$), is the set of mappings $f : G\rightarrow R$ of finite support\footnote{Let $X$ be an arbitrary set and $R$  be a ring with zero $0_R$. Suppose that $f : X\rightarrow R$ is a  function whose domain is $X$. The support of $f$, which is denoted by $\textrm{supp}(f)$, is the set of points in $X$ in which $f$ is non-zero, i.e., $\textrm{supp}(f)=\left\{x\in X\,|\,f(x)\neq 0_R\right\}$.}.
A group ring is a free module and at the same time a ring. As a free module, its ring of scalars is the given ring $R$, and its basis is correspondence with the given group $G$.
The module scalar product $\mu f$ of a scalar $\mu$ in $R$ and a vector (or mapping) $f$ is defined as the vector $ x\mapsto \mu \cdot f(x)$, where $x\in G$ and the dot operation $(\cdot)$ represents the multiplication in $R$.  The module group sum of two vectors $f$ and $g$ is defined as the vector  $x\mapsto f(x)+g(x)$. To turn the additive group $RG$ into a ring, we define the product of $f$ and $g$ to be the vector
\begin{equation}\label{group_ring_mult}
  x\mapsto \sum_{uv=x}f(u)g(v)=\sum_{u\in G}f(u)g(u^{-1}x).
\end{equation}
The summation is admissible because $f$ and $g$ are of finite support.
If the given ring $R$ is commutative, a group ring is also referred to as a \emph{group algebra}, for it is indeed an algebra over the given ring.
The mappings such as $f: G\rightarrow R$ are  \emph{formal linear combinations} of the elements of $G$, with coefficients in $R$, i.e., $\sum_{g\in G}f(g)g$, or simply $\sum_{g\in G}f_gg$.
Thus, the group ring $RG$ is a ring consisting of the set
of all summations $u=\sum_{g\in G}\alpha_g g$, where $\alpha_g\in R$. If $v=\sum_{g\in G}\beta_g g$, then the addition is defined term-by-term,
$$u+v=\sum_{g\in G}(\alpha_g+\beta_g) g,$$
while the multiplication is a convolution-like operation,
$$uv=\sum_{g,h\in G}(\alpha_g\beta_h) gh=\sum_{g\in G}\left(\sum_{h\in G}\alpha_{h}\beta_{h^{-1}g}\right)g.$$
The group $G$ acts as a basis for the module $RG$ over the ring $R$. Indeed, by considering an order over the elements of $G$, every element of
$RG$ is a vector composed of elements of $R$, where the $i^{th}$ component is associated
with the group element $g_i$. Treating a group ring as the space of functions mapping a group to a ring, the multiplication in the group ring is the convolution of two functions therein. More details about group rings and their properties can be found in \cite{29}.
%Addition is componentwise  and multiplication is obtained from the group multiplication together with the distributive law.
\begin{Example}
Consider $\mathbb{F}_q$ to be the finite field with $q$ elements.  Let $G = \mathbb{Z}_3$ be the cyclic group of three elements with generator $a$ and identity element $1_G$. An element $r$ of $\mathbb{F}_qG$ may be written as $r = z_0 1_G+ z_1a + z_2a^2$, where $z_0$, $z_1$ and $z_2$ are in $\mathbb{F}_q$. Writing a different element $s$ as $s=w_{0}1_{G}+w_{1}a+w_{2}a^{2}$, their sum is
$r+s=(z_{0}+w_{0})1_{G}+(z_{1}+w_{1})a+(z_{2}+w_{2})a^{2}$, and their product is
$rs=(z_{0}w_{0}+z_{1}w_{2}+z_{2}w_{1})1_{G}+(z_{0}w_{1}+z_{1}w_{0}+z_{2}w_{2})a+(z_{0}w_{2}+z_{2}w_{0}+z_{1}w_{1})a^{2}$. The identity element $1_G$ of $G$ induces a canonical embedding of the coefficient ring $\mathbb{F}_q$ into $\mathbb{F}_qG$ and the multiplicative identity element of $\mathbb{F}_qG$ is $(1)1_G$ where the first $1$ comes from $\mathbb{F}_q$ and the second from $G$. The additive identity element is  zero.
$\hfill \square$\end{Example}
\subsection{Group rings and the ring of matrices}
From now on, we only consider group algebras over a finite group, which are denoted by $G$ and $H$ in most cases. Let $\left\{g_1,g_2 ,\ldots ,g_n \right\}$ be a fixed listing of the elements of $G$. We have the following definition from \cite{2}.
\begin{Definition}
The $RG$-matrix of an element $w=\sum_{i=1}^n \alpha_{g_i}g_i$ in the group ring $RG$ is an element in $M_n(R)$, the ring of $n \times n$ matrices over $R$,  defined as
\begin{equation}\label{eq1}
   \mathbf{M}(RG,w)= \left[
      \begin{array}{cccc}
        \alpha_{g_1^{-1}g_1} & \alpha_{g_1^{-1}g_2} & \cdots & \alpha_{g_1^{-1}g_n} \\
        \alpha_{g_2^{-1}g_1} & \alpha_{g_2^{-1}g_2} & \cdots & \alpha_{g_2^{-1}g_n} \\
        \vdots & \vdots & \ddots & \vdots \\
        \alpha_{g_n^{-1}g_1} & \alpha_{g_n^{-1}g_2} & \cdots & \alpha_{g_n^{-1}g_n} \\
      \end{array}
    \right].
\end{equation}
\end{Definition}
It is obvious that each row and each column is a permutation, determined by the group multiplication, of the initial row.
\begin{theorem}[\emph{\cite[Theorem 1]{1}}]
Given a listing of the elements of a group $G$ of order $n$, there is a bijective
ring homomorphism $\sigma :w \mapsto \mathbf{M}(RG,w)$ between $RG$ and the $RG$-matrices
over $R$.
\end{theorem}
\section{ Definitions and Basic Concepts of quasi-cyclic LDPC Codes}\label{sec3}
Let $t$ and $b$ be two positive integers. A $b\times b$ \emph{circulant} is a $b\times b$ matrix for which each row is a right cyclic-shift of the row above it and the first row is the right cyclic-shift of the last row. The top row (or the leftmost column) of a circulant is called the \emph{generator} of the circulant.
%============================================================
%Let $v=(v_0,v_1,\ldots ,v_{t-1})$ be a $tb$-tuple over $GF(2)$  that consists of $t$ sections of $b$ bits each. For $0 \leq j <t$, the $j^{th}$ section of $v$ is a $b$-tuple over $GF(2)$ like $v_j=(v_{j,0},v_{j,1},\ldots ,v_{j,b-1})$.
%
%Let $v_j^{(1)}$ be the $b$-tuple over $GF(2)$ obtained by cyclically shifting each component of $v_j$ one place to the right. We call $v_j^{(1)}$ the (right) cyclic-shift of $v_j$. Then, $tb$-tuple $(v_0^{(1)},v_1^{(1)},\cdots ,v_{t-1}^{(1)})$ is called the $t$-sectionized cyclic-shift of $v$.
%
%\begin{Definition}Let $b$, $k$, and $t$ be positive integers such that $k<tb$. A $(tb,k)$-linear block code $\mathcal{C}_{qc}$ over $GF(2)$ is called a quasi-cyclic (QC) code if the following conditions hold:
%\begin{enumerate}
%  \item Each codeword in $\mathcal{C}_{qc}$ consists of $t$ sections of $b$ bits each.
%  \item Every $t$-sectionized cyclic-shift of a codeword in $\mathcal{C}_{qc}$  is also a codeword in $\mathcal{C}_{qc}$.
%\end{enumerate}
%Such a QC code is called a $t$-section QC code.
%\end{Definition}
%\begin{Definition}
%A $b\times b$ circulant is a $b\times b$ matrix for which each row is a right cyclic-shift of the row above it and the first row is the right cyclic-shift of the last row. The top row (or the leftmost column) of a circulant is called the generator of the circulant.
%\end{Definition}
%=======================================================
A binary QC code $\mathcal{C}_{qc}$ is commonly specified by a parity-check matrix, which is a $(t - c)\times t$ array of $b\times b$ circulants over $\mathbb{F}_2$.
\subsection{Construction of QC-LDPC codes based on finite fields}\label{sec3_2}
Consider the Galois field $\mathbb{F}_q$, where $q$ is a power of a prime number $p$. Let $\alpha$ be a primitive element of $\mathbb{F}_q$. Then, $\left\{\alpha^{-\infty}=0,\alpha^0=1,\alpha.\cdots ,\alpha^{q-2}\right\}$ form all the $q$ elements of $\mathbb{F}_q$.
For each nonzero element $\alpha^i$, with $0\leq i<q-1$, we define a $(q-1)$-tuple over $\mathbb{F}_2$, $\mathbf{z}(\alpha^i)=(z_0,\ldots ,z_{q-2})$, where the $i^{th}$ component $z_i$ is $1$ and all the other $q-2$  components are set to zero. The binary vector, $\mathbf{z}(\alpha^i)$, is referred to as the \emph{location-vector} of $\alpha^i$, with respect to the multiplicative group of $\mathbb{F}_q$, or the $M$-location-vector of $\alpha^i$. The location-vector of the $0$ element of $\mathbb{F}_q$ is defined as the all-zero $(q-1)$-tuple, $(0, 0, \ldots, 0)$.

For a given $\delta\in\mathbb{F}_q$, we form a $(q - 1)\times (q - 1)$ matrix $\mathbf{A}$ over $\mathbb{F}_2$ with the $M$-location-vectors of $\delta,\alpha\delta,\ldots \alpha^{q-2}\delta$
as the rows. Then, $\mathbf{A}$ is a $(q - 1)\times (q - 1)$ \emph{circulant permutation matrix} (CPM), i.e., $\mathbf{A}$ is a permutation matrix for which each row is the right cyclic-shift of the row above it and the first row is the right cyclic-shift of the last row. The matrix $\mathbf{A}$ is referred to as the $(q-1)$-\emph{fold matrix dispersion} (or expansion) of the field element $\delta$ over $\mathbb{F}_2$.
The construction of the parity-check matrices of the QC-LDPC codes starts with an $m\times n$ matrix $\mathbf{W}$ over $\mathbb{F}_q$ given by
\begin{IEEEeqnarray}{rCl}\label{eq2}
    \mathbf{W}&=&\left[
        \begin{array}{cccc}
          \mathbf{W}_0^t &
          \mathbf{W}_1^t &
          \cdots &
          \mathbf{W}_{m-1}^t \\
        \end{array}
      \right]^t\nonumber\\
      &=&\left[
                \begin{array}{cccc}
                  w_{0,0} & w_{0,1} & \cdots & w_{0,n-1} \\
                  w_{1,0} & w_{1,1} & \cdots & w_{1,n-1} \\
                  \vdots & \vdots & \ddots & \vdots \\
                  w_{m-1,0} & w_{m-1,1} & \ldots & w_{m-1,n-1} \\
                \end{array}
              \right],
\end{IEEEeqnarray}
whose rows satisfy the following two constraints:
\begin{enumerate}
  \item for $0\leq i<m$ and $0\leq k,l<q-1$, with $k\neq l$, $\alpha^k \mathbf{W}_i$ and $\alpha^l \mathbf{W}_i$ have at most one position in which both of them have the same symbol from $\mathbb{F}_q$ (i.e., they differ in at least $n - 1$ positions);
  \item for $0\leq i,j<m$, and $i\neq j$, with $0\leq k,l<q-1$, $\alpha^k \mathbf{W}_i$ and $\alpha^l \mathbf{W}_j$ differ in at least $n-1$ positions.
\end{enumerate}
The above two constraints on the rows of the matrix $\mathbf{W}$ are referred to as $\alpha$-\emph{multiplied row constraints} $1$ and $2$, respectively. These conditions are the sufficient conditions or the design criteria \cite{3} for constructing the parity-check matrix of the QC-LDPC codes.
In order to complete the construction, it is enough to replace each entry of $\mathbf{W}$ by its $(q-1)$-fold matrix dispersion over $\mathbb{F}_2$.
For an $m\times n$ matrix $\mathbf{W}$ over $\mathbb{F}_q$, in which the entries are written as the powers of the primitive element $\alpha$ of $\mathbb{F}_q$, we define the \emph{base matrix} or the \emph{exponent matrix} as the $m\times n$ matrix $\mathbf{B}$ over $\left\{-\infty,0,1,\ldots,q-2\right\}$, so that $B_{i,j}=\lambda$, for $1\leq i\leq m$ and $1\leq j\leq n$, if and only if $W_{i,j}=\alpha^{\lambda}$. To simplify the notation, the matrices $\mathbf{B}$ and $\mathbf{W}$ can be used interchangeably in the construction of codes.
\section{Construction of QC-LDPC Codes Based on Difference Sets and Cyclic Group Algebras}\label{sec4}
In this section, we present our key theorem based on the notation used in the preceding sections.
\begin{theorem}\label{theorem2}
Let $G=\left\{g_0=1_{G},g_1,g_2,\ldots g_m\right\}$ be a finite group and $\mathbb{F}_q$ be the finite field of order $2^{m+1}$,  with  primitive element $\alpha$. Then, the $\mathbb{F}_qG$-matrix corresponding to the element $w=\sum_{i=0}^{m}\alpha^{2^i}g_i$ gives an $(m+1)\times (m+1)$ matrix  $\mathbf{W}$ that satisfies  the $\alpha$-multiplied constraints $1$ and $2$. Replacing each component of $\mathbf{W}$ by its corresponding $(q-1)\times(q-1)$ CPM in $\mathbb{F}_q$ gives the  parity-check matrix of a  QC-LDPC code.
%\begin{proof}
%The proof is given in Appendix \ref{appA}
%\end{proof}
\end{theorem}
\begin{IEEEproof}
Let $\mathbf{W}$ be the corresponding $RG$-matrix of $w$ and let $\mathbf{W}_0=\left[
                                                         \begin{array}{ccccc}
                                                           \alpha & \alpha^2 & \alpha^{4} & \cdots & \alpha^{2^m} \\
                                                         \end{array}
                                                       \right]$
be its first row. We can consider the other rows as permutations of the first row. Thus, if we check the constraint $1$ for $\mathbf{W}_0$, the other rows also fulfill the constraint $1$. Let $0\leq k,l<q-1$,  be two integers, with $k\neq l$, and consider the vectors $\alpha^k \mathbf{W}_0$ and $\alpha^l \mathbf{W}_0$. Then, having the same values in position $i_1$ is equivalent to $2^{i_1}+k=2^{i_1}+l$ that implies $k=l$, which is a contradiction. Now, we check the second condition. Consider two different permutations of $\mathbf{W}_0$ as follows
                                                       \begin{eqnarray*}
                                                         \mathbf{W}_i &=& \left[
                                                         \begin{array}{ccccc}
                                                           \alpha^{2^{i_0}} & \alpha^{2^{i_1}}  & \cdots & \alpha^{2^{i_m}} \\
                                                         \end{array}
                                                       \right]  \\
                                                        \mathbf{W}_j &=& \left[
                                                         \begin{array}{ccccc}
                                                           \alpha^{2^{j_0}} & \alpha^{2^{j_1}}  & \cdots & \alpha^{2^{j_m}}\\
                                                         \end{array}
                                                       \right].
                                                       \end{eqnarray*}
Without loss of generality, assume $\alpha^k \mathbf{W}_i$ and $\alpha^l \mathbf{W}_j$ have the same values in the first and second positions. Then,
\begin{eqnarray*}
  2^{i_0}+k &=& 2^{j_0}+l,\quad i_0\neq j_0, \\
  2^{i_1}+k &=& 2^{j_1}+l,\quad i_1\neq j_1.
\end{eqnarray*}
Consequently, $2^{i_0}+2^{j_1}=2^{j_0}+2^{i_1}$, where $i_0\neq i_1$ and $j_0\neq j_1$. We consider three different cases.
\setdefaultleftmargin{0cm}{2cm}{}{}{}{}
\begin{enumerate}[]
\item{\textbf{Case 1}:} If $i_0<j_1$, then
  $$2^{i_0}(1+2^{j_1-i_0})=\left\{ \begin{array}{ll}
    2^{i_1}(1+2^{j_0-i_1}), & i_1<j_0 \\
    2^{j_0}(1+2^{i_1-j_0}), & i_1>j_0 \\
    2^{j_0+1}, & i_1=j_0.
  \end{array}\right.$$
  The first equation implies $i_0=i_1$, which is a contradiction. Using the second equation we conclude that $i_0=j_0$ and $j_1-i_0=i_1-j_0$, which implies that $i_1=j_1$. Thus, $k=l$ which is a contradiction. Now we consider the third equation. In the left side of this equation we have a product of an even and an odd number and in the right side we have an even number. The only possibility is to have $j_1=i_0$ and $i_0+1=j_0+1$, that imply $j_0=j_1$, which is a contradiction.
\item{\textbf{Case 2}:} If $i_0=j_1$, we can show conveniently that $i_1=j_0$ and the equation $2^{i_0}-2^{i_1}=2^{i_1}-2^{i_0}$ leads to a contradiction.
\item{\textbf{Case 3}:} If $i_0>j_1$, the proof is the same as the first case.
\end{enumerate}
Thus, $\mathbf{W}$ fulfills the $\alpha$-multiplied constraints $1$ and $2$.
\end{IEEEproof}

It follows from Theorem~\ref{theorem2} that the Tanner graph associated with the matrix $\mathbf{W}$ has girth at least $6$.

%-----------------------------------------------------------------------------------
Theorem \ref{theorem2} is also valid for other prime numbers instead of $2$. Now, we present some examples using the construction method of Theorem \ref{theorem2}.
\begin{Example}
We consider the groups of order $8$ which give the codes of length $8\times 255 =2040$.
The  groups with $8$ elements are $\mathbb{Z}_8$, $\mathbb{Z}_2\times \mathbb{Z}_2\times \mathbb{Z}_2$,  $Q_8$, $D_8$ and $\mathbb{Z}_2\times \mathbb{Z}_4$. We represent the corresponding matrix of the element $w$ in Theorem \ref{theorem2} by a matrix that contains the powers of $\alpha$, which is referred to as the \emph{exponent matrix}. For example, $W_{ij}=2$ for $1\leq i,j \leq 8$, means $W_{ij}=\alpha^2$, where $\alpha$ is the primitive element of $\mathbb{F}_{2^{8}}$.
%Let $F=GF(2^8)$ and $G=\mathbb{Z}_8$. Then,
The $\mathbb{F}_{2^{8}}\mathbb{Z}_8$-matrix of $w$ in Theorem \ref{theorem2} is as follows
\begin{IEEEeqnarray}{rCl}\label{eq3}
%\scriptstyle\scriptsize
  \mathbf{W}_{\mathbb{F}_{2^{8}}\mathbb{Z}_8}=\left[
                          \begin{array}{cccccccc}
                            1 & 2 & 4 & 8 & 16 & 32 & 64 & 128 \\
                            128 & 1 & 2 & 4 & 8 & 16 & 32 & 64 \\
                            64 & 128 & 1 & 2 & 4 & 8 & 16 & 32 \\
                            32 & 64 & 128 & 1 & 2 & 4 & 8 & 16 \\
                            16 & 32 & 64 & 128 & 1 & 2 & 4 & 8 \\
                            8 & 16 & 32 & 64 & 128 & 1 & 2 & 4 \\
                            4 & 8 & 16 & 32 & 64 & 128 & 1 & 2 \\
                            2 & 4 & 8 & 16 & 32 & 64 & 128 & 1 \\
                          \end{array}
                        \right].
\end{IEEEeqnarray}
%\normalsize
If we consider $G=D_8$, the dihedral group of order $8$, which is defined as
$$D_8=\langle r,s|r^4=1,s^2=1,s^{-1}rs=r^{-1}\rangle,$$
then the $\mathbb{F}_{2^{8}}D_8$-matrix of $w$ is
%\footnotesize
\begin{IEEEeqnarray}{rCl}\label{eq4}
%\scriptstyle\scriptsize
   \mathbf{W}_{\mathbb{F}_{2^{8}}D_8}=\left[
                          \begin{array}{cccccccc}
                            1 & 2 & 4 & 8 & 16 & 32 & 64 & 128 \\
                            8 & 1 & 2 & 4 & 128 & 16 & 32 & 64 \\
                            4 & 8 & 1 & 2 & 64 & 128 & 16 & 32 \\
                            2 & 4 & 8 & 1 & 32 & 64 & 128 & 16 \\
                            16 & 32 & 64 & 128 & 1 & 2 & 4 & 8 \\
                            128 & 16 & 32 & 64 & 8 & 1 & 2 & 4 \\
                            64 & 128 & 16 & 32 & 4 & 8 & 1 & 2 \\
                            32 & 64 & 128 & 16 & 2 & 4 & 8 & 1 \\
                          \end{array}
                        \right].
\end{IEEEeqnarray}
%\normalsize
The last example that we present here is the quaternion group. The quaternion group is a non-Abelian group of order eight which is denoted by $Q$ or $Q_8$, and is given by the following group presentation
$$Q_8=\langle-1,i,j,k|(-1)^2=1,i^2=j^2=k^2=ijk=-1\rangle,$$
where $1$, the identity element of the group, and $-1$ commute with the other elements of the group. The $\mathbb{F}_{2^8} Q_8$-matrix of $w$ is
%\footnotesize
\begin{IEEEeqnarray}{rCl}\label{eq5}
%\scriptstyle\scriptsize
  \mathbf{W}_{\mathbb{F}_{2^8} Q_8}=\left[
                          \begin{array}{cccccccc}
                            1 & 2 & 4 & 8 & 16 & 32 & 64 & 128 \\
                            2 & 1 & 8 & 4 & 32 & 16 & 128 & 64 \\
                            8 & 4 & 1 & 2 & 128 & 64 & 16 & 32 \\
                            4 & 8 & 2 & 1 & 64 & 128 & 32 & 16 \\
                            32 & 16 & 64 & 128 & 1 & 2 & 8 & 4 \\
                            16 & 32 & 128 & 64 & 2 & 1 & 4 & 8 \\
                            128 & 64 & 32 & 16 & 4 & 8 & 1 & 2 \\
                            64 & 128 & 16 & 32 & 8 & 4 & 2 & 1 \\
                          \end{array}
                        \right].
\end{IEEEeqnarray}
%\normalsize
By replacing each element of the matrices given in (\ref{eq3}), (\ref{eq4}) and (\ref{eq5}), with their corresponding CPMs and choosing some rows of the obtained binary matrices, we get regular QC-LDPC codes, as the null space of these matrices, with different rates and with girth at least $6$.
$\hfill\square$\end{Example}
%In this paper, we consider only regular codes.

In order to increase the rate of the constructed codes based on  group algebras, we replace the set of powers in  Theorem \ref{theorem2}, which is a set of the form $\left\lbrace1,2,\ldots ,2^{|G|-1}\right\rbrace$ for a group $G$, by other sets that are introduced in the next section.
\section{An Algebraic Framework for Constructing QC-LDPC Codes Based on Group Rings}\label{sec5}
In this section, we generalize the construction of Section \ref{sec4} and relate the design of a QC-LDPC code to the selection of two elements in two different group rings. Let $G$ be a group of order $v'$. A $k'$-subset $D$ of $G$ is  a $(v', k', \lambda)$-difference set if the list of differences $d_1d_2^{-1}$, $d_1,d_2 \in D$, contains each non-identity element of $G$ exactly $\lambda$ times. The number $n' = k'-\lambda$ is the order of the difference set. A difference set $D$ in $G$ is  non-Abelian, Abelian or cyclic provided $G$ is non-Abelian, Abelian or cyclic, respectively. The difference sets can be defined by using an algebraic approach.

Let $\mathbb{F}$ be a field and $G$ be  a multiplicative group. A subset $D\subseteq G$ is identified with an element $D=\sum_{g\in D} g\in \mathbb{F}G$. Moreover, $D^{(-1)}=\sum_{g\in D} g^{-1}$.
\begin{theorem}[\emph{\cite[Theorem 18.19]{17}}]\label{theorem3}
A $k'$-subset $D \subseteq G$ of a group $G$ of order $v'$ with identity element $1_G$ is a $(v', k', \lambda)$-difference set of order $n'$, if and only if $D \cdot D^{(-1)} = n' \cdot 1_G + \lambda \cdot G$ in $\mathbb{C}G$.
\end{theorem}

In the construction of QC-LDPC codes, we need  difference sets with $\lambda=1$  to avoid $4$-cycles in the Tanner graph of the code. Let $G$ and $H$ be two finite Abelian groups.
%and let $\mathbb{F}$ be a finite field with the primitive element $\alpha$.
In the sequel, we introduce a new ring $R$, which is related to $H$, and the construction of QC-LDPC codes  using the elements of the group algebra $RG$. Let $d=\sum_{h\in D} h\in \mathbb{C}H$ be an element satisfying Theorem \ref{theorem3}. Consider $D=\left\lbrace d_1,d_2,\ldots ,d_{k'}\right\rbrace$ as the set of  indices that appear in $d$, which is a subset of $H$. Then, we define an element $w$ in $RG$ and proceed like Theorem \ref{theorem2}. It should be noted that both $G$ and  $H$ can affect the performance of the constructed code, as will be shown in the simulation results. The structure of the group $G$ can also affect the encoding complexity.
%===============================================
%       Edited until here
%===============================================
\subsection{Construction of group ring based QC-LDPC codes using cyclic groups}\label{Cyclic groups}
The first and the obvious case is to consider both $G$ and $H$ as cyclic groups. The existence and the construction of the appropriate difference sets, in some classes of cyclic groups, is  guarantied by using the following theorems.
\begin{theorem}[\emph{\cite[Construction 18.28]{17}}]\label{theorem4}
Let $\alpha$  be the generator of the multiplicative group of $\mathbb{F}_{q^m}$. Then, the set of integers $\left\lbrace 0 \leq i <\frac {q^m -1}{q-1}: \mathrm{trace}_{m/1}(\alpha^i) = 0\right\rbrace$ modulo $(q^m-1)/(q-1)$ forms a (cyclic) difference set with parameters
\begin{eqnarray}\label{eqsin}
\left(\frac{q^m-1}{q-1},\frac{q^{m-1}-1}{q-1},\frac{q^{m-2}-1}{q-1}\right).
\end{eqnarray}
Here, the $\mathrm{trace}$ denotes the usual trace function $\mathrm{trace}_{m/1}(\beta) =\sum_{i=0}^{m-1}\beta^{q^i}$  from $\mathbb{F}_{q^m}$ onto $\mathbb{F}_q$ . These difference sets are \emph{Singer difference sets}.
\end{theorem}

\begin{theorem}[\emph{\cite[Construction 18.29]{17}}]\label{theorem5}
Let $f(x)=x^m+\sum_{i=1}^{m}a_i x^{m-i}$ be a primitive polynomial of degree $m$ in $\mathbb{F}_q$. Consider the recurrence relation $\gamma_n=-\sum_{i=1}^{m}a_i \gamma_{n-i}$ and take arbitrary start values. Then the set of integers $\left\lbrace 0 \leq i < \frac{q^m-1}{q-1} : \gamma_i = 0\right\rbrace$ is a Singer difference set.
\end{theorem}

By using Theorem \ref{theorem4} and \ref{theorem5}, we can construct a $(q^2 +q+1, q+1, 1)$-difference
set in the additive group $\mathbb{Z}_{q^2+q+1}$, when $q$ is a power of a prime number.  The constructed codes in this case are just the same as the constructed codes in \cite{15}. The authors of \cite{15}, have proposed the construction of $4$-cycle free QC-LDPC codes using the cyclic difference sets. For a given difference set $D=\{d_1,d_2,\ldots,d_{k'}\}$ in $\mathbb{Z}_{v'}$, with $ d_1<d_2<\cdots <d_{k'}$,  they considered the finite field  $\mathbb{F}_q$ in $3$ different cases: $q=v'+1$, or $q\geq 2d_{k'}+1$ or $q\geq 2d_{k'-1}+1$. The table of the difference sets with small parameters, is available in \cite[pp. 427--430]{17}, which can  be used in the construction of QC-LDPC codes as above.

Now, we consider the  general case where the group $H$ is non-cyclic. As we saw in Section \ref{sec4}, the structure of $G$ has no effect on our design procedure and our concentration here is on the structure of $H$. Hence, to simplify, we assume that $G$ is a cyclic group. Let $H$ be an Abelian group. We have the following theorem for Abelian groups.
\begin{theorem}[\emph{\cite[p.~193]{21}}]\label{theorem6}
Every finite Abelian group $H$ is isomorphic to a direct product of cyclic groups of the form
$$\mathbb{Z}_{p_1^{\beta_1}}\times \mathbb{Z}_{p_2^{\beta_2}}\times \cdots \times \mathbb{Z}_{p_t^{\beta_t}} ,$$
where $p_i$'s are primes (not necessarily distinct) and  $\beta_i$'s are some positive integers.
\end{theorem}

We can see from the proof of Theorem \ref{theorem2} that  existence of a difference set in $H$, is not necessary for constructing QC-LDPC codes and we can replace this assumption by a weaker condition. In the next section, we introduce a new method for constructing QC-LDPC codes using Abelian groups.
\section{Construction of Group Ring Based QC-LDPC codes Using Non-cyclic Abelian Groups}\label{Abelian groups}
In \cite{15}, QC-LDPC codes were constructed based on difference sets in cyclic groups. Difference sets do not exist in every cyclic group and furthermore, there is no efficient  algorithm to find them. Thus, the rate and length of the constructed codes based on cyclic difference sets will be limited. In this section, we introduce some combinatorial structures in arbitrary Abelian groups, which are close to difference sets and also enough for our application in the construction of QC-LDPC codes. Then, we propose a new method to construct QC-LDPC codes based on the  proposed structures and the group algebras. As we see in the sequel, these structures give  codes with the highest possible rate of a given length.
\subsection{Combinatorial structures beyond the difference sets in the construction of QC-LDPC codes}
In the construction of QC-LDPC codes based on Abelian groups, the first candidates  are difference sets in the Abelian groups. If instead of $\mathbb{Z}_{v'}$, we consider a group $A$ with $v'$ elements, which is written multiplicatively, the condition for a set $D\subset A$ with $k'$ distinct elements to be a difference set is exactly like the cyclic case. While much is known about the difference sets in the cyclic groups, little systematic work has been done for non-cyclic groups. As in the cyclic case, we are interested in difference sets with $\lambda=1$. We say that two difference sets $D$ and $D'$ in $A$ are equivalent if there exists an automorphism $\tau$ of
$A$ and an element $g\in  A$ such that $D' = D^{\tau}g$.   Non-cyclic difference sets for $k' < 20$, are enumerated in \cite{22}. We have summarized non-equivalent and  non-cyclic difference sets, which are proper for the construction of QC-LDPC codes, in \tablename~\ref{table1}.
\begin{table}[ht]
\small
\caption{Non-cyclic difference sets of size less than $20$ }
\centering
\begin{tabular}{c|c|c}
\hline
$(v',k',\lambda)$  & Underlying group & Difference set $D$\\
                         &                            &                             \\
  \hline
  \hline
$(21, 5, 1)$  &  $a^7=b^3=1,ba=a^2b$   & $1,a,a^3,b,a^2b$  \\

$(57,8,1)$ & $a^{19}=b^3=1,ba=a^7b$ & $1,a,a^3,a^8,b,a^4b,a^{13}b,$ \\
& &$ a^{18}b^2$\\

$(57,8,1)$ & $a^{19}=b^3=1,ba=a^7b$ & $1,a,a^3,a^8,b,a^5b^2,a^{9}b^2,$ \\
& &$ a^{18}b^2$\\
$(183,14,1)$ & $a^{61}=b^3=1,ba=a^{13}b$ & $1,a,a^3,a^{20}, a^{26}, a^{48},a^{57}, b, $\\
& & $a^8b, a^{18}b,a^{29}b,a^{17}b^2,a^{32}b^2,$\\
& & $a^{44}b^2$\\
$(183,14,1)$ & $a^{61}=b^3=1,ba=a^{13}b$ & $1,a,a^3,a^{20}, a^{26}, a^{48},a^{57}, b, $\\
& & $a^{12}b, a^{46}b,a^{9}b^2,a^{17}b^2,a^{27}b^2,$\\
& & $a^{38}b^2$\\

$(273,17,1)$ & $a^{13}=b^7=c^3=1,$ & $a,b,a^2b,a^4b^2,a^{11}b^2,a^5b^4,$ \\
& $ca=a^3c,cb=b^2c,$ & $a^{10}b^4,a^4c,a^9bc,a^{12}bc,$\\
& $ba=ab$& $a^2b^2c,a^6b^2c,a^{10}b^4c, a^{11}b^4c, $ \\
 & & $b^3c^2, b^5c^2, b^6c^2$\\

 $(273,17,1)$ & $a^{13}=b^7=c^3=1,$ & $a,b,a^2b,a^4b^2,a^{11}b^2,a^5b^4,$ \\
& $ca=a^{12}c,cb=bc,$ & $a^{10}b^4,a^4c,a^9bc,a^{12}bc,$\\
& $ba=ab$& $a^2b^2c,a^6b^2c,a^{10}b^4c, a^{11}b^4c, $ \\
 & & $b^3c^2, b^5c^2, b^6c^2$\\ \hline
 \end{tabular}
 \label{table1}
 \end{table}
We can see from \tablename~\ref{table1} that none of the non-cyclic difference sets with $\lambda=1$ and $k'<20$ are Abelian. Hence, considering Abelian difference sets in our framework may not be accomplished simply. In the sequel, other structures will be introduced  which can be taken into account instead of difference sets. Such structures can be found in the packing problems of  finite Abelian groups \cite{18,19}.
%\begin{Definition}
%A subset $S$ of an Abelian group $G$, where $|S| = k'$, is an $S_t$ -set of size $k$ if all sums of $t$ different elements in $S$ are distinct in the group $G$.
%\end{Definition}
\begin{Definition}
For a given natural number $t$, an $S_t$-set of size $k'$ in the Abelian group $A$ is a subset $S$ of $A$ with $k'$ elements such that all the sums of $t$ different elements in $S$ are distinct in the group $A$.
\end{Definition}

Let $s(A)$ denote the cardinality of the largest $S_2$-set in $A$.  In the study of $S_2$-sets, two central functions are $v(k')$ and $v_{\gamma}(k')$, which give the order of the smallest Abelian group and cyclic group $A$, respectively, for which $s(A)\geq k'$. Since cyclic groups are special cases of the Abelian groups, clearly $v(k') \leq v_{\gamma}(k')$, and any upper bound on $v_{\gamma}(k')$  is also an upper bound on $v(k')$. This is an important point in our work. Indeed, this inequality is equivalent to say that for a given rate $r$, the group ring based QC-LDPC codes from Abelian groups are shorter than cyclic QC-LDPC codes with the same rate. In \cite{18,19}, the values of $v_{\gamma}(k')$ and $v(k')$, for $k' \leq 15$, were determined.

%We see from the proof of Theorem~\ref{theorem2} that when the considered set of the powers of $\alpha$, have the property that the difference of any pair of them are distinct, then the constructed QC-LDPC code based on such a set is $4$-cycle free. Let $D\subset A$ denote the set of the powers. Then, we have the following result.

Due to the proof of Theorem~\ref{theorem2}, when the differences of any pair of elements in a set of powers are distinct,  the constructed QC-LDPC code based on such set is $4$-cycle free. Let $D\subset A$ denote the set of the powers. Then, we have the following result.
\begin{Proposition}\label{prop1}
The difference set of $D$ which is defined as $D-D=\left\lbrace d_1-d_2 |d_1,d_2\in D \right\rbrace$, contains no repetitive element, if and only if $D$ is $S_2$-set.
\end{Proposition}
\begin{IEEEproof}
The proof follows from the fact that if $d_1-d_2=d_3-d_4$, then $d_1+d_4=d_3+d_2$.
\end{IEEEproof}

In the construction of QC-LDPC codes, we are looking for the largest $S_2$-set in an Abelian group, which is denoted by $H$ in the sequel, with cardinality $s(H)$. Our experimental results show that the error performance of the group ring based QC-LDPC codes is related to the following two conditions: 1) increasing the cardinality of the set of the powers, which is equivalent to increasing the row weight of the constructed code, and 2) decreasing the size of the Abelian group $H$, which is equivalent to decreasing the block size $b$, which is also known as the \emph{lifting degree}. These two conditions are fulfilled  if we choose  the largest $S_2$-set in a given Abelian group $H$. Such structures have other applications in the coding theory \cite{26,27,28}. To find an $S_2$-set of maximum size in a given group, symmetries in the structures of $S_2$-sets should be considered. This is the motivation behind considering the concepts of group automorphism and subset equivalence. Several general bounds for the size of $S_2$-sets and  exhaustive computer search results for $s(H)\leq 15$ are presented in \cite{18,19}.

Two subsets $S$ and $S'$ of an Abelian group $H$ are equivalent, if $S = \psi(S)$, where $\psi :H\rightarrow H$ is a function of the form $\psi(x) = \rho(x) + h_0$, in which $\rho \in \mathrm{Aut}(H)$ is an automorphism of $H$, and $h_0 \in H$ is a constant. The \emph{equivalence mappings} $\psi$ form a group which is denoted by $E(H)$ under function composition. They also preserve the property that all sums of the pairs are
distinct.  For a given length, the following theorem helps us to estimate the maximum possible rate of the constructed QC-LDPC code using our method.
\begin{theorem} [\emph{\cite[Theorem 2]{19}}]\label{theorem7}
For a given finite Abelian group $H$, let $S$ be an $S_2$-set in $H$ with $k'$-elements. Then,
\begin{eqnarray}
|H|\geq \left(1- \frac{1}{n_2(H)+1} \right)(k^{\prime 2}-3k'+2),
\end{eqnarray}
where $n_2(H)$ is the index of the subgroup $I(H)$ of $H$ formed by involutions (an involution of $H$ is defined as an element $x$ of $H$ with order 2, i. e., $x^2=1_H$).
\end{theorem}

An algorithm for finding an $S_2$-set with maximum size in an Abelian group, namely the \emph{backtrack search with isomorph rejection}, was proposed in \cite{18,19}. We summarize  the results of \cite{18,19} in \tablename~\ref{table2} and \tablename~\ref{table3}.
%---------------------------------------------------------------------------
\begin{table}[ht]
%\scriptsize
\small
\caption{$S_2$-sets of size  $k'\leq 15$ in cyclic groups}
\centering
\begin{tabular}{c|c|l}
\hline
$k'$  & $v_{\gamma}(k')$ & $S_2$-set\\
  \hline
  \hline
1 & 1 & $\left\lbrace 0\right\rbrace$ \\
2 & 2 & $\left\lbrace 0, 1\right\rbrace$ \\
3 & 3  &$\left\lbrace 0, 1, 2\right\rbrace$ \\
4 & 6  & $\left\lbrace 0, 1, 2, 4\right\rbrace$\\
5 & 11 & $\left\lbrace 0, 1, 2,4, 7\right\rbrace$ \\
6 & 19 & $\left\lbrace 0, 1, 2, 4, 7, 12\right\rbrace$ \\
7 & 28 & $\left\lbrace 0, 1,2, 4, 8,15, 20\right\rbrace,$ \\
 & & $ \left\lbrace 0, 1, 2, 5, 9, 17, 23\right\rbrace$ \\
8 & 40 & $\left\lbrace 0, 1, 5, 7, 9, 20, 23, 35\right\rbrace$\\
9 & 56 &$ \left\lbrace 0, 1, 2, 4, 7, 13, 24, 32, 42\right\rbrace$ \\
10 & 72  & $\left\lbrace 0, 1, 2, 4, 7, 13, 23, 31, 39, 59\right\rbrace$ \\
11 & 96  & $\left\lbrace 0, 1, 2, 4, 10, 16, 30, 37, 50, 55, 74\right\rbrace,$ \\
& & $\left\lbrace 0, 1, 2, 4, 11, 21, 40, 52, 70, 75, 83\right\rbrace,$\\
& & $\left\lbrace 0, 1, 2, 4, 13, 26, 34, 40, 50, 55, 78\right\rbrace,$\\
& & $\left\lbrace 0, 1, 2, 4, 16, 22, 27, 35, 52, 59, 69\right\rbrace$ \\

12 & 114 & $\left\lbrace 0, 1, 4, 14, 22, 34, 39, 66, 68, 77, 92, 108\right\rbrace$ \\
13 & 147 &  $\left\lbrace 0, 1 ,2, 4, 7, 29, 40, 54, 75, 88, 107, 131, 139\right\rbrace$ \\
14  & 178 & $ \left\lbrace 0, 1, 2, 4, 16, 51, 80, 98, 105, 111, 137, 142, 159, 170\right\rbrace$\\
15 & 183 & $\left\lbrace 0, 1, 2, 14 ,18, 21, 27, 52, 81, 86, 91, 128, 139, 161, 169\right\rbrace$ \\
\hline
 \end{tabular}
 \label{table2}
 \end{table}
%---------------------------------------------------------------------------
%----------------------------------------------------------------------------
\begin{table}[ht]
\small
\caption{$S_2$-sets of size  $k'\leq 15$ in Abelian non-cyclic  groups}
\centering
\begin{tabular}{c|c|l|l}
\hline
$k'$  & $v(k')$ & $H$ & $S_2$-set\\
  \hline
  \hline
6 & 16 & $\mathbb{Z}_2^4$& $\left\lbrace (0, 0, 0, 0), (0, 0, 0, 1), (0, 0, 1, 0), \right. $ \\
& &  & $\left. (0, 1, 0, 0),(1, 0, 0, 0), (1, 1, 1, 1) \right\rbrace$\\

6 & 16 &  $\mathbb{Z}_2^2\times \mathbb{Z}_4$ & $\left\lbrace  (0, 0, 0), (0, 0, 1), (0, 0, 2), (0, 1, 1),\right. $ \\
& &  & $\left. (1, 0, 1), (1, 1, 3) \right\rbrace$\\

6 & 16  & $\mathbb{Z}_4^2$ &$\left\lbrace (0, 0), (0, 1), (0, 2), (1, 0), (2, 3), (3, 0) \right\rbrace$ \\
& &  & \\
7 & 24 & $\mathbb{Z}_2^3\times \mathbb{Z}_3$  & $\left\lbrace (0, 0, 0, 0), (0, 0, 1, 1), (0, 0, 0, 1),\right. $\\
& & & $ (0, 1, 0, 0), (1, 0, 0, 0), (1, 1, 1, 0),$ \\
& & &  $\left. (0, 0, 0, 2)\right\rbrace $\\

8 & 40 & $\mathbb{Z}_2\times \mathbb{Z}_4\times \mathbb{Z}_5$ & $\left\lbrace (0, 0, 0), (0, 1, 1), (0, 0, 2), (0, 2, 1), \right.$\\
 & & & $\left. (0, 3, 3), (0, 3, 4), (1, 0, 0), (1, 2, 0) \right\rbrace$ \\

9 & 52 & $\mathbb{Z}_2^2\times \mathbb{Z}_{13}$ & $\left\lbrace (0, 0, 0), (0, 1, 1), (0, 0, 1), (0, 0, 2), \right.$ \\
& & & $ (0, 1, 7), (1, 0, 0), (1, 0, 4), (1, 0, 9), $ \\
& & & $\left.(0, 1, 4)\right\rbrace$\\

11 & 96 & $\mathbb{Z}_2\times \mathbb{Z}_{16}\times \mathbb{Z}_3$ & $\left\lbrace (0, 0, 0), (0, 1, 1), (0, 0, 1), (0, 0, 2),  \right.$\\
& & & $(0, 2, 0),(0, 4, 0), (0, 8, 0), (0, 11, 0), $  \\
& & & $\left. (1, 0, 0), (1, 10, 1), (1, 13, 2) \right\rbrace,$ \\

11 & 96 & $\mathbb{Z}_2^2\times \mathbb{Z}_{8}\times \mathbb{Z}_3$ & $\left\lbrace (0, 0, 0, 0), (0, 0, 1, 1), (0, 0, 0, 1), \right. $\\
& & & $(0, 0, 4, 0),(0, 0, 7, 2), (0, 1, 0, 0),$ \\
& & & $ (0, 1, 3, 0), (0, 1, 6, 0), (1, 0, 0, 2),$\\
& & & $\left.  (1, 0, 2, 0), (1, 0, 5, 1) \right\rbrace,$ \\

13  & 147 & $\mathbb{Z}_3\times \mathbb{Z}_7^2$ & $ \left\lbrace (0, 0, 0), (1, 0, 1), (0, 0, 1), (0, 0, 4), \right.$\\
& & & $  (0, 1, 0),(0, 2, 0), (0, 4, 2), (0, 5, 0), $\\
 & & & $ (1, 1, 2), (1, 6, 4),(2, 0, 1), (2, 3, 2), $ \\
 & & & $\left. (2, 4, 4) \right\rbrace$\\
\hline
 \end{tabular}
 \label{table3}
 \end{table}
 \begin{table}[ht]
\small
\caption{Modified $S_2$-sets of size  $k'\leq 13$ in Abelian non-cyclic  groups}
\centering
\begin{tabular}{c|c|l|l}
\hline
$k'$  & $v(k')$ & $H$ & Modified $S_2$-set\\
  \hline
  \hline
4 & 18 & $\mathbb{Z}_3\times \mathbb{Z}_3\times \mathbb{Z}_2$ & $\left\lbrace (1,2,2),( 1,3 , 2),(2, 2 , 2),(2,3 ,1) \right\} $ \\

5  & 27 & $\mathbb{Z}_3\times \mathbb{Z}_3\times \mathbb{Z}_3$ & $ \left\lbrace (1, 1, 2), (1, 1, 1), (1, 2, 2), (2, 1, 2), \right.$\\
& & & $ \left. (2, 2, 1)\right\}$\\

6  & 48 & $\mathbb{Z}_3\times \mathbb{Z}_4\times \mathbb{Z}_4 $ & $ \left\lbrace (2, 2, 3), (2, 2, 4), (2, 3, 3), (2, 1, 2), \right.$\\
& & & $\left.  (3, 2, 3),(1, 3, 1) \right\rbrace$\\

7 & 72 & $\mathbb{Z}_3\times \mathbb{Z}_4\times \mathbb{Z}_6$ & $\left\lbrace (1, 2, 3), (1, 2, 5), (1, 1, 3), (1, 1, 2), \right.$\\
 & & & $\left. (2, 2, 3), (2, 1, 5), (3, 3, 6)\right\rbrace$ \\

 8 & 84 & $\mathbb{Z}_3\times \mathbb{Z}_4\times \mathbb{Z}_7$ & $\left\lbrace (3, 2, 2), (3, 2, 1), (3, 2, 4), (3, 4, 7), \right.$\\
 & & & $\left. (3, 3, 7), (1, 2, 2), (1, 3, 4),(2,4,1)\right\rbrace$ \\
 9 & 108 & $\mathbb{Z}_3\times\mathbb{Z}_6\times  \mathbb{Z}_{6}$ & $\left\lbrace (2, 4, 1), (2,4,6), (2,5,1), (2,5,5), \right.$ \\
& & & $ (2,1,5), (2,3,4), (3,4,1), (3,2,5), $ \\
& & & $\left.(1,5,2)\right\rbrace$\\
 10 & 144 & $\mathbb{Z}_3\times\mathbb{Z}_6\times  \mathbb{Z}_{8}$ & $\left\lbrace (2, 2, 5), (2,2,6), (2,2,8), (2,4,5), \right.$ \\
& & & $ (2,3,6), (2,1,3), (3,2,5), (3,4,6), $ \\
& & & $\left.(3,1,1),(1,4,1)\right\rbrace$\\
 11 & 168 & $\mathbb{Z}_4\times\mathbb{Z}_6\times  \mathbb{Z}_{7}$ & $\left\lbrace (2, 3, 1), (2,3,3), (2,3,7), (2,1,1), \right.$ \\
& & & $ (2,4,3), (1,3,1), (1,1,3), (1,4,2), $ \\
& & & $\left.(1,6,5),(3,1,4),(4,2,3)\right\rbrace$\\
 12 & 196 & $\mathbb{Z}_4\times\mathbb{Z}_7\times  \mathbb{Z}_{7}$ & $\left\lbrace (3, 7, 5), (3,7,4), (3,7,2), (3,2,5), \right.$ \\
& & & $ (3,3,5), (3,6,3), (1,7,6), (1,4,2), $ \\
& & & $\left.(1,1,3),(4,7,1),(4,2,5),(2,6,1)\right\rbrace$\\
13  & 256 & $\mathbb{Z}_8\times \mathbb{Z}_8\times \mathbb{Z}_4$ & $ \left\lbrace (6,6,4), (6,6,1), (6,4,4), (6,7,4), \right.$\\
& & & $  (6,1,3),(8,6,4), (8,4,2), (7,6,1), $\\
 & & & $ (7,7,2), (4,4,3),(2,3,3), (3,8,4), $ \\
 & & & $\left. (1,1,1) \right\rbrace$\\
\hline
 \end{tabular}
 \label{table4}
 \end{table}
%-------------------------------------------------------------------------------

In the construction of our QC-LDPC codes, we use a subset of an $S_2$-set which has the maximum size and has the following additional condition.
\begin{Definition}
Let $S$ be an $S_2$-set in the Abelian group $H$. A subset $D\subset S$ is called a \emph{modified $S_2$-set} if $2D\cap (D+D)=\varnothing$, where $2D=\left\{2\times d\,\,|\,\, d\in D\right\}$ and $D+D=\left\{d_1+d_2\,\,|\,\,d_1,d_2\in D, d_1\neq d_2\right\}$.
\end{Definition}

From now on, by an $S_2$-set, we mean a  \emph{modified $S_2$-set}.  \tablename~\ref{table4} contains all modified  $S_2$-sets of order less than or equal to $13$. We have presented the results of the Abelian groups which are the direct product of $3$ cyclic groups. Note that the cyclic groups of order $p^{\delta}-1$, for some prime number $p$ and a positive integer $\delta$, can only be used as the components of the direct product. In Section \ref{simulation}, some examples are given indicating the higher rates of the constructed codes based on Abelian group algebras compared  to the constructed codes based on cyclic difference sets.

The results of \tablename~\ref{table4} are not optimized with respect to the group size, since we have considered only the groups which are direct product of three cyclic groups. For example in the group $\mathbb{Z}_8\times \mathbb{Z}_8\times \mathbb{Z}_4$ of size $256$ there is a modified $S_2$-set of size $13$, however, if we use the group $\mathbb{Z}_4^4$ of order $256$, we have the following $S_2$-set $S$ of size $16$
%\small
\begin{IEEEeqnarray}{rCl}\label{S_2-set}
\scriptsize
S &= & \left\lbrace   ( 3   ,  4  ,   1 ,    4),
     (3  ,   4   ,  1   ,  3),
     (3 ,    4 ,    2 ,    4),
     (3 ,    4 ,    4 ,    1),
    ( 3  ,   3 ,    1  ,   4),\right. \nonumber \\
  &&\,\,\, ( 3 ,    1  ,   2 ,    1),
    ( 1  ,   3 ,    1 ,    3),
    ( 1 ,    1 ,    4 ,    4),
   (  4 ,    4 ,    1 ,    4),
   (  4 ,    3 ,    2 ,    4),\nonumber\\
  && \,\,\,  (4 ,    1,     1,     2),
     (4,     2 ,    3,     1),
     (2 ,    4 ,    4,     2),
     (2 ,    3 ,    3 ,    3),
    ( 2    , 1    , 2    , 3),\nonumber\\
&& \,\, \left.  ( 2  ,   2 ,    4 ,    1) \right \rbrace .
\end{IEEEeqnarray}
%\normalsize
\subsection{Construction method}
Now, we are ready to present the method of constructing QC-LDPC codes based on Abelian $S_2$-sets and group rings.
Let $A$ be an Abelian group. By a result attributed to Gauss (Theorem~\ref{theorem6}),  $A$ can be expressed as a direct product of a finite number of cyclic groups of prime power order. Let $A \cong \mathbb{Z}_{p_1^{\beta_1}}\times \mathbb{Z}_{p_2^{\beta_2}}\times \cdots \times \mathbb{Z}_{p_t^{\beta_t}}$ and consider $\mathbb{F}_{q_i}$  as the Galois field with $q_i$-elements and with the primitive element $\alpha_i$, where $q_i=p_i^{\beta_i}$ and $1\leq i\leq t$. We consider the Abelian group $H= \mathbb{Z}_{q_1-1}\times \mathbb{Z}_{q_2-1}\times \cdots \times \mathbb{Z}_{q_t-1}$ of order $b=\prod_{i=1}^t (q_i-1)$. Let $D=\left\lbrace d_1,d_2,\ldots ,d_{k'} \right\rbrace$ be an $S_2$-set with maximum size $k'$ in $H$. Using the map $\Phi$ that maps $\mathbf{h}=(h_1,h_2,\ldots ,h_t)\in H$ to $\boldsymbol{\alpha}^{\mathbf{h}}=(\alpha_1^{h_1},\alpha_2^{h_2}\ldots, \alpha_t^{h_t})$, with $\boldsymbol{\alpha}=(\alpha_1,\ldots,\alpha_t)$, it can be checked that $H$ and $F=\mathbb{F}_{q_1}^*\times \cdots \times \mathbb{F}_{q_t}^*$ are isomorphic as Abelian groups.
We conclude that $\mathbf{h}-\mathbf{h}'\neq \mathbf{z}-\mathbf{z}'$, for $\mathbf{h},\mathbf{h}',\mathbf{z},\mathbf{z}'\in H$,  if and only if $\Phi(\mathbf{h}-\mathbf{h}')\neq \Phi(\mathbf{z}-\mathbf{z}')$. Now, we  generalize the definition of CPM, which is used in the lifting of the finite field based QC-LDPC codes, to the group ring based QC-LDPC codes.
\begin{Definition}
Let $\alpha_i$  be the primitive element of the Galois field $\mathbb{F}_{q_i}$, where $1\leq i\leq t$ and let
\begin{IEEEeqnarray}{rCl}\label{b_eq}
\small
  b= \left(\prod_{i=1}^t(q_i-1)\right).
\end{IEEEeqnarray}
The  \emph{Quasi Circulant Permutation Matrix} of $\boldsymbol{\alpha}^{(i_1,\ldots , i_t)}=(\alpha_1^{i_1},\ldots,\alpha_t^{i_t})$, which is denoted by the QCPM of $ \boldsymbol{\alpha}^{(i_1,\ldots,i_t)}$, with $0\leq i_j\leq q_j-2$ and $1\leq j\leq t$, is a $b\times b$ matrix
%$$b= \left(\prod_{i=1}^t(q_i-1)\right),$$
which is defined as
\begin{equation}\label{QCPM}
%\small
\mathrm{QCPM}(\boldsymbol{\alpha}^{(i_1,\ldots , i_t)}) :=\mathrm{CPM}(\alpha_1^{i_1},\mathbb{F}_{q_1})\otimes \cdots \otimes  \mathrm{CPM}(\alpha_t^{i_t},\mathbb{F}_{q_t}),
\end{equation}
%\normalsize
in which $\otimes$ denotes the  the Kronecker product\footnote{If $\mathbf{A}$ is an $m\times n$ matrix and $\mathbf{B}$ is a $p\times q$ matrix, then the Kronecker product $\mathbf{A}\otimes \mathbf{B}$ is the $mp\times nq$ block matrix:
\begin{equation*}%\label{Kron_prod}
  \mathbf{A}\otimes \mathbf{B}=\left[
                                 \begin{array}{ccc}
                                   a_{1,1}\mathbf{B} & \cdots & a_{1,n}\mathbf{B} \\
                                   \vdots & \ddots & \vdots \\
                                   a_{m,1}\mathbf{B} & \cdots & a_{m,n}\mathbf{B} \\
                                 \end{array}
                               \right].
\end{equation*}
} of  matrices, the matrix $\mathrm{CPM}(\alpha_j^{i_j},\mathbb{F}_{q_j})$, for $1\leq j\leq t$, is a $(q_j-1)\times(q_j-1)$ circulant matrix and its first row is the location-vector of the element $\alpha_j^{i_j}$ with respect to the multiplicative group of $\mathbb{F}_{q_j}$. We denote the zero element of $\mathbb{F}_{q_j}$ by $\alpha_j^{-\infty}$. If for some $1\leq j\leq t$ in (\ref{QCPM}), $i_j=-\infty$, then $\mathrm{QCPM}(\boldsymbol{\alpha}^{(i_1,\ldots , i_t)})$  is defined  as    the $b \times b$ zero matrix.
\end{Definition}

\begin{Example}
We want to obtain the  QCPM of $\boldsymbol{\alpha}^{(1,3)}$ in $\mathbb{F}_4\times \mathbb{F}_5$, which is a $12\times 12$ matrix as follows
\begin{IEEEeqnarray*}{rCl}
\small
\mathrm{QCPM}(\boldsymbol{\alpha}^{(1,3)})&=&\mathrm{CPM}(\alpha_1,\mathbb{F}_4)\otimes \mathrm{CPM}(\alpha_2^3,\mathbb{F}_5)\\
&=& \left[\begin{array}{cccc}
0  &   1  &   0  \\
0 &   0  &   1   \\
1 &    0 &    0
\end{array}
\right]\otimes  \left[\begin{array}{cccc}
0  &   0  &   0  &   1 \\
1&   0  &   0  &   0 \\
0 &    1 &    0  &   0 \\
0 &    0   &  1  &   0
\end{array}
\right].
%\\
%&&= \left[\begin{array}{cccccccccccc}
%     0 &   0 &    0 &    0 &    0 &    0 &    0 &    1 &    0 &    0 &    0 &    0 \\
%     0 &   0 &    0 &    0 &    1 &    0 &    0 &    0 &    0 &    0 &    0  &   0 \\
%     0 &   0 &    0  &   0 &    0  &   1 &    0 &    0  &   0    & 0    & 0    & 0 \\
%     0  &   0&     0 &    0 &    0 &    0     &1 &     0 &    0  &   0 &    0   &  0 \\
%     0   &  0  &   0 &    0 &    0  &   0  & 0 &    0  &   0  &   0 &    0 &    1 \\
%     0 &    0 &    0  &   0    & 0 &    0   &  0  &   0  &   1   &  0 &    0   &  0 \\
%     0   &  0   &  0  &   0   &  0  &   0   &  0 &    0 &    0 &    1  &   0   &  0 \\
%     0  &   0   &  0   &  0 &   0 &    0   &  0   &  0  &   0  &   0  &   1   &  0 \\
%     0 &    0   &  0 &    1 &    0 &    0  &   0  &   0  &   0  &   0   &  0  &   0 \\
%     1 &    0  &   0   &  0  &   0  &   0  &   0  &   0 &    0  &   0  &   0  &   0 \\
%     0  &   1 &    0  &   0  &   0   &  0  &   0   &  0   &  0   &  0  &   0   &  0 \\
%     0  &   0 &    1    & 0 &    0   &  0   &  0    & 0  &   0   &  0  &   0  &   0
%\end{array}\right],
\end{IEEEeqnarray*}
%\normalsize
By replacing each component $a_{i,j}$ of $\mathrm{CPM}(\alpha_1,\mathbb{F}_4)$, with $a_{i,j}\times \mathrm{CPM}(\alpha_2^3,\mathbb{F}_5)$, we obtain the following matrix
\begin{eqnarray*}
%\scriptstyle\scriptsize
\mathrm{QCPM}(\boldsymbol{\alpha}^{(1,3)})= \left[\begin{array}{cccccccccccc}
     0 &   0 &    0 &    0 &    0 &    0 &    0 &    1 &    0 &    0 &    0 &    0 \\
     0 &   0 &    0 &    0 &    1 &    0 &    0 &    0 &    0 &    0 &    0  &   0 \\
     0 &   0 &    0  &   0 &    0  &   1 &    0 &    0  &   0    & 0    & 0    & 0 \\
     0  &   0&     0 &    0 &    0 &    0     &1 &     0 &    0  &   0 &    0   &  0 \\
     0   &  0  &   0 &    0 &    0  &   0  & 0 &    0  &   0  &   0 &    0 &    1 \\
     0 &    0 &    0  &   0    & 0 &    0   &  0  &   0  &   1   &  0 &    0   &  0 \\
     0   &  0   &  0  &   0   &  0  &   0   &  0 &    0 &    0 &    1  &   0   &  0 \\
     0  &   0   &  0   &  0 &   0 &    0   &  0   &  0  &   0  &   0  &   1   &  0 \\
     0 &    0   &  0 &    1 &    0 &    0  &   0  &   0  &   0  &   0   &  0  &   0 \\
     1 &    0  &   0   &  0  &   0  &   0  &   0  &   0 &    0  &   0  &   0  &   0 \\
     0  &   1 &    0  &   0  &   0   &  0  &   0   &  0   &  0   &  0  &   0   &  0 \\
     0  &   0 &    1    & 0 &    0   &  0   &  0    & 0  &   0   &  0  &   0  &   0
\end{array}\right].
\end{eqnarray*}
%\normalsize
$\hfill\square$\end{Example}

In order to complete our construction method, we consider the ring $R$ which is formed by the $\mathbb{Z}$-linear combination of the basis elements of the form $\alpha_1^{i_1}\otimes \cdots \otimes \alpha_t^{i_t}$, where $0\leq i_j\leq q_j-2$ and $1\leq j\leq t$. Alike the previous cases, the construction of QC-LDPC codes from Abelian group rings needs an $m\times n$ matrix over $R$, whose rows $\mathbf{W}_1,\ldots,\mathbf{W}_m$, satisfy the following two constraints:
\begin{enumerate}
  \item For $0\leq i\leq m-1$ and $\mathbf{k}=(k_1,\ldots,k_t),\mathbf{l}=(l_1,\ldots,l_t)\in\mathbb{Z}_{q_1-1}\times\cdots \times \mathbb{Z}_{q_t-1}$ with  $k_j\neq l_j$ and $j=1,\ldots,t$, the two vectors $\boldsymbol{\alpha}^{\mathbf{k}} \mathbf{W}_i$ and $\boldsymbol{\alpha}^{\mathbf{l}} \mathbf{W}_i$ have at most one position where both of them have the same symbol from $R$, (i.e., they differ in at least $n - 1$ positions).
  \item For $0\leq i,j\leq m-1$,  $i\neq j$, and $\mathbf{k}=(k_1,\ldots,k_t),\mathbf{l}=(l_1,\ldots,l_t)\in\mathbb{Z}_{q_1-1}\times\cdots \times \mathbb{Z}_{q_t-1}$, the two vectors $\boldsymbol{\alpha}^{\mathbf{k}} \mathbf{W}_i$ and $\boldsymbol{\alpha}^{\mathbf{l}} \mathbf{W}_j$  differ in at least $n - 1$ positions.
\end{enumerate}
We call these conditions the  $\boldsymbol{\alpha}$-multiplied constraints $1$ and $2$, respectively.
Note that the multiplication of $\alpha_1^{i_1}\otimes \cdots \otimes \alpha_t^{i_t}$ and $\alpha_1^{j_1}\otimes \cdots \otimes \alpha_t^{j_t}$ is defined as $\alpha_1^{k_1}\otimes \cdots \otimes \alpha_t^{k_t}$, where $k_r=i_r+j_r \pmod{q_r-1}$, for $1\leq r \leq t$.
We also denote $\alpha_1^{i_1}\otimes \cdots \otimes \alpha_t^{i_t}$ by $\boldsymbol{\alpha}^{(i_1,\ldots,i_t)}$. Based on the aforementioned notations and definitions, we have the following theorem.
\begin{theorem}\label{main_theorem}
Let $D=\left\lbrace \mathbf{d}_0,\mathbf{d}_2,\ldots,\mathbf{d}_{n-1} \right\rbrace$ be an $S_2$-set in the  Abelian group $H=\mathbb{Z}_{q_1-1}\times \cdots, \times \mathbb{Z}_{q_t-1}$, $G=\left\{g_0=1_G,g_1,g_2,\cdots g_{n-1}\right\}$ be a finite group of order $n$ and $\mathbb{F}_{q_i}$ be the Galois field of order $q_i$  with the primitive element $\alpha_i$. Let $\boldsymbol{\alpha}=(\alpha_1,\ldots, \alpha_t)$ and consider the ring $R$ which is formed by $\mathbb{Z}$-linear combination of the basis elements of the form $\alpha_1^{i_1}\otimes \cdots \otimes \alpha_t^{i_t}$, where $0\leq i_j\leq q_j-2$ and $1\leq j\leq t$. If $(q_i-1)$'s, for  $1\leq i\leq t$, are odd numbers, then the $RG$-matrix corresponding to the element of the form $w=\sum_{i=0}^{n-1}\boldsymbol{\alpha}^{\mathbf{d}_i}g_i$ gives  an $n\times n$ matrix  $\mathbf{W}$ that satisfies the $\boldsymbol{\alpha}$-multiplied constraints $1$ and $2$. By replacing each component of $\mathbf{W}$ by its corresponding $\prod_{i=1}^t(q_i-1)\times\prod_{i=1}^t(q_i-1)$ QCPM and choosing a subarray of $\mathbf{W}$,   we obtain the parity-check matrix of a $4$-cycle free QC-LDPC code.
\end{theorem}
\begin{IEEEproof}
Since all the rows $\mathbf{W}_1,\ldots,\mathbf{W}_{m-1}$ are obtained from the permutations of the first row $\mathbf{W}_0$, it is enough to show that the first $\boldsymbol{\alpha}$-multiplied constraint is fulfilled for  $\mathbf{W}_0=(\boldsymbol{\alpha}^{\mathbf{d}_1},\boldsymbol{\alpha}^{\mathbf{d}_2},\ldots,\boldsymbol{\alpha}^{\mathbf{d}_k})$. Let $\mathbf{k}=(k_1,\ldots,k_t),\mathbf{l}=(l_1,\ldots,l_t)\in\mathbb{Z}_{q_1-1}\times\cdots \times \mathbb{Z}_{q_t-1}$, with $\mathbf{l}\neq \mathbf{k}$, be such that $\boldsymbol{\alpha}^{\mathbf{l}}\mathbf{W}_0$ and $\boldsymbol{\alpha}^{\mathbf{k} }\mathbf{W}_0$ have more than one position in common. Then, for some $1\leq i\leq n$ and $1\leq  j\leq n$, $\boldsymbol{\alpha}^{\mathbf{l}}\boldsymbol{\alpha}^{\mathbf{d}_i}=\boldsymbol{\alpha}^{\mathbf{k}} \boldsymbol{\alpha}^{\mathbf{d}_{i}}$ and  $\boldsymbol{\alpha}^{\mathbf{l}}\boldsymbol{\alpha}^{\mathbf{d}_j}=\boldsymbol{\alpha}^{\mathbf{k}} \boldsymbol{\alpha}^{\mathbf{d}_{j}}$ which yield the following equations
\begin{eqnarray*}
\mathbf{l}+\mathbf{d}_{i}&=&\mathbf{k}+\mathbf{d}_{i},\\
\mathbf{l}+\mathbf{d}_{j}&=&\mathbf{k}+\mathbf{d}_{j}.
\end{eqnarray*}
Consequently, it follows that $\mathbf{k}=\mathbf{l}$, which is a contradiction. For the second constraint, assume that $\boldsymbol{\alpha}^{\mathbf{l}}\mathbf{W}_0$ and $\boldsymbol{\alpha}^{\mathbf{k}} \mathbf{W}_i$, where $1\leq i\leq m-1$, have more than one position in common.
Then, for some $1\leq i<j\leq n$ and $1\leq  i'< j'\leq n$, $\boldsymbol{\alpha}^{\mathbf{l}}\boldsymbol{\alpha}^{\mathbf{d}_i}=\boldsymbol{\alpha}^{\mathbf{k}} \boldsymbol{\alpha}^{\mathbf{d}_{i'}}$ and  $\boldsymbol{\alpha}^{\mathbf{l}}\boldsymbol{\alpha}^{\mathbf{d}_j}=\boldsymbol{\alpha}^{\mathbf{k}} \boldsymbol{\alpha}^{\mathbf{d}_{j'}}$ which imply the following equations
 \begin{eqnarray*}
\mathbf{l}+\mathbf{d}_{i}&=&\mathbf{k}+\mathbf{d}_{i'},\\
\mathbf{l}+\mathbf{d}_{j}&=&\mathbf{k}+\mathbf{d}_{j'}.
\end{eqnarray*}
Thus, we have $\mathbf{d}_i-\mathbf{d}_{i'}=\mathbf{d}_j-\mathbf{d}_{j'}$. We also know that $i\neq i'$ and $j\neq j'$. If $i\neq j'$ and $i'\neq j$, we get a contradiction with the assumption that $D$ is an $S_2$-set. Note that $\mathbf{l}\neq \mathbf{k}$ implies that $i\neq i'$ and $j\neq j'$. If $i=j'$ and $i'\neq j$ or if $i'=j$ and $i\neq j'$ we conclude that $2\mathbf{d}_i=\mathbf{d}_i'+\mathbf{d}_j$ or $2\mathbf{d}_j=\mathbf{d}_i+\mathbf{d}_j'$ which is a contradiction, since $D$ is an $S_2$-set. If $i=j'$ and $j=i'$, we have
$
  2\mathbf{d}_i = 2\mathbf{d}_j \bmod (q_1-1,\ldots, q_t-1),$
where, $\mathbf{d}_i=(d_{i,1},\ldots,d_{i,t})$, $\mathbf{d}_j=(d_{j,1},\ldots,d_{j,t})$ and the $\bmod$ operation is applied componentwise. Since $(q_i-1)$ is an odd number for $1\leq i\leq t$,  it follows that $\mathbf{d}_i=\mathbf{d}_j$ which is a contradiction.
Thus, if we replace the components of $\mathbf{W}$  with their corresponding QCPMs, we obtain the parity-check matrix of a $4$-cycle free QC-LDPC code.
\end{IEEEproof}

If we remove the repetitive members in the set $2D$, then we obtain an $S_2$-set which satisfies the conditions of Theorem~\ref{main_theorem} without requiring that $(q_i-1)$'s should be odd numbers.
Similar to the cyclic case, we present the parity-check matrix of the code by an array which consists of the powers of QCPMs. For example $\alpha_1^{i_1}\otimes \cdots \otimes \alpha_t^{i_t}$ in the parity-check matrix is denoted by $(i_1,\ldots,i_t)$. We present the steps of Theorem~\ref{main_theorem} in the next example. Although, all the assumptions of Theorem~\ref{main_theorem} are not fulfilled in this example, it is a useful example to see our method for constructing QC-LDPC codes.
\begin{Example}
Consider the Abelian  $S_2$-set $D=\left\lbrace (0, 0, 0, 0), \right. \\ \left. (0, 0, 0, 1), (0, 0, 1, 0), (0, 1, 0, 0),(1, 0, 0, 0), (1, 1, 1, 1) \right\rbrace$ in  $\mathbb{Z}_2^4$. There are two groups of order $6$, namely $\mathbb{Z}_6$ and $S_6$. Consider $G=\mathbb{Z}_6$ as the underling group. The parity-check matrix of the QC-LDPC code  based on Theorem~\ref{main_theorem}, can be constructed  by taking some rows of following matrix and replacing its components with their corresponding QCPMs:
%\tiny
%\begingroup\makeatletter\def\f@size{5}\check@mathfonts
\begin{eqnarray*}
%\scriptscriptstyle\tiny
\mathbf{W}=
\left[ \begin{array}{cccccc}
(0, 0, 0, 0) & (0, 0, 0, 1) & (0, 0, 1, 0) & (0, 1, 0, 0) & (1, 0, 0, 0) & (1, 1, 1, 1) \\
 (1, 1, 1, 1) & (0, 0, 0, 0) & (0, 0, 0, 1) & (0, 0, 1, 0) & (0, 1, 0, 0) & (1, 0, 0, 0) \\
  (1, 0, 0, 0) & (1, 1, 1, 1) & (0, 0, 0, 0) & (0, 0, 0, 1) & (0, 0, 1, 0) & (0, 1, 0, 0)\\
  (0, 1, 0, 0) & (1, 0, 0, 0) & (1, 1, 1, 1) & (0, 0, 0, 0) & (0, 0, 0, 1) & (0, 0, 1, 0) \\
 (0, 0, 1, 0) & (0, 1, 0, 0) & (1, 0, 0, 0) & (1, 1, 1, 1) & (0, 0, 0, 0) & (0, 0, 0, 1)  \\
 (0, 0, 0, 1) &  (0, 0, 1, 0) & (0, 1, 0, 0) & (1, 0, 0, 0) & (1, 1, 1, 1) & (0, 0, 0, 0)
\end{array}\right].
\end{eqnarray*}
%\endgroup
%\normalsize
As an example, the QCPM of $(0, 0, 1, 0)$ is
%\tiny
\begin{eqnarray*}
%\scriptscriptstyle\tiny
\mathrm{QCPM}(\boldsymbol{\alpha}^{(0, 0, 1, 0)})
%\scriptscriptstyle\scriptsize
=\left[ \begin{array}{cccccccccccccccc}
     0   &  0  &   1   &  0  &   0 &    0   &  0   &  0   &  0  &   0  &   0   &  0  &   0 &    0  &   0  &   0\\
     0 &    0 &    0    & 1&     0  &   0&     0 &    0 &    0   &  0  &   0 &    0 &    0 &    0  &   0   &  0\\
     1  &   0    & 0  &   0   &  0  &   0  &   0 &    0    & 0    & 0  &   0 &    0&     0  &   0   &  0 &    0\\
     0   & 1  &   0  &   0    & 0   &  0&     0 &    0  &   0 &    0   &  0   &  0  &   0   &  0   &  0  &   0\\
     0    & 0   &  0 &    0   &  0    & 0  &   1  &   0   &  0&     0   & 0  &   0   &  0&     0&     0    & 0\\
     0  &   0 &    0    & 0 &    0   &  0&     0 &    1    & 0 &    0  &   0&     0 &    0    & 0  &   0   &  0\\
     0   &  0   &  0   &  0   &  1   &  0    & 0   &  0   &  0   &  0  &   0 &    0   &  0  &   0  &   0   &  0\\
     0  &   0   &  0  &   0   &  0 &    1   &  0  &   0 &    0  &   0  &   0 &    0     &0   &  0  &   0&     0\\
     0  &   0&     0  &   0  &   0   &  0   &  0    & 0     &0  &   0   &  1  &   0   &  0  &   0  &   0   &  0\\
     0   &  0 &    0    & 0&     0   &  0  &   0 &    0   &  0    & 0   &  0&     1  &   0    & 0 &    0    & 0\\
     0   &  0  &   0   &  0 &    0   &  0   &  0   &  0   &  1  &   0  &   0   &  0  &   0 &    0&     0   &  0\\
     0  &   0   &  0   &  0   &  0   &  0   &  0   &  0   &  0  &   1 &    0   &  0&     0  &   0    & 0  &   0\\
     0   &  0   &  0  &   0    & 0  &   0  &   0   &  0  &   0  &   0 &    0   &  0  &   0   &  0   &  1   &  0\\
     0  &   0    & 0   &  0   &  0  &   0  &   0  &   0  &   0  &   0  &   0   &  0   &  0   &  0    & 0   &  1\\
     0 &    0  &   0  &   0   &  0   &  0   &  0  &   0  &   0  &   0 &    0  &   0   &  1  &   0   &  0   &  0\\
     0 &    0  &   0   &  0  &   0&     0  &   0    & 0    & 0  &   0  &   0  &   0   &  0  &   1   &  0   &  0
\end{array}\right].
\end{eqnarray*}
%\normalsize
If we consider the first three rows of $\mathbf{W}$, then we obtain a QC-LDPC code of length $96$ and rate at least $0.5$.
$\hfill\square$\end{Example}

Using Theorem~\ref{main_theorem} and the proposed method in \cite{15}, for constructing QC-LDPC codes from cyclic difference sets, we get the following theorem.
\begin{theorem}\label{main_theorem2}
Let $D=\left\lbrace \mathbf{d}_0,\mathbf{d}_2,\ldots,\mathbf{d}_{n-1} \right\rbrace$ be an $S_2$-set in the  Abelian group $H=\mathbb{Z}_{q_1-1}\times \cdots, \times \mathbb{Z}_{q_t-1}$, $G=\left\{g_0=1_G,g_1,g_2,\cdots g_{n-1}\right\}$ be a finite group of order $n$ and $\mathbb{F}_{q_i}$ be the Galois field of order $q_i$  with the primitive element $\alpha_i$. Let $\boldsymbol{\alpha}=(\alpha_1,\ldots, \alpha_t)$ and consider the ring $R$ which is formed by $\mathbb{Z}$-linear combination of the basis elements of the form $\alpha_1^{i_1}\otimes \cdots \otimes \alpha_t^{i_t}$, where $0\leq i_j\leq q_j-2$ and $1\leq j\leq t$. If $(q_i-1)$'s, for  $1\leq i\leq t$ are odd numbers, (or $2D$ contains no repetitive elements), and $\mathbf{W}$ and $\mathbf{W}^{-1}$  are the $RG$-matrices corresponding to the elements $w=\sum_{i=0}^{n-1}\boldsymbol{\alpha}^{\mathbf{d}_i}g_i$ and $w^{-1}=\sum_{i=0}^{n-1}\boldsymbol{\alpha}^{-\mathbf{d}_i}g_i$, respectively, then $\mathbf{W}'=\left[
                                             \begin{array}{c|c}
                                               \mathbf{W} & \mathbf{W}^{-1}  \\
                                             \end{array}
                                           \right]$
gives an $n\times 2n$ matrix  that satisfies  the $\boldsymbol{\alpha}$-multiplied constraints $1$ and $2$. By replacing each component of $\mathbf{W}'$ with its corresponding $\prod_{i=1}^t(q_i-1)\times\prod_{i=1}^t(q_i-1)$ QCPM and choosing a subarray of $\mathbf{W}'$,   we obtain  the parity-check matrix of a $4$-cycle free QC-LDPC code.
\end{theorem}
\begin{IEEEproof}
Since all the rows $\mathbf{W}_1',\ldots,\mathbf{W}_{m-1}'$ are obtained from the  permutations of the first row $\mathbf{W}_0'$, it is enough to show that the first $\boldsymbol{\alpha}$-multiplied constraint is fulfilled for  $\mathbf{W}_0'=(\boldsymbol{\alpha}^{\mathbf{d}_1},\ldots,\boldsymbol{\alpha}^{\mathbf{d}_n}|\boldsymbol{\alpha}^{-\mathbf{d}_1},\ldots,\boldsymbol{\alpha}^{-\mathbf{d}_n})$. Let $\mathbf{k}=(k_1,\ldots,k_t)$ and $\mathbf{l}=(l_1,\ldots,l_t)$ in $\mathbb{Z}_{q_1-1}\times\cdots \times \mathbb{Z}_{q_t-1}$, with $\mathbf{l}\neq \mathbf{k}$, be such that $\boldsymbol{\alpha}^{\mathbf{l}}\mathbf{W}_0'$ and $\boldsymbol{\alpha}^{\mathbf{k}} \mathbf{W}_0'$ have more than one position in common. Based on Theorem~\ref{main_theorem}, both of these common positions cannot be in the first $n$ positions or in the last $n$ positions. For some $1\leq i\leq n$ and $n+1\leq  j\leq 2n$, let $\boldsymbol{\alpha}^{\mathbf{l}}\boldsymbol{\alpha}^{\mathbf{d}_i}=\boldsymbol{\alpha}^{\mathbf{k}} \boldsymbol{\alpha}^{\mathbf{d}_{i}}$ and  $\boldsymbol{\alpha}^{\mathbf{l}}\boldsymbol{\alpha}^{-\mathbf{d}_j}=\boldsymbol{\alpha}^{\mathbf{k}} \boldsymbol{\alpha}^{-\mathbf{d}_{j}}$ which results in the following equations
\begin{eqnarray*}
\mathbf{l}+\mathbf{d}_{i}&=&\mathbf{k}+\mathbf{d}_{i},\\
\mathbf{l}-\mathbf{d}_{j}&=&\mathbf{k}-\mathbf{d}_{j}.
\end{eqnarray*}
Consequently, we have $\mathbf{k}=\mathbf{l}$, which is a contradiction. For the second constraint, assume that $\boldsymbol{\alpha}^{\mathbf{l}}\mathbf{W}_0'$ and $\boldsymbol{\alpha}^{\mathbf{k}} \mathbf{W}_i'$, where $1\leq i\leq m-1$, have more than one position in common.
Then, for some $1\leq i\neq i'\leq n$ and $n+1\leq  j\neq j'\leq 2n$, $\boldsymbol{\alpha}^{\mathbf{l}}\boldsymbol{\alpha}^{\mathbf{d}_i}=\boldsymbol{\alpha}^{\mathbf{k}} \boldsymbol{\alpha}^{\mathbf{d}_{i'}}$ and  $\boldsymbol{\alpha}^{\mathbf{l}}\boldsymbol{\alpha}^{\mathbf{d}_j}=\boldsymbol{\alpha}^{\mathbf{k}} \boldsymbol{\alpha}^{-\mathbf{d}_{j'}}$ which results in the following equations
 \begin{eqnarray*}
\mathbf{l}+\mathbf{d}_{i}&=&\mathbf{k}+\mathbf{d}_{i'},\\
\mathbf{l}-\mathbf{d}_{j}&=&\mathbf{k}-\mathbf{d}_{j'}.
\end{eqnarray*}
It follows that  $\mathbf{d}_i-\mathbf{d}_{i'}=\mathbf{d}_j'-\mathbf{d}_{j}$. If $i\neq j'$ and $i'\neq j$, we get a contradiction with the assumption that $D$ is an $S_2$-set. If $i=j$ and $i'\neq j'$ or if $i'=j'$ and $i\neq j$ we conclude that $2\mathbf{d}_i=\mathbf{d}_i'+\mathbf{d}_j'$ or $2\mathbf{d}_j'=\mathbf{d}_i+\mathbf{d}_j$ which is a contradiction, since $D$ is an $S_2$-set. If $i=j$ and $i'=j'$, we have
$
  2\mathbf{d}_i = 2\mathbf{d}_i' \bmod (q_1-1,\ldots, q_t-1)$. Since $(q_i-1)$ is an odd number for $1\leq i\leq t$, $\mathbf{d}_i=\mathbf{d}_i'$ which is a contradiction.
\end{IEEEproof}

To describe the construction of Theorem~\ref{main_theorem2} by group-rings representations, we can use the following theorem.
\begin{theorem}[\emph{\cite[Lemma 3.4]{29}}]\label{passman}
Let $G$ and $H$ be two groups and let $\mathbb{K}$ be a field. Then $$\mathbb{K}[G]\otimes_{\mathbb{K}} \mathbb{K}[H]\cong \mathbb{K}[G\times H].$$
\end{theorem}

We can replace the field $\mathbb{K}$ in Theorem~\ref{passman} by any commutative ring with identity. Thus, the group ring that describes the construction of Theorem~\ref{main_theorem2} is $R[G']=R[C_2\times G]$, where $C_2=\left\{1,-1\right\}$ is a multiplicative group of order $2$. In this case, we have considered the first $n$ rows of an $RG'$-matrix.
\subsection{Achievable parameters of the group ring based QC-LDPC codes}
Here, we explain the important parameters of the obtained codes using Abelian group rings. These parameters are compared with the achievable parameters of the other construction methods, namely the ones based on finite fields. The parameters that we have considered for our analysis are the length and the rate of the code. We only consider the girth $6$ QC-LDPC codes for our comparisons.  First, we consider the construction methods based on finite fields. We conclude the following result from \cite[Corollary 1]{banihashemi} and we use it to estimate the values of rate and length that can be achieved by using the designed QC-LDPC codes based on the finite fields approaches.
%\begin{Proposition}[\emph{\cite[Corollary 1]{banihashemi}}]\label{prop2}
\begin{Proposition}\label{prop2}
In constructing a QC-LDPC code with cyclic lifting degree $b$ using an $m\times n$ exponent matrix $\mathbf{B}$, with $m\leq n$, that does not contain $-\infty$ in the components, a necessary condition
for having a girth  at least $6$ in the Tanner graph is $b\geq n$.
\end{Proposition}

When we apply Proposition~\ref{prop2} to  finite field based QC-LDPC codes, we reach  the upper bound $n\leq q-1$ on the row weight of the code. The construction of QC-LDPC codes based on Latin squares over finite fields, which is proposed in \cite{Latin_square}, and the proposed construction methods in \cite{QC_cyclic}, achieve this upper bound. Thus, construction of QC-LDPC codes with lengths $\gamma(q-1)$ and rates $1-\frac{\rho}{\gamma}$ is possible, in which $q$ is a prime power  and $1\leq\gamma,\rho\leq q-1$.
%However, we  found experimentally that in the design of  low-rate (smaller than $0.6$) QC-LDPC codes based on the proposed approaches in \cite{Latin_square,QC_cyclic}, we do not necessarily come up with a good error performance. In order to improve the error performance of their low rates codes, the authors of \cite{Latin_square}  designed the codes with rate $0.5$ by using the masking techniques that generate irregular codes with acceptable performances.

As explained  above, by using the approaches based on finite fields, the lifting degree $b$ is of the form $p^{\beta}-1$, for a prime number $p$ and a positive integer $\beta$. For example, the achievable values of $b$, which are smaller than $100$, are
\begin{equation*}
\begin{array}{ccccccccccc}
     2 & 3 & 4 & 6 & 7 & 8 & 10 & 12 & 15 & 16 & 18, \\
     22 & 24 & 26 & 28 & 31 & 35 & 36 & 40 & 42 & 46 & 48, \\
     52 & 58 & 63 & 66 & 70 & 72 & 78 & 80 & 82 & 88 & 96.
   \end{array}
\end{equation*}
Thus, only $33\%$ of the possible values for $b$ can be achieved by using the approaches based on finite fields. When we employ the Abelian group rings,  the lifting degree $b$ is of the form $\prod_{i=1}^t \left(p_i^{\beta_i}-1\right)$, in which  $p_i$'s are distinct prime numbers and  $\beta_i$'s and $t$  are positive integers. In this case, the achievable values for $b$, which are smaller than $100$, are
\begin{equation*}
%\scriptsize
\begin{array}{ccccccccccc}
     2 & 3 & 4 & 6 & 7 & 8  & \Circle{$9$} & 10 & 12 & \Circle{$14$} & 15\\
     16 & 18 &\Circle{$21$} & 22 & 24 & \Circle{$27$} & 26 & 28 & \Circle{$30$}& 31 & \Circle{$32$}\\
     35 & 36 & 40 & 42 & \Circle{$45$} & 46 & 48 & \Circle{$49$}& 52 & \Circle{$54$} & \Circle{$56$}\\
     58 & \Circle{$60$} & \Circle{$62$} & 63 & \Circle{$64$} & 66 & 70 & 72 & 78 & 80 & \Circle{$81$}\\
     82 & \Circle{$84$} & 88 & \Circle{$90$}& \Circle{$92$} & \Circle{$93$} & 96  & \Circle{$98$}\mbox{\large ,} & & &
   \end{array}
\end{equation*}
where the circled values are the ones that cannot be obtained by using the approaches based on finite fields. This indicates $57\%$ increase in the number of achievable values for $b$, compared to the one for finite fields.

Using Theorem~\ref{theorem7}, we  find an upper bound on the row weight of the group ring based QC-LDPC codes. Let $H$ be an Abelian group used in our construction. Let $s(H)$ denote the cardinality of the largest $S_2$-set in $H$, which gives $2s(H)$ as the maximum achievable row weight of the group ring based QC-LDPC codes based on $H$. Then, by using Theorem~\ref{theorem7},  it follows that
%\begin{equation}\label{s_H_upper_bound}
%  s(H)\leq \frac{3+\sqrt{9-4(2-h_y(H))}}{2}=\frac{3+\sqrt{1+4h_y(H)}}{2},
%\end{equation}
\begin{IEEEeqnarray}{rCl}\label{s_H_upper_bound}
  s(H)&\leq& \left\lfloor \frac{3+\sqrt{9-4(2-h_y(H))}}{2}\right\rfloor\nonumber\\
  &=&\left\lfloor\frac{3+\sqrt{1+4h_y(H)}}{2}\right\rfloor,
\end{IEEEeqnarray}
where $\left\lfloor x \right\rfloor$, for a real number $x$, denotes the largest integer less than or equal to $x$ and
\begin{equation}\label{h_y}
  h_y(H)=\frac{\left|H\right|\left(n_2(H)+1\right)}{n_2(H)}.
\end{equation}
Finding the Abelian groups that achieve this upper bound, is an interesting problem.  When $\left|H\right|$ is large enough, we can assume $h_y(H)\approx  \left|H\right|$ and $(\ref{s_H_upper_bound})$ will be an upper bound in terms of $\left|H\right|$. Consequently, construction of QC-LDPC codes with lengths $\gamma\prod_{i=1}^t \left(p_i^{\beta_i}-1\right)$ and rates $1-\frac{\rho}{\gamma}$ is possible, in which $\gamma,\rho$ are integers with $1\leq\gamma\leq 2s(H)$ and $1\leq\rho\leq s(H)$,  the $p_i$'s are distinct prime numbers and the $\beta_i$'s and $t$  are positive integers.
\section{A New Encoding of QC-LDPC Codes Based on the Multiplication of Group Algebras}\label{sec5_2}
In general, the quasi-cyclic codes  are encoded by multiplying a message vector $\mathbf{m}$ of length $kb$ by a $(kb\times nb)$ generator matrix $\mathbf{G}$, where $\mathbf{G}$  is usually in systematic form, i.e., $\mathbf{G} = \left[
                                            \begin{array}{cc}
                                              \mathbf{I}_{kb} & \mathbf{P}_{kb\times (n-k)b} \\
                                            \end{array}
                                          \right]$, where
$\mathbf{I}_{kb}$ is  the $kb\times kb$ identity matrix. There are two difficulties in the implementation of this encoding procedure. First, the generator matrix $\mathbf{G}$ is usually a dense matrix and requires a large number of memory units, i.e., $b^2(n-k)k$ units, to store $\mathbf{P}$. Second, although the encoding of QC codes can be partially parallelized so that the computation units are reduced by a factor of $b$, the total number of symbol operations is still $b^2(n-k)k$, which is the same as that for general linear codes. In \cite{4}, an efficient encoding method has been proposed for QC-LDPC codes. The authors of \cite{4}  computed a generator matrix $\mathbf{G}$ with quasi-cyclic structure benefiting from the quasi-cyclic structure of the parity-check matrix and the Gaussian elimination method. Another method was proposed in \cite{1} that uses the structure of group algebras to obtain the generator matrix. This method can be used for the unit elements of a group algebra. Let $w$ be a unit element in the group algebra $RG$. Let $\mathbf{W}$ be its corresponding $RG$-matrix. Then $\mathbf{W}$ is an invertible matrix over $R$. Without loss of generality, we consider the parity-check matrix $\mathbf{H}$ of the code as the first $n-k$ rows of $\mathbf{W}$. We divide the matrices $\mathbf{W}$ and $\mathbf{W}^{-1}$ as follow
\begin{eqnarray*}
  \mathbf{I}_{n\times n}&=&\mathbf{WW}^{-1}=\left[
                          \begin{array}{c}
                            \mathbf{H}_{(n-k)\times n} \\
                            \mathbf{J}_{k\times n}\\
                          \end{array}
                        \right]\left[
                                 \begin{array}{cc}
                                   \mathbf{A}_{n\times(n-k)} & \mathbf{B}_{n\times k} \\
                                 \end{array}
                               \right] \\
   &=& \left[
         \begin{array}{cc}
           \mathbf{H}_{(n-k)\times n}\mathbf{A}_{n\times(n-k)}  & \mathbf{H}_{(n-k)\times n}\mathbf{B}_{n\times k}  \\
           \mathbf{J}_{k\times n}\mathbf{A}_{n\times(n-k)} & \mathbf{J}_{k\times n}\mathbf{B}_{n\times k} \\
         \end{array}
       \right].
\end{eqnarray*}
The above equation gives $\mathbf{H}_{(n-k)\times n}\mathbf{B}_{n\times k}=\mathbf{0}_{n-k\times k}$ and consequently, it follows that  $\mathbf{B}_{n\times k}^t$ is a generator matrix for the given code.

Although, we  used the group rings to construct our QC-LDPC codes, our  construction method is completely different from the presented method in \cite{1}.  We design our codes over a group ring $RG$, where $|G|=n$. Then, based on the available connection  between the $RG$-matrices in $M_n(R)$ and the elements of the group ring, we replace the components of the $RG$-matrix by their corresponding QCPMs.
In both cases that we considered,  i.e., when $R$ is a finite field or when $R$ is the tensor product of multiple fields, the map that sends the elements of $R$ to their corresponding QCPMs, is a multiplicative group isomorphism. Indeed, it preserves  the multiplication but not necessarily the addition. Thus, we may have $\mathrm{QCPM} (r_1+r_2)\neq \mathrm{QCPM}(r_1)+\mathrm{QCPM}(r_2)$, for some $r_1,r_2\in R$. In fact, an element $\mathbf{U}$ in $M_n(R)$  can be invertible in $M_n(R)$  but after replacing its components with their corresponding QCPMs,  the obtained matrix can be a non-singular binary matrix (i.e., its determinant can be an even number). The idea that we use here is replacing the matrix multiplication in $M_{nb}(\mathbb{F}_2)$, where $b=\prod_{i=1}^t(q_i-1)$, by a convolution like operation in the group ring $R'G$, where
$$R'=\frac{\mathbb{F}_2[x_1]}{\left\langle x_1^{q_1-1}-1\right\rangle} \otimes \cdots\otimes \frac{\mathbb{F}_2[x_t]}{\left\langle x_t^{q_t-1}-1 \right\rangle},$$
and $x_1,\ldots, x_t$ are independent variables. We define the QCPM of $x_1^{i_1}\otimes \cdots\otimes  x_t^{i_t}\in R'$, where $0\leq i_j\leq q_j-2$ and $1\leq j\leq t$, as $\mathrm{QCPM}(\boldsymbol{\alpha}^{(i_1,\ldots,i_t)})$, which is a $b\times b$ matrix.
\begin{theorem}\label{th_11}
A matrix $\mathbf{M}\in M_n(R')$ is a unit (a zero-divisor) if and only if the matrix which is obtained by replacing the components of $\mathbf{M}$ with their corresponding QCPMs, is a unit (a zero-divisor) in $M_{nb}(\mathbb{F}_2)$.
\end{theorem}
\begin{IEEEproof}
The proof follows from the fact that the map $\Phi$ that sends $x_i$ to a circulant matrix $X_i$ of size $(q_i-1)\times (q_i-1)$ and with the first row of the form $(0,1,0,\ldots,0)$, is an isomorphism between two rings $M_n(R')$ and $\textrm{Im}(\Phi)\subset M_{nb}(\mathbb{F}_2)$. We  define $\Phi$ over other elements of $R'$ naturally.
\end{IEEEproof}
\subsection{Mathematical description of encoding for group ring based QC-LDPC codes}
Let $w$ be an element in the group algebra $R'G$ and $\mathbf{W}$ be its corresponding  $R'G$-matrix of size $n\times n$.
Let $\mathbf{H}$  be the $rb\times nb$ parity-check matrix of a group ring based QC-LDPC code $\mathcal{C}$, with $r<n$. The matrix $\mathbf{H}$ is obtained by choosing some rows from the array matrix $\mathbf{W}$, which is denoted by $\mathbf{H}_{arr}$, and replacing the components of $\mathbf{H}_{arr}$ by their corresponding QCPMs. The group ring $R'G$ is a finite ring with identity and $w$  is either a unit or a zero-divisor. Consequently, $\mathbf{W}$ is either a unit matrix or a zero-divisor matrix in $M_n(R')$.  To simplify our notation, we state the following theorem.
\begin{theorem}\label{th_12}
Let $G$ be a finite Abelian group and $R'G$ be the aforementioned group ring. Then, $R'G$ is isomorphic to the group algebra $\mathbb{F}_2 G'$, where $G'=C_{q_1-1}\times \cdots \times C_{q_t-1}\times G$ and  $C_{q_i-1}$ is the multiplicative cyclic group of order $q_i-1$, for $i=1,\ldots, t$.
\end{theorem}
\begin{IEEEproof}
The proof follows from the following isomorphisms
$$\frac{\mathbb{F}_2[x_i]}{\langle x_i^{q_i-1}-1 \rangle}\cong \mathbb{F}_2C_{q_i-1},\quad i=1,\ldots, t.$$
Consequently,
$R'G\cong  (\mathbb{F}_2C_{q_1-1}\otimes\cdots \otimes \mathbb{F}_2C_{q_t-1})[G]$.
Based on Theorem \ref{passman},  $\mathbb{F}_2C_{q_1-1}\otimes\cdots \otimes \mathbb{F}_2C_{q_t-1}$ is isomorphic to $\mathbb{F}_2\left[C_{q_1-1}\times \cdots \times C_{q_t-1} \right]$. Let $H=C_{q_1-1}\times \cdots \times C_{q_t-1}$. We show that $\mathbb{F}_2[H][G]$ is isomorphic to $\mathbb{F}_2[H\times G]$. To this end, it can be checked easily that the map $\Phi$ which is given by
$$\Phi\left(\sum_{g\in G}\sum_{h\in H}\alpha_{(g,h)}(g,h)\right)=\sum_{g\in G}\beta_g g,$$
is an isomorphism between $\mathbb{F}_2[H\times G]$ and $\mathbb{F}_2[H][G]$, where
$\beta_g=\sum_{h\in H}\alpha_{(g,h)} h\in \mathbb{Z}_2[H]$.
\end{IEEEproof}

Now, we describe the encoding approach in  both cases.
%\setdefaultleftmargin{0cm}{2cm}{}{}{}{}
%\begin{enumerate}[]
%%\begin{itemize}
%\item{\textbf{Case 1}:}

\textbf{Case 1}: Let $\mathbf{W}$ be the $R'G$-matrix of a unit element $w\in R'G$, $\mathbf{H}_{arr}$ be a subarray of $\mathbf{W}$ and $\mathbf{H}$ be its corresponding $(rb\times nb)$ binary matrix after replacing the QCPMs. Consider $\mathbf{U}$ and $u$ as the inverses of $\mathbf{W}$ and $w$ in $M_n(R')$ and $R'G$, respectively.  We want to encode a binary vector $\mathbf{m}\in \mathbb{F}_2^{kb}$, where $k=n-r$. We divide the input vector $\mathbf{m}$ into $k$ sections of size $b$ as $\mathbf{m}=(\mathbf{m}_1,\ldots, \mathbf{m}_k)$.
    %For $i=1,\ldots ,k$, and $\mathbf{m}_i=(m_{i,0},\ldots ,m_{i,b-1})$,
    Then, we map the following vector to $\mathbf{m}$
\begin{eqnarray*}
(\mathbf{m}_1,\ldots ,\mathbf{m}_k)\mapsto \left(m_1(x_1,\ldots ,x_t),\ldots , m_k(x_1,\ldots ,x_t) \right),
\end{eqnarray*}
where $m_j(x_1,\ldots ,x_t)=\sum_{i=0}^{b-1} m_{j,i} X^{\Psi^{-1} (i)}$, $X=x_1\otimes \cdots \otimes x_t$ and $\Psi$ is a bijection map from $\mathbb{Z}_{q_1-1}\times \cdots \times \mathbb{Z}_{q_t-1}$ to $\mathbb{Z}_b= \mathbb{Z}_{(q_1-1)\times \cdots \times (q_t-1) }$, such that
\begin{eqnarray}\label{Psi_def}
\Psi \left(i_1,\ldots, i_t \right)=\sum_{j=1}^{t-1} \left(i_j-1\right)\left(\prod_{s=j+1}^t (q_s-1)\right)+i_t.
\end{eqnarray}
Let $\mathbf{H}_{arr}$ be the subarray of $\mathbf{W}$ corresponding to the list $\mathcal{L}=\left\lbrace g_{i_1}^{-1},\ldots ,g_{i_r}^{-1}\right\rbrace$ in $G$. Consider the list $\mathcal{L}'=G-\left\{  g_{i_1},\ldots ,g_{i_r} \right\}=\left\{  g_{j_1},\ldots ,g_{j_k} \right\}$, and $m_{\mathcal{L}'}=\sum_{i=1}^{k}m_i(x_1,\ldots,x_t)g_{j_i}$. Then, the encoding of $\mathbf{m}$ can be done by using the flowing group ring multiplication $c=m_{\mathcal{L}'}u$ and replacing the components of $c$ with their corresponding QCPM-generators. The QCPM-generator of an element $x_1^{i_1}\otimes \cdots \otimes x_t^{i_t}$ is the vector $\mathbf{e}_{i_1}^{q_1-1} \otimes \cdots \otimes \mathbf{e}_{i_t}^{q_t-1} $, where $\mathbf{e}_{i_j}^{q_j-1}$ is a vector of length $(q_j-1)$ in which the $i_j^{th}$ position is $1$ and the other components are $0$, for $j=1,\ldots,t$. We check the validity of this statement in Theorem~\ref{unit_drive}.
\begin{theorem}\label{unit_drive}
Let $\mathbf{H}_{arr}$ be the subarray of $\mathbf{W}$ corresponding to a list $\mathcal{L}=\left\lbrace g_{i_1}^{-1},\ldots ,g_{i_r}^{-1}\right\rbrace$ in $G$, (i.e., $\mathbf{H}_{arr}$ is formed by rows $i_1,\ldots,i_r$ of $\mathbf{W}$). If we construct the parity-check matrix of the code $\mathcal{C}$ from $\mathbf{H}_{arr}$, based on \cite[Theorem 5.1]{1}, the matrix $\mathbf{G}_{arr}$  formed by the rows of $\mathbf{W}^{-1}$ with indices given in $\mathcal{L}'=\left\{g_{j_1},\ldots , g_{j_k}\right\} =G-\left\{g_{i_1},\ldots ,g_{i_r}\right\}$, can be used to construct the generator matrix of $\mathcal{C}$. Replace the components of $\mathbf{H}_{arr}$ and $\mathbf{G}_{arr}$ with their corresponding QCPMs and denote the obtained matrices by $\mathbf{H}_1$ and $\mathbf{G}_1$, respectively. The encoding of a vector $\mathbf{m}$ of length $kb$, that means the calculation of $\mathbf{m}\mathbf{G}_1$,   can be done by computing the group ring multiplication $c=m_{\mathcal{L}'}u$ and replacing the components of $c$ with their corresponding QCPM-generators.
\end{theorem}
\begin{IEEEproof}
The encoding of $\mathbf{m}=(\mathbf{m}_1,\ldots ,\mathbf{m}_k)$, where $\mathbf{m}_i=(m_{i,0},\ldots ,m_{i,b-1})$ and $i=1,\ldots,k$, means the calculation of $\mathbf{c}=\mathbf{m}\mathbf{G}_1$. Using block matrices, we have
\begin{eqnarray}
\mathbf{c}=\left[\begin{array}{ccc}
\mathbf{m}_1 & \cdots &\mathbf{m}_k \end{array} \right]\left[ \begin{array}{ccc}
\mathbf{W}'_{g_{j_1}^{-1}g_0} & \cdots & \mathbf{W}'_{g_{j_1}^{-1}g_{n-1}}\\
\vdots & \ddots & \vdots\\
\mathbf{W}'_{g_{j_k}^{-1}g_0} & \cdots & \mathbf{W}'_{g_{j_k}^{-1}g_{n-1}}
\end{array}
\right],
\end{eqnarray}
where $\mathbf{W}'_{g_{j_z}^{-1}g_{s}}$ is the QCPM of the $(j_z,s)^{th}$ component of $\mathbf{W}^{-1}$. For $z=1,\ldots,n$, the sub-block $\mathbf{c}_z$, which corresponds to the indices $(z-1)b+1$ to $zb$ of $\mathbf{c}$, is obtained by $\left( \sum_{i=1}^{k}\mathbf{m}_i\mathbf{W}'_{g_{j_i}^{-1}g_z} \right) g_z$. The group element $g_z$ in the right side of this equation indicates the location of the sub-vector $\sum_{i=1}^{k}\mathbf{m}_i\mathbf{W}'_{g_{j_i}^{-1}g_z}$ in the given codeword $\mathbf{c}$. In our proposed encoding method, we compute the group ring multiplication
$$c=\left(\sum_{i=1}^{k}m_i(x_1,\ldots ,x_t)g_{j_i} \right)(w'_{g_{0}}g_{0}+\cdots +w'_{g_{n-1}}g_{n-1}).$$
Using the properties of the group ring multiplication, the coefficient of $g_z$ in the above multiplication is $\sum_{i=1}^{k}m_i(x_1,\ldots ,x_t) g_{j_i} w'_{g_{k_i}}g_{k_i}$, where $ g_{j_i}g_{k_i}=g_z$, for $i=1,\ldots ,k$. Since $G$ is an Abelian group, we have $g_{k_i}=g_{j_i}^{-1}g_z$ and the coefficient of $g_z$ is $\left(\sum_{i=1}^{k}m_i(x_1,\ldots ,x_t)w'_{g_{j_i}^{-1}g_z}\right)g_{z}$.  After replacing the QCPM-generators, we reach the same result as the usual encoding.
\end{IEEEproof}

%\item{\textbf{Case 2}:}
\textbf{Case 2}: Let $w$ be a zero divisor in $R'G$  and $\mathbf{W}\in M_n(R')$ be its corresponding $R'G$-matrix such that for a matrix $\mathbf{U}\in M_n(R')$ and $u\in R'G$, $\mathbf{WU}^{t}=\mathbf{0}_{n\times n}$ and $wu^{t}=0$\footnote{For a given group $G=\{g_0=1_G,g_1,\ldots,g_{n-1}\}$ and a group ring $RG$, let $u=\sum_{i=0}^{n-1}\beta_{g_i} g_i$ be an element in $RG$. Then we define $u^t$ as  $\sum_{i=0}^{n-1}\beta_{g_i} g_i^{-1}$.}. Let $\mathbf{H}_{arr}$  be the subarray of $\mathbf{W}$ corresponding to the list $\mathcal{L}=\left\lbrace g_{i_1}^{-1},\ldots ,g_{i_r}^{-1}\right\rbrace\subset G$ and $\mathbf{H}$ be its corresponding $(rb\times nb)$ binary matrix which is obtained by replacing the elements of $\mathbf{H}_{arr}$  with their corresponding  QCPMs. Let $\mathcal{C}$ be a code with parity-check matrix $\mathbf{H}$. Then, the generator matrix of $\mathcal{C}$ is a $k'\times nb$ binary matrix such that $\mathbf{GH}^{t}=\mathbf{0}_{k'\times rb}$, where $k'=nb-\mathrm{rank}(\mathbf{H})$. If the matrix  $\mathbf{H}$ has full rank, then instead of finding the generator matrix of $\mathcal{C}$, we consider $\mathbf{H}$ as the  generator matrix of the code $\mathcal{C}^{\perp}$ and we find its parity-check matrix. Similar to  the method used in Theorem~\ref{unit_drive}, every codeword $\mathbf{c}'$ in  $\mathcal{C}^{\perp}$ is obtained as follows: first, we have an element $c'$ in $R'G$ of the form $c'=mw$, where $m\in M'$ and $M'$ is the $R'$-submodule of $R'G$ generated by  the list $\mathcal{L}$ of $G$. Then, the  vector $\mathbf{c}'$ is obtained by replacing the components of $c'$ with their corresponding QCPM-generators.  If $\mathcal{L}w$ is linearly independent, $\mathrm{rank}(\mathbf{W})=|\mathcal{L}|=r$ and $\mathrm{rank}(\mathbf{U})=n-r$, then we find $n-r$ independent rows of $\mathbf{U}$  which are in accordance with a list in $G$ like $\mathcal{L}'=\left\{g_{j_1}^{-1},\ldots , g_{j_{n-r}}^{-1}\right\}$. We put these $n-r$ independent rows of $\mathbf{U}$  in a matrix which is denoted by $\mathbf{U}_{arr}$. Then, a necessary and sufficient condition to have a single check element, is obtaining the rank $(n-r)b$  after replacing the elements of  $\mathbf{U}_{arr}$ by their corresponding QCPMs \cite[Theorem 4.9]{1}. Hence, the encoding of a vector $\mathbf{m}$  can be done by using the group ring multiplication $c=m_{\mathcal{L}'}u$, where $m_{\mathcal{L}'}=\sum_{i=1}^{n-r}m_i(x_1,\ldots,x_t)g_{j_i}$, and replacing the components of $c$ with their corresponding QCPM-generators.
    Now, let  both $\mathbf{H}$ and $\mathbf{H}_{arr}$ have full rank, $\dim M'=|\mathcal{L}|=r<\omega=\mathrm{rank}(\mathbf{W})$ and $\mathbf{W}\mathbf{U}^t=\mathbf{0}_{n\times n}$, with $\mathrm{rank}(\mathbf{U})=n-\omega$. In addition, let the substituting of QCPMs in $\mathbf{U}$ admit a matrix of rank $(n-\omega)b$. In this case, we obtain the generator matrix of $\mathcal{C}$ by adding $\omega-r$ extra vectors to the independent rows of $\mathbf{U}$ or $\mathbf{U}_{arr}$. Moreover, the rows of $\mathbf{W}$ corresponding to the list $\mathcal{L}$ of $G$, are independent and we can extend $ \mathcal{L}$ to a subset $T=\left\{g_{i_1}^{-1},\ldots ,g_{i_r}^{-1},g_{i_{r+1}}^{-1},\ldots, g_{i_{\omega}}^{-1}\right\}$ of $G$,  corresponding to the independent rows of $\mathbf{W}$, and put all these row vectors in a matrix $\mathbf{W}_{\omega}$. Since, $\mathrm{rank}(\mathbf{W})=\omega$, there exists an $n\times \omega$ matrix $\mathbf{C}$ such that $\mathbf{W}_{\omega}\mathbf{C}=\mathbf{I}_{\omega}$. This implies
    \begin{equation}\label{mat_eq}
      \mathbf{W}_{\omega}\mathbf{C}=\left[
             \begin{array}{c}
               \mathbf{H}_{arr} \\
               \mathbf{H}_2 \\
             \end{array}
           \right]\left[
                    \begin{array}{cc}
                      \mathbf{C}_1 & \mathbf{C}_2 \\
                    \end{array}
                  \right]=\mathbf{I}_{\omega}.
    \end{equation}
    We conclude that $\mathbf{H}_{arr}\mathbf{C}_2=\mathbf{0}_{r\times (\omega-r)}$. Let $\mathbf{U}_{n-\omega}$ be a matrix formed by $n-\omega$ linearly independent columns of $\mathbf{U}$. Then, it can be shown \cite{1}, that the binary matrix $\mathbf{G}_b$ that is obtained by replacing the components of $\mathbf{G}_{arr}=\left[
                                                                               \begin{array}{cc}
                                                                                 \mathbf{U}_{n-\omega} &
                                                                                 \mathbf{C}_2
                                                                               \end{array}
                                                                             \right]^t$
 with their corresponding  QCPMs, is the generator matrix of $\mathcal{C}$. In Remark~\ref{rem1}, we have explained the details of our encoding method based on  multiplication of group rings.
%    When $H_{arr}$ has not full rank and other conditions are fulfilled, we choose a sub-array $H_{arr}'$ of it and continue just like above.
%\end{enumerate}
%We can check conveniently that the all of the examples in previous section are unit elements. Note that the size of matrices based on the %construction of Theorem \ref{theorem2} is very  small. For example if we want to construct a code of length $106483$ we must check a %matrix of size $(13\times 13)$. We can use softwares like GUAVA \cite{13} or MATLAB for doing this.

\begin{rem}\label{rem1}
Using the aforementioned notation, let the binary matrix $\mathbf{G}_b$ of size $(n-r)b\times nb$ be the generator matrix of $\mathcal{C}$  obtained from $\mathbf{G}_{arr}=\left[\begin{array}{cc}
 \mathbf{U}_{n-\omega} & \mathbf{C}_2
\end{array}\right]^t$ by replacing its components with their corresponding QCPMs. Let $ \mathbf{U}_{n-\omega}$ be the subarray of $R'G$-matrix $\mathbf{U}$ corresponding to the list $\mathcal{L}_u=\left\{g_{j_1},\ldots ,g_{j_{n-\omega}} \right\}\subset G$.
Let $\mathbf{U}_{n-\omega}^b$ and $\mathbf{C}_2^b$ be the matrices obtained by replacing the QCPMs in $\mathbf{U}_{n-\omega}$ and $\mathbf{C}_2$, respectively.
The encoding of $\mathbf{m}=(\mathbf{m}_1,\ldots ,\mathbf{m}_k)$, with $\mathbf{m}_i=(m_{i,0},\ldots ,m_{i,b-1})$, $i=1,\ldots,k$ and $k=n-r$, can be done as $\mathbf{c}=\mathbf{m}\mathbf{G}_b$, which is equivalent to
\begin{equation}\label{enc}
\mathbf{c}= \mathbf{m}^1\mathbf{U}_{n-\omega}^b+\mathbf{m}^2 \mathbf{C}_2^b,
\end{equation}
where $\mathbf{m}^1=(\mathbf{m}_1,\ldots,\mathbf{m}_{n-\omega})$ and $\mathbf{m}^2=(\mathbf{m}_{n-\omega+1},\ldots , \mathbf{m}_k)$.  Since $\mathbf{U}_{n-\omega}$ is a subarray of the $R'G$-matrix $\mathbf{U}$, by using the same method in the proof of  Theorem~\ref{unit_drive}, we can prove that the term $\mathbf{c}_1=\mathbf{m}^1\mathbf{U}_{n-\omega}^b$ can be obtained by performing the following group ring multiplication
\begin{equation*}%\label{group_ring_mult}
 c_1= \left(\sum_{i=1}^{n-\omega}m_i(x_1,\ldots ,x_t)g_{j_i}^{-1} \right)(u_{g_{0}}^tg_{0}+\cdots +u_{g_{n-1}}^tg_{n-1}),
\end{equation*}
where $(u_{g_{0}}^t,\ldots ,u_{g_{n-1}}^t)^t$ is the first column of $\mathbf{U}$. It is enough to show that the term $\mathbf{c}_2=\mathbf{m}^2 \mathbf{C}_2^b$ can also be obtained by using a group ring multiplication. To this end, $\mathbf{C}_2$ must be a subarray of an $R'G$-matrix. Let us consider $\mathbf{W}_{\omega}\mathbf{C}=\mathbf{I}_{\omega}$ and let $\mathbf{C}^1$ be the first column of $\mathbf{C}$. Let $\mathbf{W}_1$ be the first row of the $R'G$-matrix $\mathbf{W}$. Since all other rows of $\mathbf{W}$ can be written as a permutation of the first row, $\mathbf{W}_{\omega}$ will be of the following form
\begin{eqnarray}
  \mathbf{W}_{\omega} &=&\left[
           \begin{array}{c}
             g_{i_1}^{-1}(\mathbf{W}_1) \\
             \vdots \\
             g_{i_{\omega}}^{-1}(\mathbf{W}_1) \\
           \end{array}
         \right],
\end{eqnarray}
where $g_{i_j}^{-1}(\mathbf{W}_1)$, for $j=1,\ldots ,\omega$, is the $i_j^{th}$ row of $\mathbf{W}$. Note that  $\mathbf{W}_{\omega}\mathbf{C}=\mathbf{I}_{\omega}$ implies $\left\langle g_{i_j}^{-1}(\mathbf{W}_1),(\mathbf{C}^1)^{t}\right\rangle =\delta_{j,1}$, for $j=1,\ldots ,t$, in which $\delta_{j,1}=1\,(j=1)$, $0\,(j\neq 1)$ is the Keronecker's delta and $\left\langle ,\right\rangle$ denotes the inner product in $\mathbb{F}_2^n$. We also have the following trivial result.
\begin{lemma}\label{permute}
Let $\mathbb{F}$ be an arbitrary field and let $\left\langle ,\right\rangle$ denote the inner product over $\mathbb{F}^n$, for a positive integer $n$. Then, for every $\mathbf{x},\mathbf{y}\in \mathbb{F}^n$, and every permutation $\sigma$ on $\left\{1,\ldots , n\right\}$,  $\left\langle \mathbf{x},\mathbf{y} \right\rangle=\left\langle \sigma(\mathbf{x}),\sigma(\mathbf{y})\right\rangle$.
\end{lemma}

If there exists a set $\mathcal{I}=\left\{g_1,\ldots ,g_{\omega}\right\}\subset G$ such that $g_l^{-1}g_{i_s}^{-1}\in T=\left\{g_{i_1}^{-1},\ldots ,g_{i_r}^{-1},g_{i_{r+1}}^{-1},\ldots, g_{i_{\omega}}^{-1}\right\}$, for $l,s=1,\ldots ,\omega$, then based on Lemma~\ref{permute}, there exists a reordering on $\mathcal{I}$, like $\left\{g_1',\ldots ,g_{\omega}'\right\}$, that gives  an $R'G$-matrix $\mathbf{C}'$ as $$\mathbf{C}'=\left[
                         \begin{array}{ccc}
                           g_1'(\mathbf{C}^1) & \cdots & g_{\omega}'(\mathbf{C}^1) \\
                         \end{array}
                       \right],$$
such that $\mathbf{W}_{\omega}\mathbf{C}'=\mathbf{I}_{\omega}$.
Thus, instead of $\mathbf{C}_2$, we can consider a submatrix of $\mathbf{C}'$ which is  an $R'G$ matrix. Consequently, the second part of encoding, i.e., $\mathbf{m}^2\mathbf{C}_2^b$, can be obtained by substituting the QCPM-generators in the components of the following group ring multiplication
%\scriptsize
\begin{eqnarray}
c_2= \left(\sum_{i=n-\omega+1}^{n-r}m_i(x_1,\ldots ,x_t)(g_{i+\omega-n+r}')^{-1} \right)c_{R'G}',
\end{eqnarray}
where $c_{R'G}'=C_{1,1}'g_{0}+\cdots +C_{n,1}'g_{n-1}$ and $(C_{1,1}',\ldots ,C_{n,1}')^t$ is the first column of $\mathbf{C}'$. In this case, the codewords of $\mathcal{C}$ cannot be obtained by using a single generator. Removing each one of the aforementioned  conditions makes the encoding highly complicated.
$\hfill\square$\end{rem}

Due to the mathematical complexity of the  encoding method proposed in Case $2$, finding an elements $w$ in $R'G$ that satisfy the assumptions of Case $1$ is our desire.  In Proposition~\ref{inv_prop}, we specify some conditions under which the obtained array matrix $\mathbf{W}$, remains an invertible matrix over $\mathbb{F}_2$ after replacing the components of $\mathbf{W}$ with their corresponding QCPMs. We need the following results and definitions to establish this result.

Let $\boldsymbol{\Gamma}$ denote the $n\times n$ cyclic shift matrix whose entries are $\Gamma_{i,j} = 1$ if $j-i\equiv 1 \pmod{n}$, and $0$, otherwise. An $n\times n$ circulant matrix $\mathbf{A}$ over  the ring of integers modulo $m$, which is denoted by $\mathbb{Z}_m$ for a positive integer $m$,  can be written as $\mathbf{A} =\sum_{i=0}^{n-1}a_i{\boldsymbol{\Gamma}}^i$, where $a_i \in \mathbb{Z}_m$ and $i=0,\ldots,n-1$. We associate with  the circulant matrix $\mathbf{A}$ the polynomial $ f(x) =\sum_{i=0}^{n-1}a_ix^i$ in the ring $\mathbb{Z}_m[x]$. The following theorem states the necessary and sufficient conditions for  $\mathbf{A}$ being an invertible matrix over $\mathbf{Z}_m$.
\begin{theorem}[\emph{\cite[Theorem 2.2]{30}}]\label{inv_circ}
Let $m = p_1^{k_1} p_2^{k_2}\cdots p_h^{k_h}$ denote the prime powers factorization of $m$ and let $f$ denote the polynomial over $\mathbb{Z}_m$ associated to a circulant matrix $\mathbf{A}$. The matrix $\mathbf{A}$ is invertible over $\mathbb{Z}_m$ if and only if, for $i = 1,\ldots , h$, we have
\begin{eqnarray*}
\gcd \left( f(x),x^n-1\right)=1 \quad \mbox{in}\,\, \mathbb{Z}_{p_i}[x].
\end{eqnarray*}
\end{theorem}

Let $q$ be a power of an odd prime $p$ and let $\zeta_n$ denote a primitive $n^{th}$ root of unity. The $n^{th}$ cyclotomic polynomial $\Phi_n(x)$ is
\begin{eqnarray*}
\Phi_n(x)=\prod_{0<i<n,(i,n)=1}(x-\zeta_n^i).
\end{eqnarray*}
\begin{theorem}[\emph{\cite[Theorem 2.47]{31}}]\label{cyc_th}
If $\gcd(q, n)=1$, then $\Phi_n(x)$ factors into $\varphi(n)/d$ distinct monic irreducible polynomials in $\mathbb{F}_q[x]$ of the same degree $d$, where $d$ is the least positive integer such that $q^d \equiv 1\pmod{n}$.
\end{theorem}

\begin{Definition}
A number $\vartheta$ is a primitive root modulo $n$ if every number coprime to $n$ is congruent to a power of $\vartheta$ modulo $n$. In other words, $\vartheta$ is a generator of the multiplicative group of integers modulo $n$ \cite{32}.
\end{Definition}

Let us denote  the inverse of $a\pmod{b}$ by $[a^{-1}]_b$. Based on these statements, we have the following proposition.
\begin{Proposition}\label{inv_prop}
Let $q=2^m$ and  $G=\left\{g_0,g_1,\ldots,g_{n-1}\right\}$ be a cyclic group of order $n$, where  $n$ and $q-1$ are distinct  odd prime numbers. Consider $D=\left\{d_0,\ldots , d_{n-1} \right\}\subset \mathbb{Z}_{q-1}$ as a modified $S_2$-set  such that  $\max (D')-\min (D')\leq \varphi(n(q-1))$, where $D'=\left\{d'_0,\ldots, d'_{n-1}\right\}$ and $d'_i=nd_i[n^{-1}]_{q-1}+i(q-1)[(q-1)^{-1}]_{n} \pmod{n(q-1)}$.  Let $\alpha$ be the primitive element of $\mathbb{F}_q$, $w=\sum_{i=0}^{n-1}\alpha^{d_i}g_i\in\mathbb{F}_q G$ and $\mathbf{W}$ be its corresponding $\mathbb{F}_q G$-matrix. If $2$ is the primitive root modulo $n(q-1)$ and $(f(x),\Phi_{n}(x)\Phi_{(q-1)}(x))=1$ over $\mathbb{F}_2[x]$, where $f(x)=\sum_{i=0}^{n-1}x^{d'_{i}}$, then, the matrix obtained from $\mathbf{W}$ by replacing the components with their corresponding CPMs, is an invertible matrix over $\mathbb{F}_2$.
\end{Proposition}
\begin{IEEEproof}
We use Theorem~\ref{th_11} in the case $t=1$ and $R'[G]=\mathbb{F}_2 C_{q-1}[G]$. Instead of considering  $\mathbf{W}\in M_n(\mathbb{F}_q)$ and replacing the components of $\mathbf{W}$ by their corresponding CPMs, we consider the image of  $w$ as an element $w'$ in $\mathbb{F}_2 C_{q-1}[G]$. It is enough to show that $w'$ is an invertible element. Based on Theorem~\ref{th_12}, $R'G\cong \mathbb{F}_2[C_{q-1}\times G]$ and since $(n,q-1)=1$, $R'G$ and $\mathbb{F}_2[C_{n(q-1)}]$ are isomorphic  via an isomorphism, namely $\gamma$. It should be noted that  the support of $w'$  is the subset  $S=\left\{(d_i,g_i)|i=0,\ldots,n-1 \right\}$ of $C_{q-1}\times G$. Consider an element $g$ to be the generator of $G$, then $(\alpha,g)$ is the generator of the cyclic group $C_{q-1}\times G$, which is isomorphic to $C_{n(q-1)}$. We show that $D'$ is the support of $\gamma(w')$. To this end, we find an element  $\beta_i$ in $\mathbb{Z}_{n(q-1)}$ such that $(\alpha,g)^{\beta_i}= \alpha^{d_i}g^i$, for $i=0,\ldots , n-1$. This implies the following system of congruent equations
\begin{IEEEeqnarray*}{rCl}
\begin{array}{rcl}
  \beta_i  \equiv & d_i & (\bmod\,\, (q-1)), \\
  \beta_i \equiv & i & (\bmod\,\, n).
\end{array}
\end{IEEEeqnarray*}
Based on Chinese remainder theorem \cite[Section 31.5]{33}, the solution of this system of equations is $d'_{i} =nd_i[n^{-1}]_{q-1}+i(q-1)[(q-1)^{-1}]_{n}\pmod{n(q-1)}$. Thus, the $\mathbb{F}_2[C_{n(q-1)}]$-matrix of $\gamma(w')$ is an $n(q-1)\times n(q-1)$ cyclic matrix and its corresponding polynomial is $f(x)= \sum_{i=0}^{n-1}x^{d'_{i}}$. It is enough to show that $(f(x),x^{n(q-1)}-1)=1$ in $\mathbb{F}_2[x]$. We know that $x^{n(q-1)}-1=\prod_{d|n(q-1)}\Phi_d(x)=\Phi_1(x)\Phi_n(x)\Phi_{q-1}(x)\Phi_{n(q-1)}(x)$, \cite{34}. The degree of $f(x)$ is $\max(D')$ and it has $0$ as a root with multiplicity $\min(D')$. Since, $(2,n(q-1))=1$ and  $2$ is the primitive root modulo $n(q-1)$, based on Theorem~\ref{cyc_th},  $\Phi_{n(q-1)}(x)$ is an irreducible polynomial. The condition $\max (D')-\min (D')\leq \varphi(n(q-1))$ implies $(\Phi_{n(q-1)}(x), f(x))=1$. Based on the assumptions, $(f(x),\Phi_{n}(x)\Phi_{(q-1)}(x))=1$ and since $f(1)=n\not\equiv 0\pmod{2}$, we have $(f(x),x^{n(q-1)}-1)=1$ and the result holds.
\end{IEEEproof}

Based on Theorem~\ref{inv_circ} and Theorem~\ref{cyc_th}, construction of invertible circulant matrices over $\mathbb{F}_q$ is an straightforward job, but when we replace the CPMs, finding the sufficient conditions for remaining invertible over $\mathbb{F}_2$ is a complicated task and results in the conditions of  Theorem~\ref{inv_prop}.
\begin{Example}
Consider the  group $G=\left\{g_0=1_G,g_1,g_2\right\}$ as the cyclic group of order $3$ and $R'=\mathbb{F}_2C_7$. Let $w=xg_0+x^2g_1+x^4g_2$, in which $x$ is generator of the multiplicative cyclic group of order $7$, $C_7=\left\{1,x,\ldots ,x^6\right\}$. Then, the $R'G$-matrix of $w$ is
\begin{equation}\label{W_ex}
\mathbf{W}=\left[
    \begin{array}{ccc}
      x & x^2 & x^4 \\
      x^4 & x & x^2 \\
      x^2 & x^4 & x \\
    \end{array}
  \right].
\end{equation}
Choose  the first row of $\mathbf{W}$ as $\mathbf{H}_{arr}$, replace  $x,x^2,x^4$ by their corresponding CPMs and denote the obtained binary matrix by $\mathbf{H}$. We use this matrix as the parity-check matrix of the QC-LDPC code $\mathcal{C}$. The element $w$ is a zero divisor in $R'G$ and we can find $u\in R'G$ such that $uv=0$. We find $u$ as follows. It is easy to check that $\det (\mathbf{W})=x^6+x^5+x^3+1=(x+1)^3(x^3+x+1)$. Dividing $x^7+1$ by $x^3+x+1$ gives $f(x)=x^4+x^2+x+1$. It is also easy to check that the adjoint matrix of $\mathbf{W}$ is
\begin{equation}\label{adj}
\textrm{adj}\left(\mathbf{W}\right)=\left[
    \begin{array}{ccc}
      x^2+x^6 & x^4+x^5 & x+x^3 \\
      x+x^3 & x^2+x^6 & x^4+x^5 \\
      x^4+x^5 & x+x^3 & x^2+x^6 \\
    \end{array}
  \right],
\end{equation}
and $\mathbf{W}\mathbf{W}^{*}=\det(W)\mathbf{I}_3$, where $\mathbf{W}^{*}=\textrm{adj}\left(\mathbf{W}\right)^t$. Then, put $\mathbf{U}=f(x)\mathbf{W}^{*}$ which is the following matrix after simplifications
\begin{equation*}
\mathbf{U}=(x^4+x^2+x+1)\left[
    \begin{array}{ccc}
       1 & 1 & 1 \\
      1 & 1 &1 \\
      1 & 1 & 1 \\
    \end{array}
  \right].
\end{equation*}
It is clear that $\mathbf{WU}=\mathbf{0}_3$. We can check that $\textrm{rank}(\mathbf{W})=16$ and $\textrm{rank}(\mathbf{U})=3$, over $\mathbb{F}_2$. Thus, $\mathbf{W}$ does not have the conditions of Remark~\ref{rem1}, but we explain the encoding method by using these matrices with some modifications. Replace the CPMs in $\mathbf{W}$, and denote the obtained matrix by $\mathbf{W}^b$. Then, the  rows in the list $\mathcal{L}=\left\{1-13, 15, 16, 17\right\}$ of $\mathbf{W}^b$ are linearly independent over $\mathbb{F}_2$. Let $\mathbf{H}'$ be the submatrix of $\mathbf{W}$ corresponding to $\mathcal{L}$.  There is a $16\times 21$ binary matrix $\mathbf{C}$ such that $\mathbf{H}'\mathbf{C}=\mathbf{I}_{16}$
%\scriptsize
\begin{equation*}
%\scriptstyle
\scriptsize
\mathbf{C}=\left[
    \begin{array}{cccccccccccccccc}
   1  &   0  &   0  &   1   &  1  &   0  &   0  &   0   &  1  &   0  &   1  &   1   &  1   &  1  &   0   &  1\\
   0  &   0  &   1  &   0   &  0  &   0  &   1  &   1   &  0  &   0  &   1  &   1   &  0   &  0  &   1   &  1\\
   0  &   0  &   1  &   1   &  0  &   0  &   1  &   1   &  0  &   0  &   1  &   0   &  1   &  0  &   1   &  0\\
   1  &   0  &   1  &   1   &  1  &   0  &   0  &   0   &  1  &   0  &   0  &   1   &  1   &  1  &   0   &  1\\
   0  &   0  &   0  &   0   &  0  &   0  &   0  &   0   &  0  &   0  &   0  &   0   &  0   &  0  &   0   &  1\\
   1  &   0  &   1  &   0   &  0  &   0  &   1  &   1   &  1  &   0  &   1  &   1   &  0   &  0  &   1   &  1\\
   0  &   1  &   1  &   1   &  0  &   0  &   1  &   1   &  0  &   1  &   1  &   0   &  1   &  0  &   1   &  0\\
   1  &   1  &   1  &   0   &  1  &   1  &   0  &   0   &  1  &   1  &   0  &   1   &  0   &  1  &   1   &  0\\
   1  &   0  &   0  &   1   &  1  &   0  &   1  &   0   &  1  &   0  &   1  &   1   &  1   &  1  &   0   &  1\\
   1  &   0  &   1  &   0   &  0  &   0  &   1  &   1   &  0  &   0  &   1  &   1   &  0   &  0  &   1   &  1\\
   0  &   1  &   1  &   1   &  0  &   0  &   1  &   1   &  0  &   0  &   1  &   0   &  1   &  0  &   1   &  0\\
   1  &   0  &   0  &   1   &  1  &   0  &   0  &   0   &  1  &   0  &   0  &   1   &  1   &  1  &   0   &  1\\
   0  &   0  &   1  &   0   &  0  &   0  &   1  &   1   &  0  &   0  &   1  &   0   &  0   &  0  &   1   &  1\\
   0  &   0  &   0  &   0   &  0  &   0  &   0  &   0   &  0  &   0  &   0  &   0   &  0   &  0  &   0   &  0\\
   0  &   0  &   1  &   1   &  0  &   0  &   1  &   1   &  0  &   0  &   1  &   0   &  0   &  0  &   1   &  0\\
   1  &   0  &   1  &   0   &  1  &   0  &   1  &   1   &  1  &   0  &   1  &   1   &  0   &  0  &   1   &  1\\
   1  &   0  &   0  &   1   &  1  &   0  &   1  &   1   &  1  &   0  &   1  &   1   &  1   &  1  &   0   &  0\\
   0  &   0  &   0  &   0   &  0  &   0  &   0  &   0   &  0  &   0  &   0  &   0   &  0   &  0  &   0   &  0\\
   0  &   0  &   0  &   0   &  0  &   0  &   0  &   0   &  0  &   0  &   0  &   0   &  0   &  0  &   0   &  0\\
   0  &   0  &   0  &   0   &  0  &   0  &   0  &   0   &  0  &   0  &   0  &   0   &  0   &  0  &   0   &  0\\
   0  &   0  &   0  &   0   &  0  &   0  &   0  &   0   &  0  &   0  &   0  &   0   &  0   &  0  &   0   &  0
    \end{array}
  \right].
\end{equation*}
%\normalsize
Replace the CPMs in $\mathbf{U}$ and  choose  the first $3$ columns of the obtained matrix and denote it by $\mathbf{U}_1$. Since $\mathbf{H}$ is corresponding to the first $7$ rows of $\mathbf{W}$, we choose the columns $8-16$ of $\mathbf{C}$ and denote the obtained matrix by $\mathbf{C}_1$. The generator matrix of $\mathcal{C}$ is a $14\times 21$ binary matrix like $\mathbf{G}$ such that $\mathbf{GH}^t=\mathbf{0}_{14\times 7}$ and $\textrm{rank}(\mathbf{G})=14$. Put $\mathbf{G}_1=\left[
                                                                                \begin{array}{cc}
                                                                                  \mathbf{U}_1 & \mathbf{C}_1 \\
                                                                                \end{array}
                                                                              \right]^t$ which is of rank  $12$. We only consider the message vectors with $0$ in the last two coordinates. Then, $\mathbf{G}_1$ can be used for encoding the  messages of this form.   Note that $\mathbf{U}_1^t$ is a submatrix of $\mathbf{U}^t$ which is the following matrix
\begin{equation}\label{UT}
  \mathbf{U}^t=(x^6+x^5+x^3+1)\left[
    \begin{array}{ccc}
     1 & 1 & 1 \\
      1 &1 & 1 \\
      1 & 1 & 1 \\
    \end{array}
  \right].
\end{equation}
Let $u=\sum_{i=0}^2f_i(x)g_i$, where $f_i(x)=\sum_{j=0}^6c_{i,j}x^j$, for $i=0,1,2$. Then, $u^t=\sum_{i=0}^2f_i^{-1}(x)g_i^{-1} $, where $f_i^{-1}(x)=\sum_{j=0}^6c_{i,j}x^{7-j}$. For example, the encoding of the vector $\mathbf{m}=[\mathbf{m}_1 \,\, \mathbf{m}_2]$, with $\mathbf{m}_1=[1\,\,0\,\,1]$ and
                                                                              $\mathbf{m}_2=[ 0  \,\,   1  \,\,   0  \,\,   0  \,\,   1   \,\,  1  \,\,   0 \,\,    1   \,\,  1
]$, can be done as $\mathbf{c}=\mathbf{m}_1\mathbf{U}_1^t+\mathbf{m}_2\mathbf{C}_1^t$. We have
$$\mathbf{m}_1\mathbf{U}_1^t=[0\,\, 1\,\, 1\,\, 1\,\, 0\,\, 0\,\, 1\,\, 0\,\, 1\,\, 1\,\, 1\,\, 0\,\, 0\,\, 1\,\, 0\,\, 1\,\, 1\,\, 1\,\, 0\,\, 0\,\, 1].$$
This multiplication can also be done as follows
\begin{eqnarray*}
 c_1(x) &=& m_1(x)u^t=(1+x^2)(x^6+x^5+x^3+1)\sum_{i=0}^2g_i \\
   &=&(x+x^2+x^3+x^6)(g_0+g_1+g_2).
\end{eqnarray*}
Now, replace the generator of CPMs in $c_1(x)$ which gives the following binary vector
$$\mathbf{c}_2=[0\,\, 1\,\, 1\,\, 1\,\, 0\,\, 0\,\, 1\,|\, 0\,\, 1\,\, 1\,\, 1\,\, 0\,\, 0\,\, 1\,|\, 0\,\, 1\,\, 1\,\, 1\,\, 0\,\, 0\,\, 1].$$
$\hfill\square$\end{Example}

\section{Encoding Implementation}\label{implementation}
To make a better understanding about the  encoding method proposed in Section~\ref{sec5_2}, we present its naive implementation  in the case that a unit group ring element is used in the construction of code. Let $u=\sum_{j=1}^nu_j(x_1,\ldots,x_t)g_j=\sum_{j=1}^{n}\sum_{i=0}^{b-1}u_{j,i}X^{\Psi^{-1}(i)}g_j$ be the group ring element that generates the code, in which $\Psi$ is defined in (\ref{Psi_def}).
As explained in Section~\ref{sec5_2}, the encoding of a vector $\mathbf{m}=(\mathbf{m}_{1},\ldots,\mathbf{m}_k)$, with $\mathbf{m}_j=(m_{j,0},\ldots,m_{j,b-1})$ and $j=1,\ldots,k$, can be done by viewing  $\mathbf{m}$ as a group ring element of the form $m_{\mathcal{L}'}=\sum_{j=1}^km_j(x_1,\ldots,x_t)g_{i_j}=\sum_{j=1}^k\sum_{i=0}^{b-1}m_{j,i}X^{\Psi^{-1}(i)}g_{i_j}$ and computing $m_{\mathcal{L}'}u$.

\begin{figure*}
\centering
\includegraphics[width=6in]{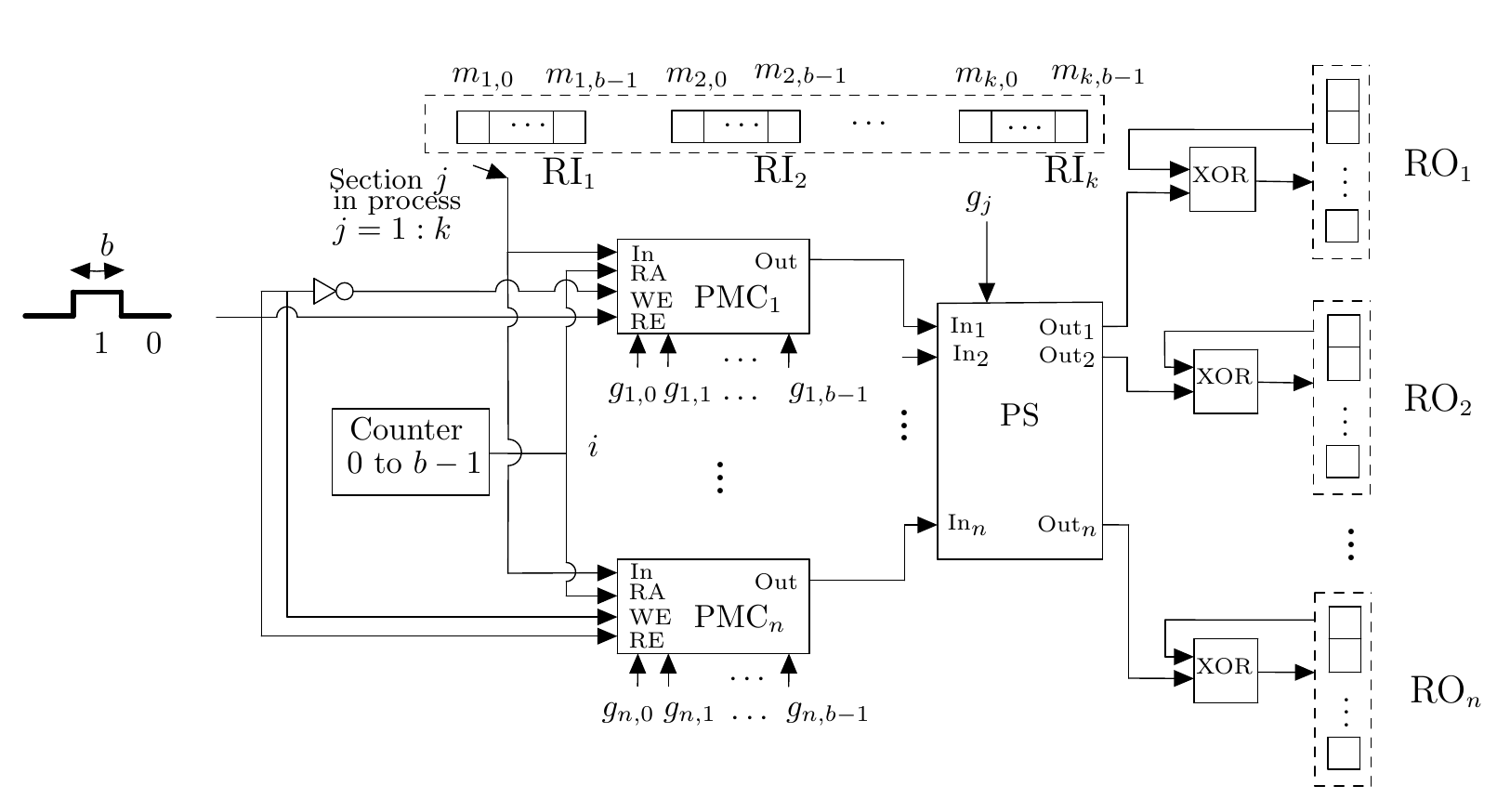}
\caption{The encoder circuit of the unit-derived group ring based QC-LDPC codes.}\label{Enc_diagram}
\hrulefill
\end{figure*}

In \figurename~\ref{Enc_diagram}, we present the encoder circuit that implements this procedure. The inputs of this encoder are loaded from $k$ input registers  (RIs) of size $b$ and the outputs are stored in $n$ output registers (ROs) of size $b$. This circuit is composed of $n$ \emph{partial multiplier circuits} (PMCs) and each one multiplies $m_{\mathcal{L}'}$ by one $u_s(x_1,\ldots,x_t)$, for $s=1,\ldots,n$. The encoding operation is performed in $kb$ clocks, after $b$ clocks delay in the beginning. In this circuit, the sub-block  denoted by PS, is a programable switch which is composed of $n$ $1$-to-$n$ demultiplexers that depending on $g_{i_j}$, (i.e., when the encoder is processing the $j^{th}$ sub-vector $\mathbf{m}_j$ of the input vector $\mathbf{m}$) it routs the inputs to their correct positions in the output registers. After routing and choosing the appropriate register, the new value of the register is the XOR of its previous value and the new routed value.

\begin{figure*}
\centering
\includegraphics[width=6in]{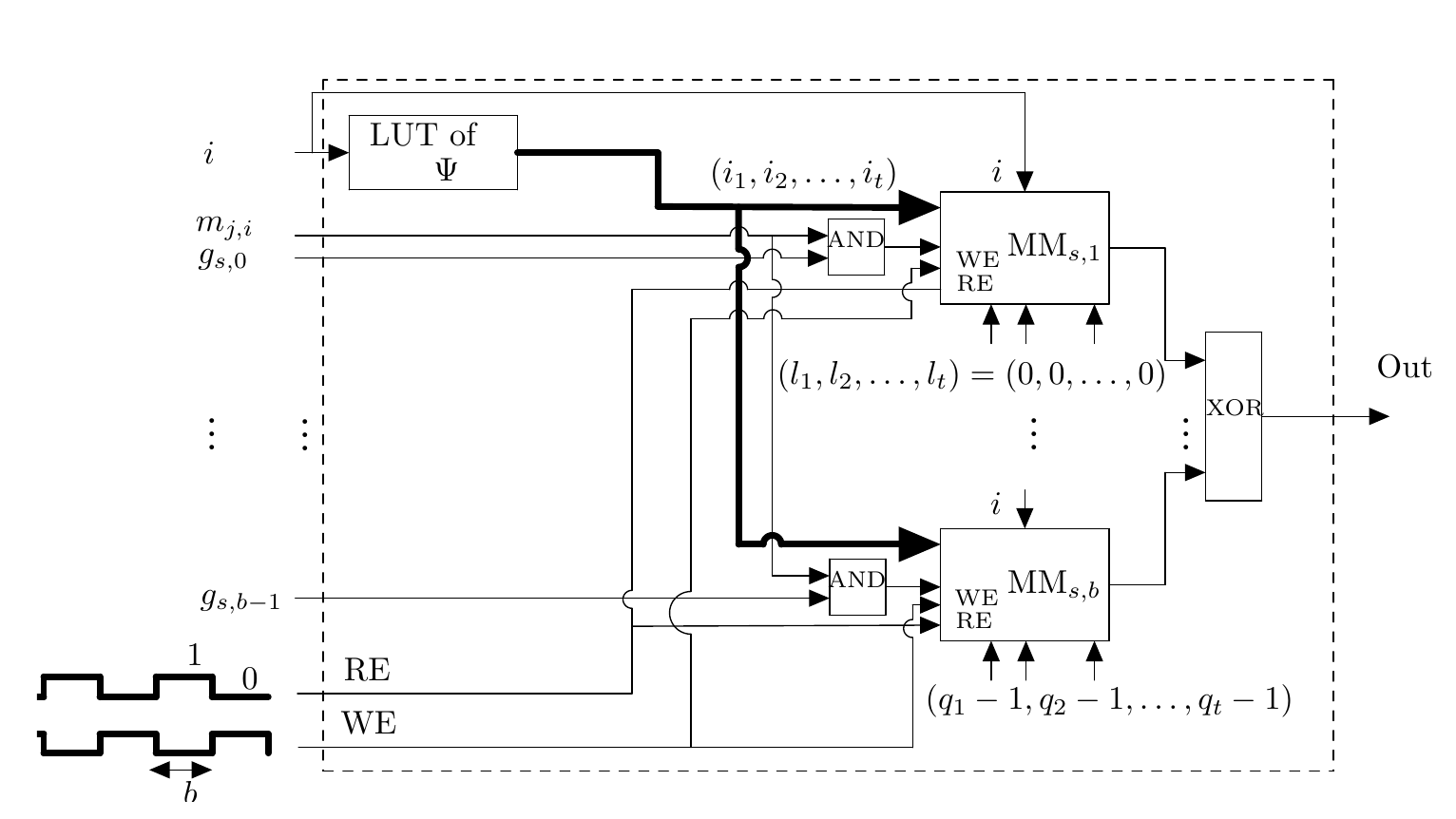}
\caption{The implementation of the $s^{th}$ partial multiplier circuit, $\textrm{PMC}_s$, for $s=1,\ldots,n$.}\label{PMCs}
\hrulefill
\end{figure*}

Each PMC itself is composed  of $b$ monomial multipliers (MMs), which are denoted by $\textrm{MM}_{s,1},\ldots,\textrm{MM}_{s,b}$ for $\textrm{PMC}_s$ and $s=1,\ldots,n$. The structure of $\textrm{PMC}_s$, for $s=1,\ldots,n$, is depicted in \figurename~\ref{PMCs}. In $\textrm{PMC}_s$, the monomial $u_s(x_1,\ldots,x_t)g_s=\sum_{j=0}^{b-1}u_{s,j}X^{\Psi^{-1}(j)}g_s$ is multiplied by $m_{\mathcal{L}'}$ in $kb$ clocks. In the implementation of this circuit, we use a lookup table (LUT) that preserves the input-output relation of the map $\Psi$, which is defined in (\ref{Psi_def}). The ports RE and WE in PMCs and MMs denote the read enable and write enable terminals, respectively, which are used to enable  the reading and writing operations on the appropriate RAMs.

\begin{figure*}
\centering
\includegraphics[width=6in]{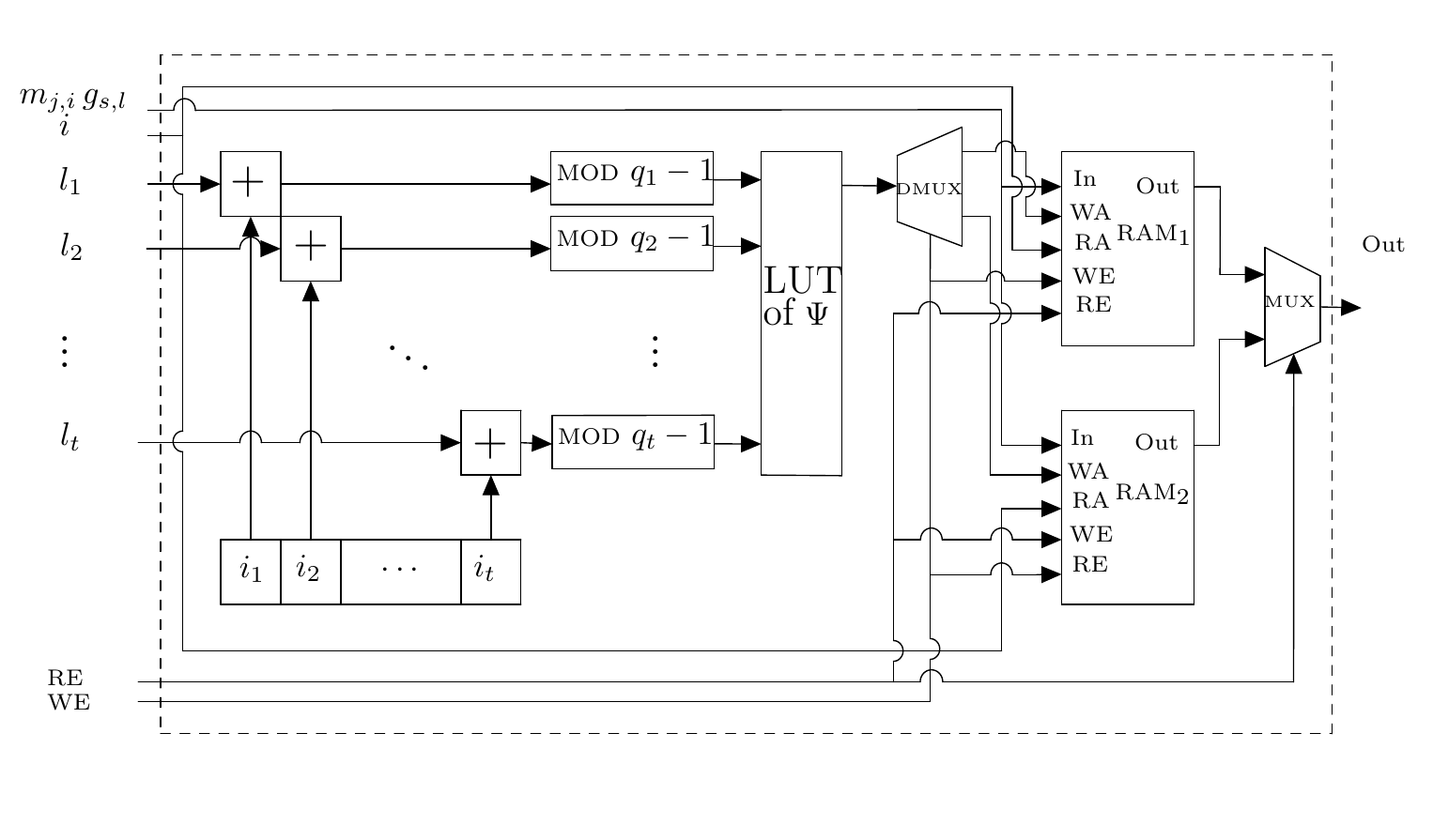}
\caption{The implementation circuit of $\textrm{MM}_{s,l}$, for $l=1,\ldots,b$.}\label{MMsl}
\hrulefill
\end{figure*}

The implementation circuit of $\textrm{MM}_{s,l}$, for $l=1,\ldots,b$, is presented in \figurename~\ref{MMsl}  that takes  $m_{j,i}X^{\Psi^{-1}(i)}g_{i_j}=m_{j,i}x_1^{i_1}\otimes \cdots\otimes x_t^{i_t}g_{i_j}$ in clock $jn+i$, for $j=0,\ldots,k$, $i=1,\ldots,b$, and multiplies it by $u_{s,l}X^{\Psi^{-1}(l)}g_{s}=u_{s,l}x_1^{l_1}\otimes \cdots \otimes x_t^{l_t}g_s$ in the group ring $R'G$. The ports RA and WA are used to specify the read address and  write address in the RAMs. After each $b$ clocks, the writing operation in one of the provided RAMs in $\textrm{MM}_{s,l}$, for all $1\leq s\leq n$ and $1\leq l\leq b$, is finished and the reading process in $\textrm{MM}_{s,l}$ starts from the beginning address of this RAM. At the same time, the  reading operation from the other RAM of each $\textrm{MM}_{s,l}$  is finished and the writing on it will be started. Thus, a rectangular waveform with $50\%$ duty cycle and the pulse width of $2b$ clocks can be used for controlling the read and write operations.

The implementation of this naive encoder can be improved in many aspects. It is presented to make the understanding of the encoding procedure  easier for the one who is not familiar with  group ring operations. However, the proposed encoder have some good properties like using RAMs instead of shift registers that decreases the implementation cost and also the  power required for encoder circuit. The implementation of this circuit requires $nb$ $2$-to-$1$ multiplexers, $nb$ $1$-to-$2$ and $n$ $1$-to-$n$ demultiplexers, $nb+1$  LUT (which itself is composed of multiplexers)  with $b$ inputs and $t$ outputs, $nb+log_2(b)-2$ AND gates, $n$ XOR gates with $b$ inputs and $n$ XOR gates with two inputs. In Addition, we require $2nb$ full adders (FAs) with $\log_2(q_1-1)$ bits,  $2nb$ FAs with  $\log_2(q_2-1)$ bits,... and $2nb$ FAs with  $\log_2(q_t-1)$ bits, which is equivalent to have $2nb$ FAs with $\log_2(b)$ bits. The memory requirements include a RAM with $2nb^2$ bits and $(tb+1)n+\log_2(b)$ flip-flops, (the terms $\log_2(b)$ in the number of registers and $\log_2(b)-2$ in the number of AND gates are due to the implementation of a counter in the encoder circuit that counts from $0$ to $b-1$ continuously). Since the encoding is performed in $kb$ clocks, all the space complexity, the time complexity and the memory requirements of this encoder remain linear in the code length $nb$.

Considering the  encoder described above, does not motivate us for implementing the encoding of group ring based QC-LDPC codes, because another encoder with linear time-space complexities  is proposed  in \cite{4} for QC-LDPC codes. In the sequel, we analyze the complexity of  encoding over group rings theoretically, that indicates the existence of a faster encoder, compared to the one in \cite{4},  for group ring based QC-LDPC codes. Our analysis requires some backgrounds about generalizing FFT convolution over group rings.
\subsection{Complexity analysis}
Mathematically looking, our encoding is the multiplication of two group ring elements $m_{\mathcal{L}'}$ and $u$ in $R'G$, where $G$ is an Abelian group of order $n$ and $R'=R_1\otimes R_2\otimes \cdots \otimes  R_t$, with
\begin{equation}
R_i=\mathbb{F}_2C_{q_i-1}\cong\frac{\mathbb{F}_2[x_i]}{\left\langle x_i^{q_i-1}-1\right\rangle},\quad i=1,\ldots,n.
\end{equation}
Treating a group ring as the space of functions mapping a group to a ring, the  multiplication in a group ring is the convolution of two functions therein. Due to the recent advances in signal processing and computer algebra \cite{group_ring_FFT}, discrete Fourier transform (DFT) has been generalized to finite rings which model quantized sequences. The widespread use of the
DFT is mostly caused by the great efficiency of the fast Fourier transform (FFT) algorithm for its computation. Thus, convolutions
can be computed over finite rings using FFT techniques. The FFT algorithm itself is independent of the ring which is used, but depends only on
the sequence length (in our case the group order of $G$)~\cite{group_ring_FFT}. Following the general case that is considered in \cite{group_ring_FFT}, let $R$ be a commutative ring with identity and $G$ be a finite
Abelian group of order $n$.
\begin{Definition}
It is said that $R$ supports a discrete Fourier transform over $G$ if $RG$ is isomorphic to $R^n$, which is the pointwise product algebra of $n$-tuples from $R$. An isomorphism $\Gamma: RG\rightarrow R^n$  is called a (generalized) discrete Fourier transform which is not necessarily
unique.
\end{Definition}

The usual cyclic convolution of sequences with length $n$ fits into this framework by choosing $G$ as $C_n$, which is the cyclic group of
order $n$.  The necessary and sufficient conditions on $R$ to make $RG$ isomorphic to $R^n$ were determined in~\cite{group_ring_FFT}. Consider $R$ as a finite commutative ring with identity. Then, $R$ can be written as a direct sum of local rings, i.e., $R=R_1\oplus\cdots  \oplus R_l$, where $R_i$'s are commutative local rings with identity \cite[Theorem VI.2]{macdonald}. A local ring is a commutative ring which has exactly one maximal ideal $M$.
It is proved  that $R$ supports a Fourier transform over $G$ if and only if each
$R_i$ supports a Fourier transform over $G$ \cite[Theorem 1]{group_ring_FFT}.
\begin{theorem}[\emph{\cite[Theorem 2]{group_ring_FFT}}]\label{FFT1}
Let $R$ be a local ring and $G$ be a finite Abelian group of order $n$ and exponent $m$, which is the maximum of the orders of the elements of $G$. Then, $R$ supports a discrete Fourier transform over $G$ if and only if
\begin{enumerate}
  \item $R$ contains a primitive $m^{th}$ root of unity\footnote{Let $R$ be a commutative ring with identity. An element $\xi\in R$ is a primitive $m^{th}$ root of unity if $\xi^m = 1$ and $\xi^k\neq 1$ for $1 < k < m$.};
  \item $m$ is a unit in $R$.
\end{enumerate}
\end{theorem}

If $M$ is the maximal ideal of a local ring $R$, then $R/M$ is a finite field which is called the \emph{residue field}. Let $R$ be a finite ring with decomposition $R = R_1\oplus \cdots \oplus R_l$, where $R_i$'s, are finite local rings with residue
fields $R_i/M_i$ of order $p_i^{k_i}$, $i=1,\ldots,l$. Then, we have the following theorem.
\begin{theorem}[\emph{\cite[Theorem 4]{group_ring_FFT}}]\label{FFT2}
Let $R = R_1\oplus \cdots \oplus R_l$, where $R_i$'s are finite local rings and define $O(R)=\mathrm{gcd}\left\{p_i^{k_i}-1,\,\,i=1,\ldots,l\right\}$. If $G$ is a finite Abelian group of exponent $m$, then $R$ supports a discrete Fourier transform over $G$ if and only if $m|O(R)$.
\end{theorem}

For any finite ring, $O(R)$  determines exactly the sequence lengths (or the group exponents) for which a
discrete Fourier transform can be defined, namely the divisors of $O(R)$ \cite{group_ring_FFT}. After giving the necessary and sufficient conditions on the finite ring $R$ for supporting DFT over $G$, the authors of \cite{group_ring_FFT} have introduced some conditions for implementing the FFT and the fast convolution methods over finite rings.

Due to  above discussions, by choosing an appropriate size $n$ for the group $G$,  the convolution of $m_{\mathcal{L}'}$ and $u$
can be done by using an FFT algorithm that involves $O(n\log_2 n)$ multiplication over $R$. Now, we analyze the complexity of multiplication over $R$. Our goal is to find an upper bound for the cost of multiplying two polynomials in $R[X]$ of degree less than $s$.
For a given ring $R$, one of the best currently
known bounds that indicates the  computational cost of multiplying two single variable polynomial over $R$  was obtained by Cantor and Kaltofen in \cite{ring_mult}. Their algorithm performs $O(s \log_2 s \log_2 \log_2 s)$ additions and subtractions and $O(s \log_2 s)$
multiplications in $R$ and it relies on suitable incarnations of the FFT. In our case, the multiplication of two elements in the ring of $t$-variates polynomials, $\mathbb{F}_2[x_1,\ldots,x_t]$, is required, where the maximum degree of the $i^{th}$ variable $x_i$ is $q_i-1$, for $i=1,\ldots,t$. We  consider $\mathbb{F}_2[x_1,\ldots,x_t]$ as $R_1[x_t]$, where $R_1=\mathbb{F}_2[x_1,\ldots,x_{t-1}]$. Thus, the multiplication of two elements in $\mathbb{F}_2[x_1,\ldots,x_t]$ involves $O\left((q_t-1)\log_2(q_t-1)\right)$ multiplications in $R_1$. Similarly, $R_1$ can be written as $R_2[x_{t-1}]$, where $R_2=\mathbb{F}_2[x_1,\ldots,x_{t-2}]$, and every multiplication in $R_1$ is equivalent to $O\left((q_{t-1}-1)\log_2(q_{t-1}-1)\right)$ multiplications in $R_2$. Consequently, the multiplication in $\mathbb{F}_2[x_1,\ldots,x_t]$ involves $O((q_t-1)(q_{t-1}-1)\log_2(q_t-1)\log_2(q_{t-1}-1))$ multiplications in $R_2$.
Using the same procedure for other $x_i$'s  yields the multiplication complexity in $\mathbb{F}_2[x_1,\ldots,x_t]$. Thus, the  number of binary operations required for multiplying two elements  in $\mathbb{F}_2[x_1,\ldots,x_t]$ is $O\left(\prod_{i=1}^t (q_i-1)\log_2 (q_i-1)\right)$, which is simplified to $O\left(b\prod_{i=1}^t \log_2 (q_i-1)\right)$.
Hence, we reach an upper bound for the cost of multiplying two  elements in $R'G$ that counts the total required number of AND gates as $O\left(nb\log_2 n\prod_{i=1}^t \log_2 (q_i-1)\right)$. It is significantly lower than $n^2b^2$ binary multiplications involved in the regular multiplication of two elements in $R'G$. Dividing $O\left(nb\log_2 n\prod_{i=1}^t \log_2 (q_i-1)\right)$ operations into $nb$ time intervals admits an encoder with linear time complexity in the code length $nb$ and logarithmic space complexity  $O\left(\log_2 n\prod_{i=1}^t \log_2 (q_i-1)\right)$, which is a significant reduction in the space complexity compared to the proposed encoder in \cite{4}.
%Due to the lower bounds of \cite{banihashemi}, for girth $6$ QC-LDPC codes, $b\geq n$.  Thus, the term $n+b$ is not more than $2b$.
We can also implement an encoder with space complexity $O(nb)$ and time complexity $O\left(\log_2 n\prod_{i=1}^t \log_2 (q_i-1)\right)$ that indicates a faster implementation of encoding for group ring based QC-LDPC codes compared to the other families. For example, if we consider $t=1$, the time complexity of the proposed encoding for group ring based QC-LDPC codes is determined as $O\left(\log_2 n\log_2 b\right)$. Due to the given lower bounds in \cite{banihashemi}, for a $4$-cycle free QC-LDPC code we have $n\leq b$. Thus, using the encoder of  group ring based QC-LDPC codes gives the time complexity of $O\left((\log_2 b)^2\right)$ while using the encoder of \cite{4} gives the time complexity of $O\left( b^2\right)$ that indicates a significant reduction in the time complexity of encoding.

%=================================================================
\textcolor{mycolor}{Consequently, the implementation of FFT for group ring based QC-LDPC codes makes a significant improvement in the complexity of encoding. Using FFT has been shown to be amenable to analysis and construction of
some QC-LDPC codes and obtaining their generator matrices \cite{C10}. The introduction of matrix transformation via the Galois Fourier transform (GFT) is an important development of  quasi-cyclic (QC) codes \cite{C10}. Galois Fourier transform was applied in \cite{FFT_encoding} for implementing two low-complexity encoding algorithms for quasi-cyclic codes. In the sequel, we give a brief introduction on GFT and the encoding methods introduced in \cite{FFT_encoding}. We also present a comparison between the proposed encoding method in this paper and the proposed methods in  \cite{FFT_encoding}.
\newline
Consider a binary QC code with $mb\times nb$ parity-check matrix $\mathbf{H}$,  which is an $m\times n$ array of binary $b\times b$ circulant matrices where $b$ is assumed to be an   odd number \cite{C10}. Using Fermat-Euler Theorem, there is a two's power number $q$ such that $q-1$ is divisible by $b$ \cite{blahut}. Let $\alpha$ be an element in $\mathbb{F}_q$ of order $b$. Let $\mathbf{a}=(a_0,\ldots,a_{b-1})$  be a vector over $\mathbb{F}_q$. Its Fourier
transform \cite{FFT_encoding}, denoted by $\mathcal{F}[\mathbf{a}]$, is given by the vector $\mathbf{f}=(f_0,\ldots,f_{b-1})$ whose $t^{th}$ component, $f_t$, for $0\leq t\leq b-1$, is given by $f_t=\sum_{l=0}^{b-1}\alpha^{tl}a_l$. The vector $\mathbf{a}$, which is the inverse Fourier transform of the vector $\mathbf{f}$, denoted by $\mathcal{F}^{-1}[\mathbf{f}]$, can be retrieved as $a_l=\sum_{t=0}^{b-1}\alpha^{-tl}f_t$. Define the following two $b\times b$ \emph{Vandermonde}  matrices over $\mathbb{F}_q$: $\mathbf{V}=\left[\alpha^{-ij}\right]$ and $\mathbf{V}=\left[\alpha^{ij}\right]$, for $0 \leq i,j < b$. The following lemma says that all circulant matrices can be diagonalized by the same similarity transformation $\mathbf{A}\mapsto \mathbf{V} \mathbf{A} \mathbf{V}^{-1}$.
\begin{lemma}[\emph{\cite[Lemma 1]{C10}}]
Let $\mathbf{A}$ be a $b\times b$ circulant matrix over $\mathbb{F}_q$ with generator $(a_0,\ldots,a_{b-1})$. Let  $\mathbf{V}=\left[\alpha^{-ij}\right]$, $0 \leq i,j < b$, be a $b\times b$ matrix, where $\alpha$ is an element in
$\mathbb{F}_q$ of order $b$. Then,  $\mathbf{V} \mathbf{A} \mathbf{V}^{-1}$ is a $b\times b$ diagonal matrix, $\mathbf{A}^{\mathcal{F}}$, whose diagonal vector is the Fourier transform of $(a_0,\ldots,a_{b-1})$, i.e.,
\begin{eqnarray}
  \mathbf{A}^{\mathcal{F}} &=& \mathrm{diag}\left(\sum_{l=0}^{b-1}a_l, \sum_{l=0}^{b-1}\alpha^l a_l,\ldots, \sum_{l=0}^{b-1}\alpha^{(b-1)l}a_l \right).
\end{eqnarray}
\end{lemma}
The fact that all circulant matrices can be diagonalized using the same similarity transformation allows to diagonalize any array of circulant matrices as follows \cite{C10}.
\begin{lemma}[\emph{\cite[Lemma 2]{C10}}]
Let $\mathbf{H}$ be an $m\times n$ array of $b\times b$ circulant matrices over $\mathbb{F}_q$, $\mathbf{H}=\left[\mathbf{A}_{i,j}\right]$, where $\mathbf{A}_{i,j}$ is a circulant matrix with generator $(a_{i,j,0},\ldots,a_{i,j,b-1})$, $0\leq i<m$, $0\leq j <n$. Define
\begin{eqnarray}
  \mathbf{H}^{\mathcal{F}} = \mathrm{diag}(\underbrace{\mathbf{V},\ldots,\mathbf{V}}_{m})\,\mathbf{H}\,\, \mathrm{diag}(\underbrace{\mathbf{V}^{-1},\ldots,\mathbf{V}^{-1}}_{n}).
\end{eqnarray}
Then, $\mathbf{H}^{\mathcal{F}}$ is an $m\times n$ array of $b\times b$ diagonal matrices. In
particular, $\mathbf{H}^{\mathcal{F}}=\left[\mathbf{A}_{i,j}^{\mathcal{F}}\right]$, where
\begin{eqnarray*}
  \mathbf{A}_{i,j}^{\mathcal{F}} = \mathrm{diag}\left(\sum_{l=0}^{b-1}a_{i,j,l}, \sum_{l=0}^{b-1}\alpha^l a_{i,j,l},\ldots, \sum_{l=0}^{b-1}\alpha^{(b-1)l}a_{i,j,l} \right),
\end{eqnarray*}
for $0\leq i<m$, $0\leq j <n$.
\end{lemma}
Next, it has been shown that some row and column permutations can be performed on $\mathbf{H}^{\mathcal{F}}$  to get a $b\times b$ diagonal array of $m\times n$ matrices. For any integer $i$, denote the nonnegative integer
less than $b$ and congruent to $i$ modulo $b$ by $(i)_b$.
\begin{lemma}[\emph{\cite[Lemma 2]{C10}}]
Let $\pi_{m}(i)=m(i)_b+\left\lfloor i/b\right\rfloor$, $0\leq i < mb$, and $\pi_{n}(j)=n(j)_b+\left\lfloor j/b\right\rfloor$, $0\leq j < nb$. Then,  $\pi_{m}$ is a permutation on $\left\{0,1,\ldots , mb-1\right\}$ and $\pi_{n}$ is a permutation on $\left\{0,1,\ldots , nb-1\right\}$. Furthermore, permuting the rows and columns of $\mathbf{H}^{\mathcal{F}}$ using $\pi_{m}$ and $\pi_{n}$, respectively, yields the  matrix $\mathbf{H}^{\mathcal{F},\pi}=\mathrm{diag}\left(\mathbf{B}_{0},\mathbf{B}_1,\ldots ,\mathbf{B}_{b-1}\right)$ which is a $b\times b$  diagonal array of $m\times n$ matrices $\mathbf{B}_t=\left[b_{i,j,t}\right]$, where $b_{i,j,t}=\sum_{l=0}^{b-1}\alpha^{tl}a_{i,j,l}$, for $0\leq i<m$, $0\leq j<n$ and $0\leq t< b$.
\end{lemma}
It is proved that the correspondence $\mathbf{H}\leftrightarrow\mathbf{H}^{\mathcal{F},\pi}$  is a one-to-one correspondence between arrays of circulant matrices and diagonal arrays of matrices \cite[Theorem 1]{C10}. The following theorem uses the same procedure for the dual code of a QC code to obtain its generator matrix.
\begin{theorem}[\emph{\cite[Theorem 5]{C10}}]
  Let $\mathbf{H}$ be an $m\times n$ array of $b\times b$ circulant matrices and $\mathbf{H}^{\mathcal{F},\pi}=\mathrm{diag}\left(\mathbf{B}_{0},\mathbf{B}_1,\ldots ,\mathbf{B}_{b-1}\right)$, where $\mathbf{B}_{0},\mathbf{B}_1,\ldots ,\mathbf{B}_{b-1}$ are $m\times n$ matrices. Let $m'$ be a positive
integer not less than
\begin{eqnarray*}
  n-\min\left\{\mathrm{rank}\left(\mathbf{B}_0\right),\mathrm{rank}\left(\mathbf{B}_1\right),\ldots,\mathrm{rank}\left(\mathbf{B}_{b-1}\right)\right\}.
\end{eqnarray*}
Let $\mathbf{D_s}$ be an $m'\times n$ matrix of rank $n-\mathrm{rank}\left(\mathbf{B}_s\right)$ such that $\mathbf{B}_s\mathbf{D}_s^t=\mathbf{0}$, for $0\leq s<b$. Let $\hat{\mathbf{G}}=\mathrm{diag}\left(\mathbf{D}_0,\mathbf{D}_1,\ldots,\mathbf{D}_{b-1}\right)$ and $\hat{\mathbf{G}}^{\pi^{-1}}$ be $\hat{\mathbf{G}}$ after applying the permutations $\pi_{m'}^{-1}$ and $\pi_{n}^{-1}$ on the rows and columns of $\hat{\mathbf{G}}$, respectively. Then,
\begin{eqnarray*}
  \mathbf{G} &=& \hat{\mathbf{G}}^{\pi^{-1},\mathcal{F}} \\
             &=& \mathrm{diag}(\underbrace{\mathbf{V},\ldots,\mathbf{V}}_{m'})\,\hat{\mathbf{G}}^{\pi^{-1}}\,\, \mathrm{diag}(\underbrace{\mathbf{V}^{-1},\ldots,\mathbf{V}^{-1}}_{n}),
\end{eqnarray*}
is an $m'\times n$ array of $b\times b$ circulant matrices such that $\mathbf{H}\mathbf{G}^t=\mathbf{0}$ and $\mathrm{rank}(\mathbf{G})=nb-\mathrm{rank}(\mathbf{H})$. Furthermore, if $\mathbf{H}$ is binary,
then the matrices $\mathbf{D}_{0},\mathbf{D}_1,\ldots ,\mathbf{D}_{b-1}$ can be selected such that $\mathbf{G}$
is binary.
\end{theorem}
The transformation introduced above, which is denoted by GFT in the sequel, was applied in \cite{FFT_encoding} to design two efficient encoding approaches. In the design of these encoding methods, it is assumed that $b=q-1$. In this case, the GFT of a $b$-tuple over $\mathbb{F}_2$ is a $b$-tuple over $\mathbb{F}_{2^r}$, i.e., $q=2^r$. It is  proved that the submatrices on the main diagonal of the block diagonal matrix $\hat{\mathbf{G}}^{\mathcal{F},\pi}$ satisfy the following condition which is known as the \emph{conjugacy constraint} \cite{blahut}
\begin{eqnarray}
  \mathbf{D}_{(2t)_b} &=& \mathbf{D}_t^{\circ 2},
\end{eqnarray}
where $\mathbf{M}^{\circ t}$ denotes the  Hadamard product\footnote{The Hadamard product of two matrices
$\mathbf{M}=\left[m_{i,j}\right]$ and $\mathbf{N}=\left[n_{i,j}\right]$ of the same size, denoted by $\mathbf{M}\circ \mathbf{N}$, is defined as their element-wise product, i.e., $\mathbf{M}\circ \mathbf{N}=\left[m_{i,j}n_{i,j}\right]$.} of $t$ copies of the matrix $\mathbf{M}$ and $t$ is a
nonnegative integer. The matrix $\mathbf{D}_{(2t)_b}$ is called a \emph{conjugate matrix} of $\mathbf{D}_t$ \cite{FFT_encoding}. Thus, we can group all the submatrices on the main diagonal $\mathbf{D}_i$, $0\leq i<b$
into $\lambda$ conjugacy classes, $\Psi_0,\Psi_1,\ldots,\Psi_{\lambda-1}$, where
\begin{eqnarray*}
  \Psi_i &=& \left\{\mathbf{D}_{t_i},\mathbf{D}_{(2t_i)_b},\ldots , \mathbf{D}_{(2^{\eta_i-1}t_i)_b}\right\} \\
    &=& \left\{\mathbf{D}_{t_i},\mathbf{D}_{t_i}^{\circ2},\ldots,\mathbf{D}_{t_i}^{\circ 2^{\eta_i-1}} \right\},
\end{eqnarray*}
in which $\mathbf{D}_{t_i}$ is the representative of the conjugacy class $\Psi_i$ and $\eta_i$ is the least integer satisfying $(2\eta_i t_i)_b = t_i$, and $\eta_i$  divides $r$ \cite{blahut}. The conjugacy classes have a key role in the  encoding methods proposed in \cite{FFT_encoding} which are given next.}

\textcolor{mycolor}{Consider an $(nb, kb)$ QC code $\mathcal{C}$ over $\mathbb{F}_{2^r}$ with generator matrix $\mathbf{G} = \left[\mathbf{W}_{i,j} \right]$, $0 \leq i < k$, $0 \leq j < n$, which is a $k\times n$ block matrix of $b\times b$ circulants. Suppose $\mathbf{m} = \left[\mathbf{m}_i\right]$ as the message vector and the resulting codeword as $\mathbf{c} =
\left[ \mathbf{c}_j \right]$, $0 \leq i < k$, $0 \leq j < n$, where both $\mathbf{m}_i$ and $\mathbf{c}_j$ are vectors
of length $b$. Since $\mathbf{W}_{i,j}$ is a circulant, $\mathbf{m}_i\mathbf{W}_{i,j}=\mathbf{m}_i\mathbf{V}\mathbf{V}^{-1}\mathbf{W}_{i,j}\mathbf{V}\mathbf{V}^{-1}=\left(\mathbf{m}_i^{\mathcal{F}}\mathbf{W}_{i,j}^{\mathcal{F}}\right)^{\mathcal{F}^{-1}}$, where $\mathbf{m}_i^{\mathcal{F}}=\mathbf{m}_i\mathbf{V}$.
As a result, $\mathbf{c}_j$ can be computed by GFT as \cite{FFT_encoding}
\begin{eqnarray*}
  \mathbf{c}_j &=& \left(\mathbf{m}_0^{\mathcal{F}}\mathbf{W}_{0,j}^{\mathcal{F}}+\mathbf{m}_1^{\mathcal{F}} \mathbf{W}_{1,j}^{\mathcal{F}}+\cdots + \mathbf{m}_{k-1}^{\mathcal{F}}\mathbf{W}_{k-1,j}^{\mathcal{F}}\right)^{\mathcal{F}^{-1}}.
\end{eqnarray*}
Due to the block diagonal structure of $\hat{\mathbf{G}}^{\mathcal{F},\pi}$, multiplying
a vector of length $kb$ by such a $kb\times nb$ matrix can be computed on $b$ submatrices $\mathbf{D}_i$'s of size $k\times n$ separately. Thus, using GFT reduces the number of operations efficiently by a factor $b$. This approach is well-known for  implementing the filtering by discrete Fourier transform. It greatly reduces the computational complexity of encoding of nonbinary QC codes \cite{FFT_encoding}. The overall computational complexity of the
GFT encoding is less than  $bk(n-k)(\log_2^2 b+ \log_2 b)+(n+k)b (\log_2b)^{\log_2 6}+(n+k)b^2(\log_2 b)^{\log_2(3/4)}$
in terms of bit operations \cite{FFT_encoding}. To compare the complexity of GFT encoding with the proposed encoding based on group ring multiplication, we consider $k=n$, and the worst case of complexity analysis is $n=b$, since $n\leq b$.
% using the upper bound $n\leq b$, the worst case of complexity analysis is  $n=b$.
Thus, the complexity of GFT encoding is upper bounded by $2b^2(\log_2b)^{\log_2 6}+2b^3(\log_2 b)^{\log_2(3/4)}$. Since $\log_2(6)=2.585$, both terms $O(b^2(\log_2b)^{\log_2 6})$ and $O(b^3(\log_2 b)^{\log_2(3/4)})$ are higher than the complexity of group ring based QC-LDPC codes, which is $O(b^2(\log_2 b)^2)$. The overall memory consumption of the GFT encoding
is $bk(n - k)$ Galois symbols, which is the same as that of the regular encoding \cite{FFT_encoding}.}

\textcolor{mycolor}{As mentioned  above, the GFT encoding can greatly reduce the complexity of encoding of nonbinary QC codes.
However, its efficiency decreases for binary codes, because it involves many Galois field multiplications in the vector-matrix multiplication \cite{FFT_encoding}.  Thus, it is suggested to encode a binary message
$\mathbf{m}$ directly in the transform domain to save these Galois field multiplications. Thus, the encoding in the transform domain (ETD) was presented in \cite{FFT_encoding} for binary QC codes. We avoid going through the detail of this encoding approach and we only present its complexity analysis. The overall computational
complexity of the ETD is less than $bk(n-k)\log_2 b+n(2b-\lambda)\log_2^2(b)+n(b-\lambda)\log_2 b+nb^2(\log_2 b)^{(\log_2(3/4))}$. Considering $k=n=b$ yields the approximation of complexity terms for ETD encoding  as $O(b^2\log_2^2(b))$ and $O(b^3(\log_2 b)^{(\log_2(3/4))})$. Similar to GFT encoding, the term $b^3(\log_2 b)^{(\log_2(3/4))}$ makes the complexity of the ETD encoding higher than the complexity of encoding based on group ring multiplication.}
\textcolor{mycolor}{\begin{rem}
In summary, using FFT in lowering the encoding complexity can be done in different manners. In fact, using an appropriate implementation of FFT that respects the algebraic structure of the code, increases the efficiency of encoding. For example,  GFT encoding and  ETD encoding cannot be employed for group ring based QC-LDPC codes, because these approaches
highly depend on the circulant structure of the sub-blocks in the generator matrix. Thus, we can employ these encoding methods for CPM-QC-LDPC codes, but not for group ring based QC-LDPC codes, since they are designed based on QCPMs. The proposed encoding based on  group ring multiplication employs an appropriate implementation of the FFT that complies the structure of underlying group ring and obtains a remarkable reduction in the encoding complexity. The main reason for outperforming of the encoding method proposed  for group ring based QC-LDPC codes on other FFT based encoding methods, is using  two different FFTs in the encoding procedure. One of these FFTs is enabled by employing Abelian groups in the structure of the base matrix of group ring based QC-LDPC codes. The second one is enabled by modelling the sub-blocks of the message vectors as multivariate polynomials and implementing the partial multiplications in the components of codewords with fast convolution methods.
\end{rem}}
%=================================================================
\section{Simulation Results}\label{simulation}
\begin{figure}[!h]
\centering
\includegraphics[width=5in]{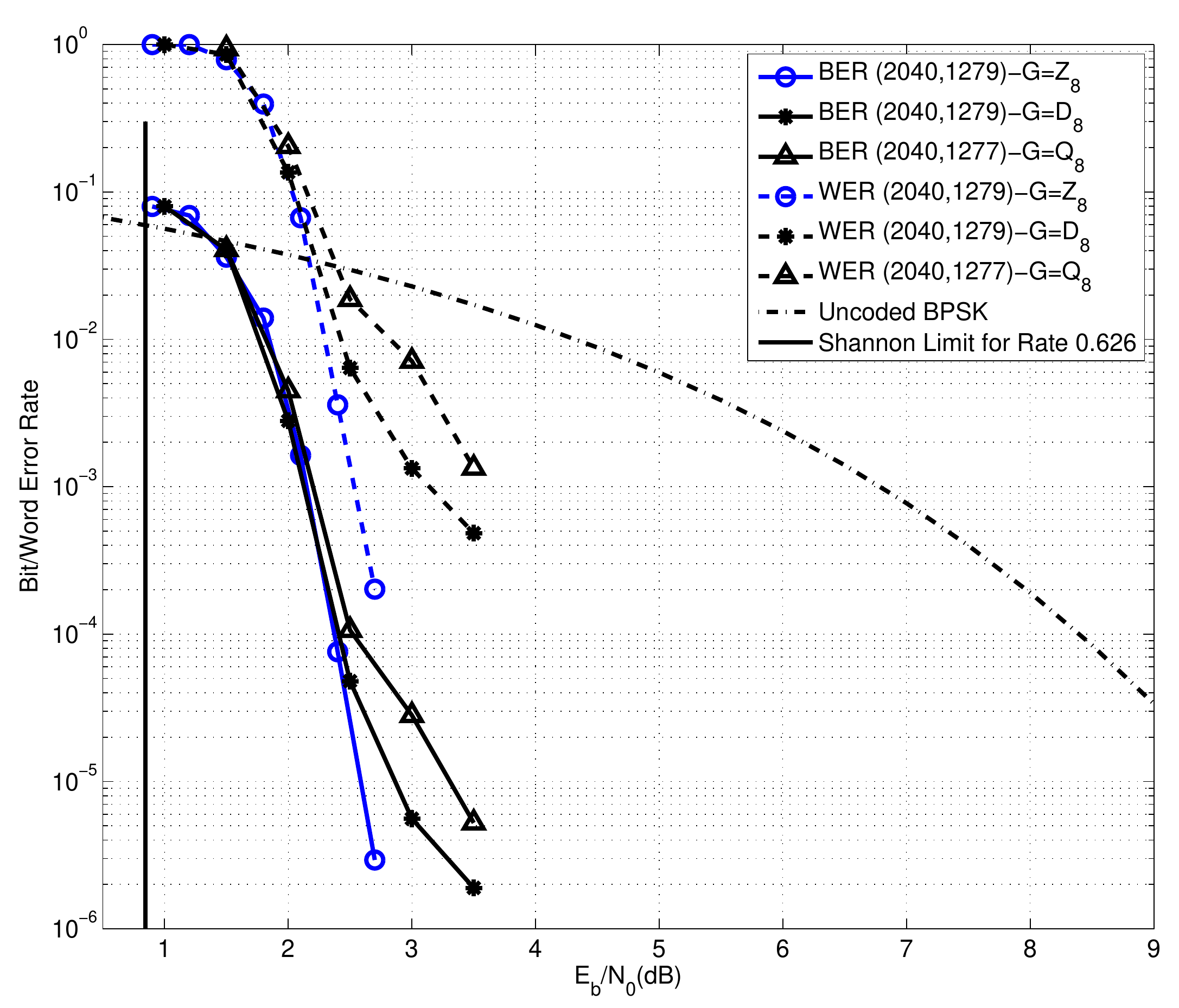}
\caption{BER and WER performances of the group ring based QC-LDPC codes based on different groups of order $8$.}
\label{figsim3}
\end{figure}

In this section, we present the numerical results that verify the efficiency of  group ring based QC-LDPC codes.   Bit error rate (BER) and word error rate (WER) performances of the  codes constructed based on groups of order $8$ and the structure of Theorem~\ref{theorem2} are presented in \figurename~\ref{figsim3}. For $G=D_8$, we use the rows $3,5,6$ of the $\mathbb{F}_{2^8}G$-matrix $\mathbf{W}$ as the parity-check matrix. For $G=Q_8$, we use the rows $4,5,6$ and for $G=\mathbb{Z}_8$ we use the first three rows of $\mathbf{W}$ as the parity-check matrix. Thus, the row and the column weights of these codes are $3$ and $8$, respectively. The null space of these matrices  are  $(1279,2040)$-regular QC-LDPC codes and their error performance using the SPA ($30$ iterations) over the AWGN channel with BPSK modulation, is illustrated in \figurename~\ref{figsim3}. The rate of these codes is $0.626$. Due to these numerical results, the error performance of the code based on $G=\mathbb{Z}_8$ is better than the error performance of the codes based on other groups with the same order. Thus, in  all  codes constructed  in the sequel, the group $G$ is considered to be a cyclic group.
%In the simulation, Code1 obtained from the rows 1,3 and 4 of (\ref{eq4}).  Code2 is obtained from the rows 6,7, and 8 of (\ref{eq4}) and Code3 corresponds to the rows 3,5 and 6 and the both of them are simulated at the same situation of Code1.
Our simulation results also indicate that in the case of using non-cyclic groups as the underlying group $G$, the error performance of the obtained code is related to the subarray of $\mathbf{W}$ which is used as the parity-check matrix of the code. In the rest of this section, the parity-check matrix of the given codes are corresponding to the subarrays of the form $\mathbf{H}(\rho,\gamma)$ at the upper left corner of $\mathbf{W}$, where $\rho$ and $\gamma$, for $1\leq\rho,\gamma\leq n$, denote the number of blocks in the rows and in the columns, respectively.

\begin{figure}[!h]
\centering
%\includegraphics[width=5in]{Performance.pdf}
%trim={<left> <lower> <right> <upper>}
\includegraphics[trim={1.5cm 0.9cm 2cm 1.2cm},clip,width=5in]{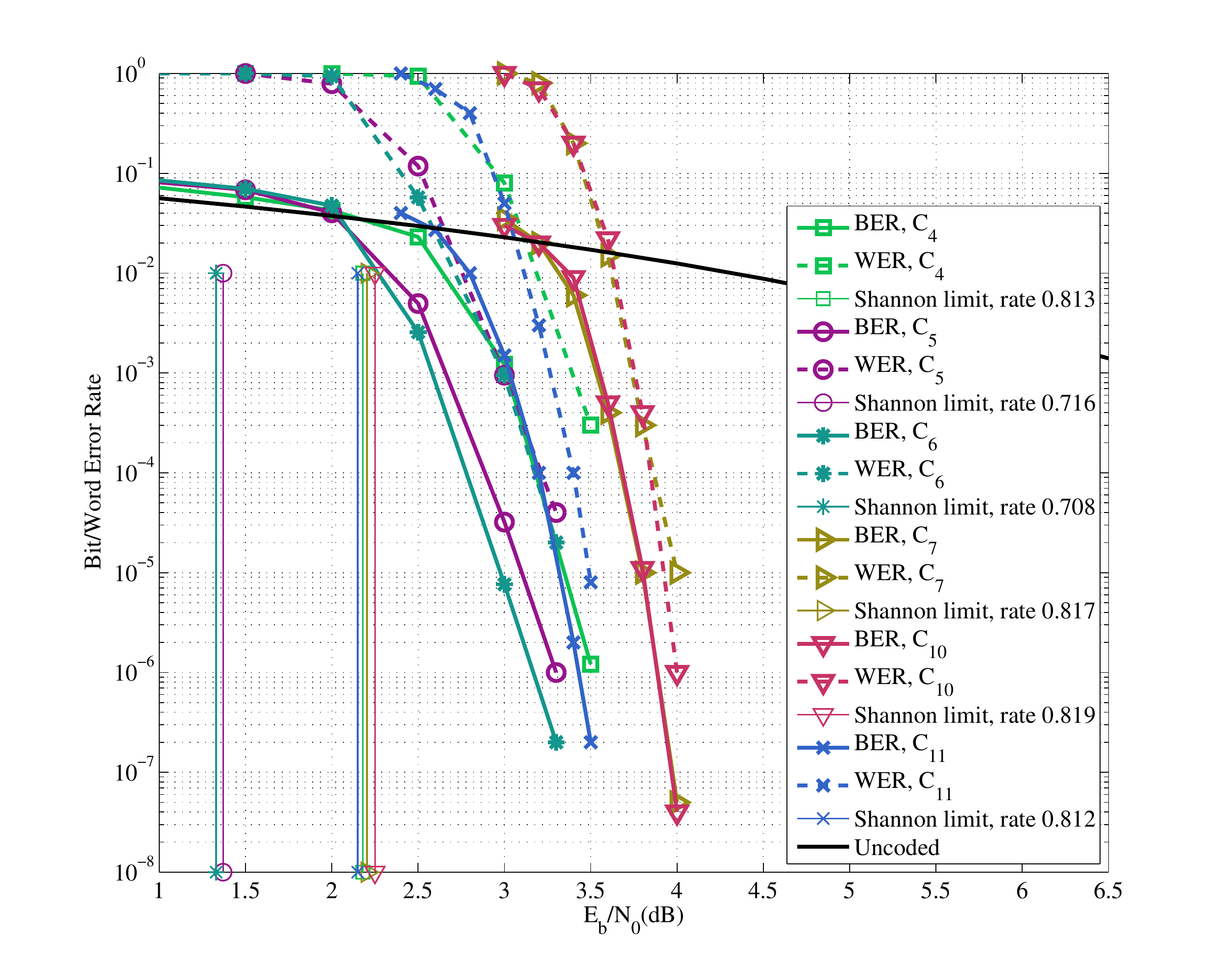}
\caption{Comparison between the error performance of the group ring based and finite field based QC-LDPC codes in \cite[Example 1]{QC_cyclic} over the AWGN channel.}
\label{figsim4}
\end{figure}

\begin{figure}[!h]
\centering
%\includegraphics[width=5in]{Performance.pdf}
%trim={<left> <lower> <right> <upper>}
\includegraphics[trim={1.5cm 0.9cm 2cm 1.2cm},clip,width=5in]{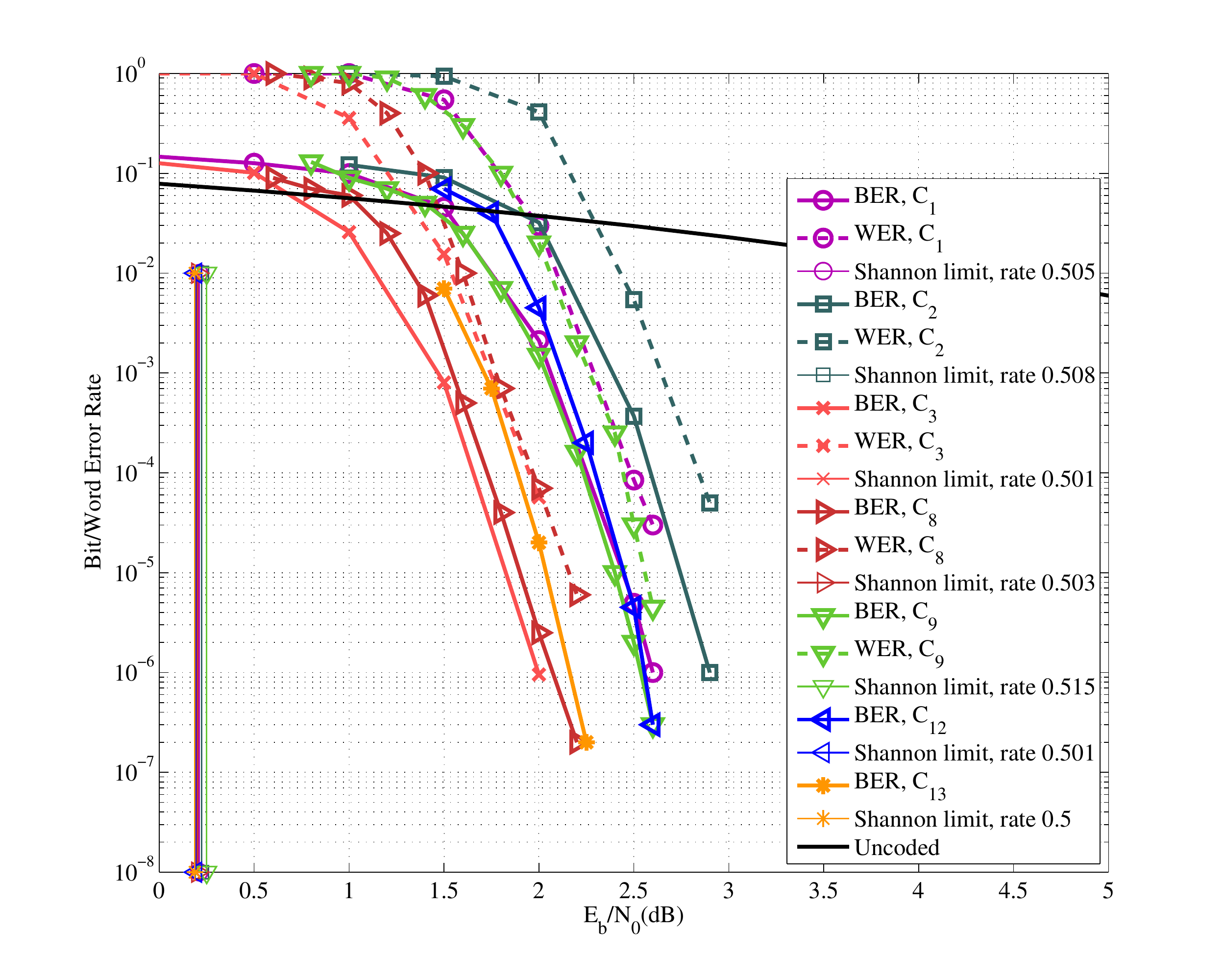}
\caption{Comparison between the error performance of group ring based and finite field based QC-LDPC codes in \cite{6,C12} over the AWGN channel.}
\label{figsim7}
\end{figure}

%================================
\textcolor{mycolor}{In \figurename~\ref{figsim4} and \figurename~\ref{figsim7}, we present the comparisons between the error performance of algebraic QC-LDPC codes constructed based on finite fields \cite{QC_cyclic,C12,C10,C11,6} and the error performance of group ring based QC-LDPC codes. Let $\mathbf{H}(3,8)$  and $\mathbf{B}$ be the parity-check matrix obtained from Theorem~\ref{theorem2} with $G=\mathbb{Z}_8$ and its corresponding base matrix, respectively. Denote the parity-check matrix obtained from the base matrix $-\mathbf{B}$ by $-\mathbf{H}(3,8)$ and   consider $\mathcal{C}_4$ as the null space of the parity-check matrix $\mathbf{H}=\left[
                     \begin{array}{c|c}
                       \mathbf{H}(3,8) & -\mathbf{H}(3,8) \\
                     \end{array}
                   \right]$. Then, $\mathcal{C}_4$ is a code with length $4080$, dimension $3319$ and rate $0.813$. Bit error and block error performances of $\mathcal{C}_4$ are illustrated in \figurename~\ref{figsim4}. At the BER of $10^{-6}$, $\mathcal{C}_4$ performs  $1.32$dB from the Shannon limit and it can be compared with $\mathcal{C}_7$, $\mathcal{C}_{10}$ and $\mathcal{C}_{11}$ which are introduced next. In \figurename~\ref{figsim4}, $\mathcal{C}_7$ is the $(3969, 3243)$ QC-LDPC code given in \cite[Example 1]{QC_cyclic} with rate $0.817$ and it performs $1.686$dB from the Shannon limit. The code  $\mathcal{C}_{10}$ in this figure is the $(4032, 3304)$ QC-LDPC code with rate $0.819$ given in \cite[Example 4]{C10} that performs $1.64$dB from the Shannon limit. The code  $\mathcal{C}_{11}$ is an algebraic irregular QC-LDPC code \cite[Example 2]{C11} with length $4032$, dimension $3276$ and rate $0.812$ that performs $1.28$dB from the Shannon limit.}

\textcolor{mycolor}{As another example, consider $\mathcal{C}_5$ as the null space of the parity-check matrix $$\mathbf{H}=\left[
                     \begin{array}{c|c}
                       \mathbf{H}(4,7) & -\mathbf{H}(4,7) \\
                     \end{array}
                   \right],$$
where $\mathbf{H}(4,7)$ is obtained from Theorem~\ref{theorem2} with $G=\mathbb{Z}_7$. This code is a $(1778,1273)$ QC-LDPC code with rate $0.716$ and it performs $1.93$dB from the Shannon limit. The error performance of this code can be compared with the error performance of $\mathcal{C}_6$ that performs $1.84$dB from the Shannon limit. This code is a $(2032,1439)$ QC-LDPC code with rate $0.708$ and it is obtained by considering $q=128$ and $\mathbf{H}(5,16)$ in \cite[Example 1]{QC_cyclic}.}

%Due to these simulation results, in the rates close to $0.5$, group ring based QC-LDPC codes outperform the QC-LDPC codes based on cyclic subgroups of finite fields. Indeed, the proposed method in \cite{QC_cyclic} generates high rate QC-LDPC codes with good error performance while, their method cannot generate good low rate QC-LDPC codes. In order to compare our codes with similar counterparts in \cite{QC_cyclic}, we took the  used base matrix in \cite[Example 1]{QC_cyclic}. Considering the subarray $\mathbf{H}(32,32)$ of the proposed base matrix in \cite[Example 1]{QC_cyclic} yields a code with rate $0.65$ which is still high for comparing with our example. Thus, we used their construction method with $q=256$ and subarrays $\mathbf{H}(4,8)$ and $\mathbf{H}(3,8)$ and both these codes admit an error floor in their error performance curve.  Using  the proposed method in \cite{QC_cyclic} for constructing low rate QC-LDPC codes generates high column weights that increases the required memory for saving their parity-check matrices, which is another disadvantage.
%\begin{figure}[!h]
%\centering
%\includegraphics[width=5in]{Performance4.pdf}
%\caption{Comparison between the error performance of the group ring based and finite field based QC-LDPC codes in \cite[Example 1]{QC_cyclic} over the AWGN channel.}
%\label{figsim4}
%\end{figure}

\textcolor{mycolor}{In \figurename~\ref{figsim7}, we consider the construction of low rate QC-LDPC codes using Theorem~\ref{theorem2} and their error performance. In this figure, $\mathcal{C}_1$ is the null space of the parity-check matrix $\mathbf{H}(4,8)$ which is obtained from Theorem~\ref{theorem2} with $G=\mathbb{Z}_8$. This code is a $(2040,1031)$ QC-LDPC code with rate $0.505$ that performs $2.38$dB from the Shannon limit. The error performance of this code can be compared with the error performance of $\mathcal{C}_9$. This code is a $(2040,1051)$ QC-LDPC code with rate $0.515$ and given in \cite[Example 2]{6} that performs $2.29$dB from the Shannon limit. The code $\mathcal{C}_2$ in \figurename~\ref{figsim7} is the null space of the parity-check matrix $\left[
                                                                                             \begin{array}{c|c}
                                                                                               \mathbf{H}(4,8) & -\mathbf{H}(4,2) \\
                                                                                             \end{array}
                                                                                           \right]
$, where $\mathbf{H}(4,8)$ and $\mathbf{H}(4,2)$  are obtained from Theorem~\ref{theorem2} with $G=\mathbb{Z}_8$. It is a $(2550,1297)$ QC-LDPC code with rate $0.508$ that performs $2.676$dB from the Shannon limit. It can be compared with $\mathcal{C}_{12}$ and $\mathcal{C}_{13}$. The former is a $(4, 8)$-regular QC-LDPC code
with length $2640$, dimension $1323$ and rate $0.501$; the latter is obtained by masking the base matrix of   $\mathcal{C}_{12}$ with the following $4\times 8$ matrix \cite[Example 1]{C12}:
\begin{IEEEeqnarray}{rCl}
  \mathbf{Z}(4,8) &=& \left[\begin{array}{cccccccc}
                        1 & 0 & 1 & 0 & 1 & 1 & 1 & 1 \\
                        0 & 1 & 0 & 1 & 1 & 1 & 1 & 1 \\
                        1 & 1 & 1 & 1 & 1 & 0 & 1 & 0 \\
                        1 & 1 & 1 & 1 & 0 & 1 & 0 & 1
                      \end{array}
  \right],
\end{IEEEeqnarray}
that gives a $(3,6)$-regular $(2640,1320)$ QC-LDPC code with rate $0.5$ and with higher girth compared to $\mathcal{C}_{12}$. Both these codes have a low error floor in their performance curves. At the BER of $10^{-6}$, $\mathcal{C}_{12}$ and $\mathcal{C}_{13}$  perform $2.3675$dB and $1.972$dB from the Shannon limit, respectively.
In comparing $\mathcal{C}_{12}$, $\mathcal{C}_{13}$ and $\mathcal{C}_2$, it is evident that  $\mathcal{C}_2$ has a weak error performance. The reason of this weak error performance arises from the high column degree of this code compared to its length. For example,  $\mathcal{C}_1$ is a code with shorter length and its performance is almost similar to $\mathcal{C}_{12}$. Another instance of group ring based QC-LDPC codes is $\mathcal{C}_3$ which is a $(3066,1538)$ code with rate $0.501$. This code is obtained from Theorem~\ref{theorem2} with $G=\mathbb{Z}_9$ and its parity-check matrix is $\mathbf{H}(3,6)$. At the BER of $10^{-6}$, it performs $1.8075$dB from the Shannon limit. This code can be compared with $\mathcal{C}_8$ which is a  $(3066,1544)$ QC-LDPC code with rate $0.503$ and it is given in \cite[Example 1]{6}. At the BER of $10^{-6}$, $\mathcal{C}_8$ performs $1.87$dB from the Shannon limit.}

%======================================

\begin{figure}[!h]
\centering
\includegraphics[width=5in]{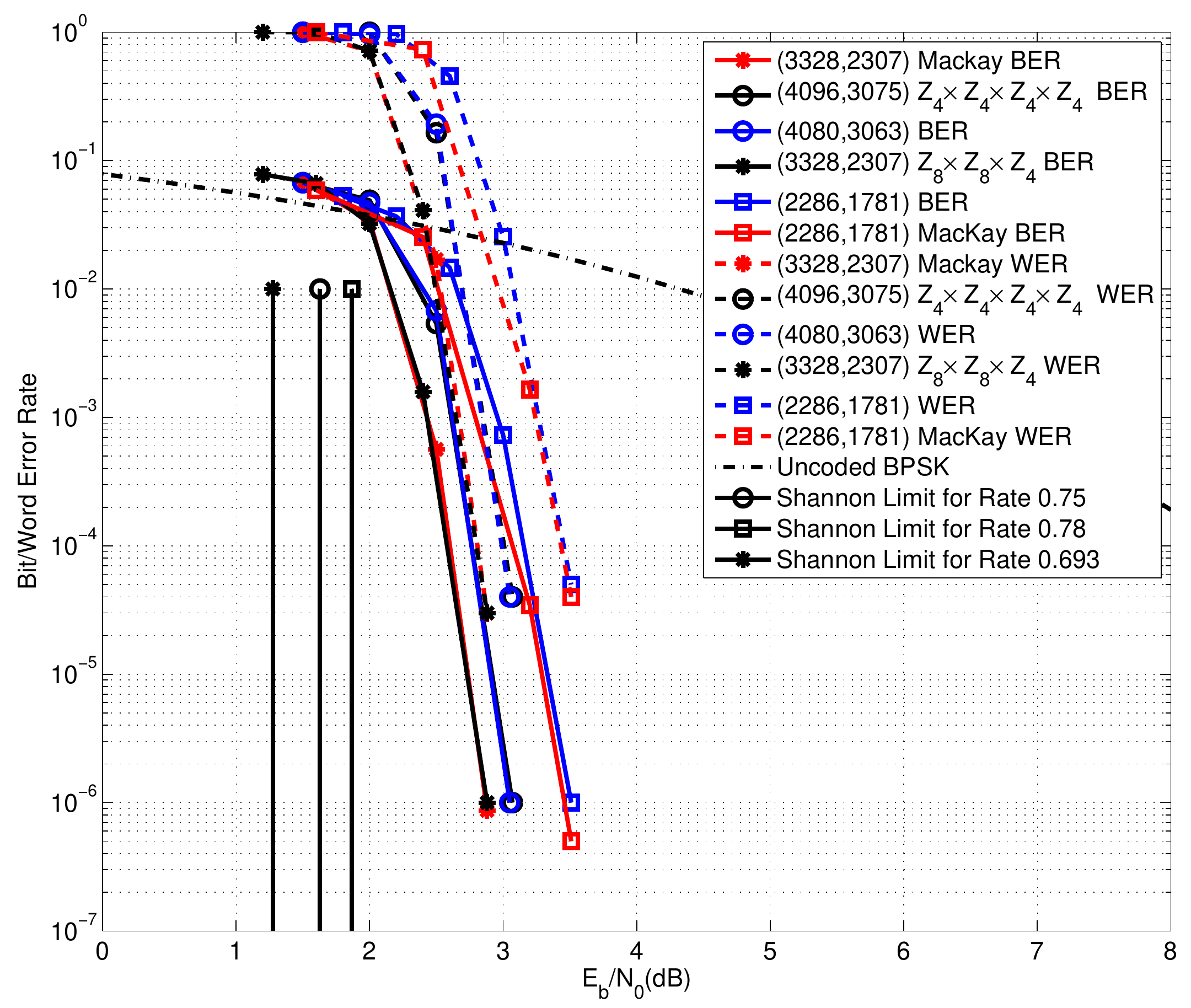}
\caption{BER and WER performances of the QC-LDPC codes based on modified $S_2$-sets over the AWGN channel.}
\label{figsim1}
\end{figure}

In \figurename~\ref{figsim1}, we present the simulation results of the QC-LDPC codes obtained from Theorem~\ref{main_theorem} and the  codes constructed in~\cite{15}.
The addressed codes in \cite{15} can also be obtained from Theorem~\ref{main_theorem2} by choosing the group $G$ as a cyclic group and $D$ as a difference set in $\mathbb{Z}_{q-1}$.
We also compared our codes with the random MacKay LDPC codes \cite{38}.
The codes with lengths $4080$ and $2286$ are chosen from  \cite{15} for comparing the results. Simulation results show that the performance of the code with length $2286$ is $0.04$dB away from the performance of a random MacKay LDPC code with the same rate and length and average column degree $3.5$.      The code with length $4096$ and dimension $3075$ is obtained from the modified $S_2$-set found in $\mathbb{Z}_4\times \mathbb{Z}_4\times \mathbb{Z}_4\times \mathbb{Z}_4$. The group $G$ is considered to be the cyclic group of order $16$. We consider the first $4$ rows of the corresponding $R'G$ matrices in both of these cases. Thus, the row and the column degrees of both codes are $4$ and $16$, respectively. Both  of these codes have the rate $0.75$ and nearly the same error performance. The code of length $3328$ and dimension $2307$ is obtained from the modified $S_2$-set  found in $\mathbb{Z}_8\times \mathbb{Z}_8\times \mathbb{Z}_4$, which is given in \tablename~\ref{table4}. The row and the column degrees of this code are $4$ and $13$, respectively.
Another MacKay LDPC code with these parameters and the average column degree $4$ is constructed.  The error performance of all these codes using the SPA ($50$ iterations) over the AWGN channel, with BPSK modulation, is illustrated in \figurename~\ref{figsim1}.
In this figure, it can be seen  that our code and the MacKay LDPC code both have the same error performance.
A performance comparison has already been done between the binary QC-LDPC codes of \cite{15} and their random-like binary girth $6$ non-QC and QC counterparts. These non-QC and QC codes were generated by software \cite{36} and using the method given in \cite{35}, respectively. According to the results of \cite{15}, the QC-LDPC codes based on difference sets have a considerably better performance than the random-like QC and non-QC girth $6$ LDPC codes. Due to the similar performance of group ring based QC-LDPC codes and the QC-LDPC codes of \cite{15}, the group ring based QC-LDPC codes of Theorem~\ref{main_theorem} outperform the LDPC codes of \cite{36} and \cite{35}.

\begin{figure}[!h]
\centering
\includegraphics[width=5in]{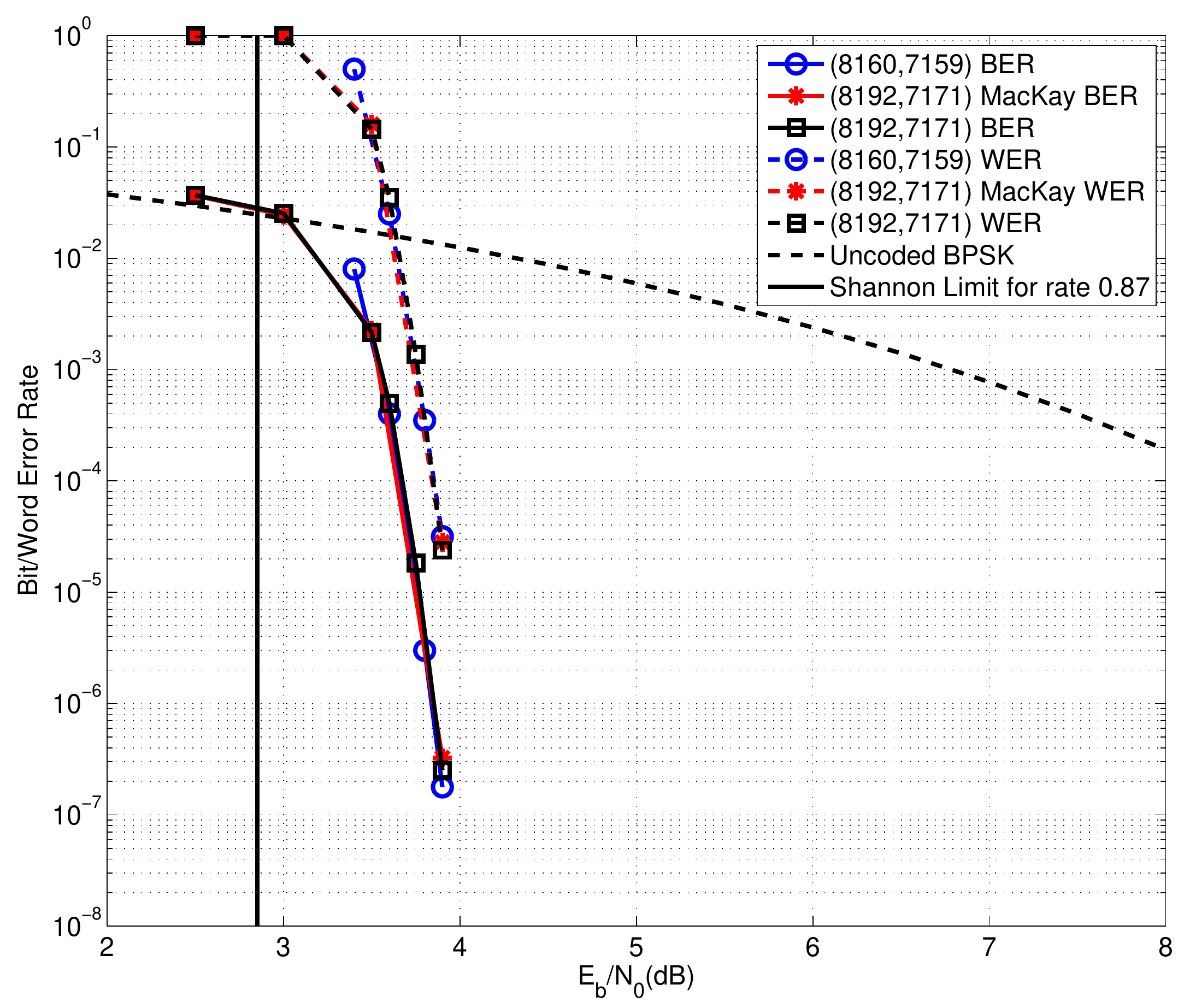}
\caption{Performance comparison between the finite field based QC-LDPC code in \cite[Example 11.9]{37}, the group ring based QC-LDPC code from $\mathbb{Z}_4\times \mathbb{Z}_4\times \mathbb{Z}_4\times \mathbb{Z}_4$ and a random MacKay  code \cite{38}  over the AWGN channel.}
\label{figsim2}
\end{figure}

\begin{figure}[!h]
\centering
\includegraphics[width=5in]{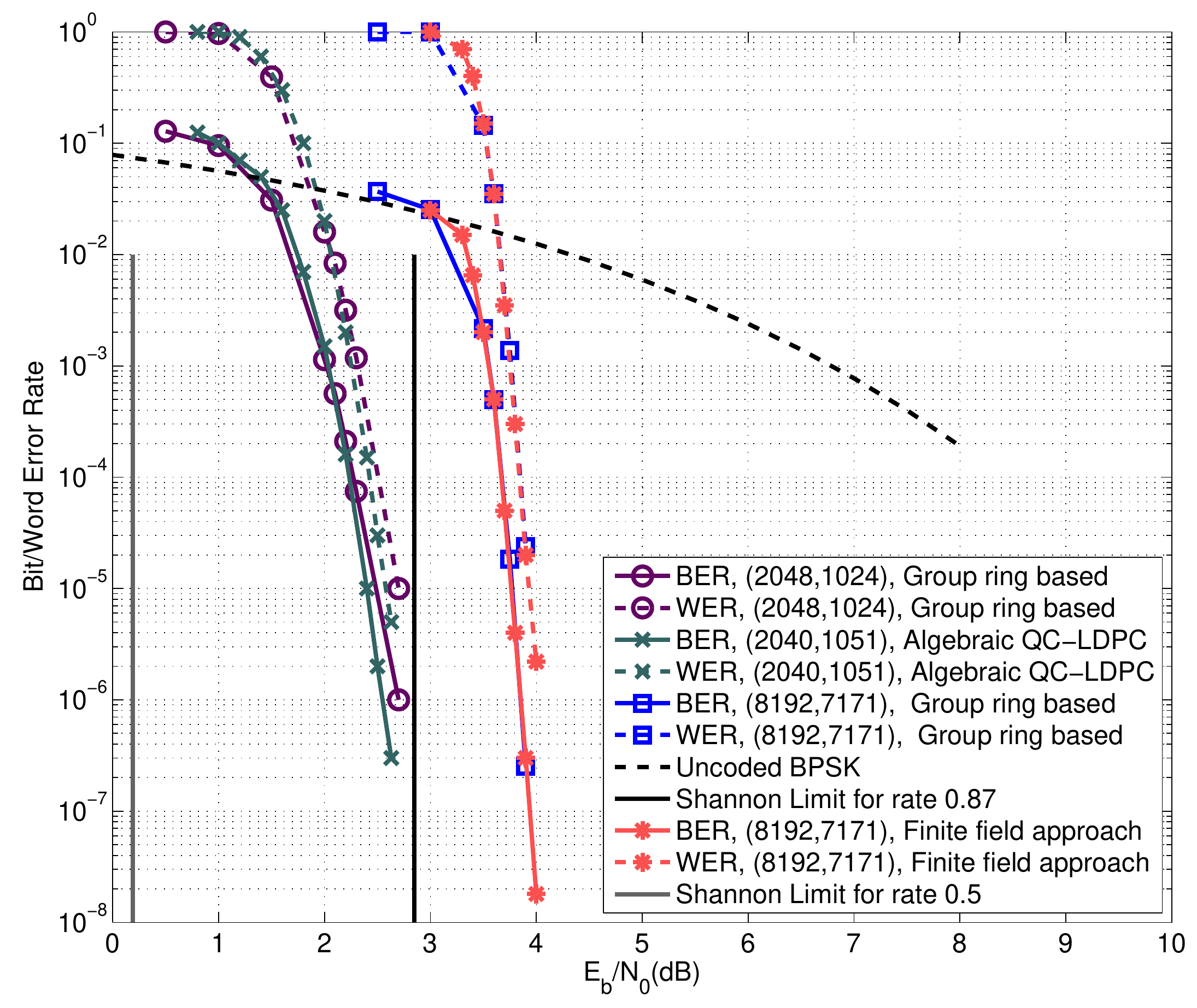}
\caption{Performance comparison between the finite field based QC-LDPC codes in \cite[Example 2]{12}, \cite[Example 2]{6} and the group ring based QC-LDPC codes from $\mathbb{Z}_4\times \mathbb{Z}_4\times \mathbb{Z}_4\times \mathbb{Z}_4$  over the AWGN channel. }
\label{figsim5}
\end{figure}

We use the given $S_2$-set in (\ref{S_2-set}) and the construction method of Theorem~\ref{main_theorem2} to generate another group ring based code as follows. Considering the subarray $\mathbf{H}(4,32)$  in our base matrix gives a QC-LDPC code with rate $0.87$ and length $8192$ that can be compared with the constructed code in~\cite[Example 11.9]{37}.  Both of these codes have the row degree $4$ and the column degree $32$ and the rate $0.87$.  In \figurename~\ref{figsim2}, we present the simulation results of these two codes.
Another code that is presented in \figurename~\ref{figsim2}, is a random MacKay LDPC code \cite{38} with length $8192$ and rate $0.87$ and average column degree $4$.  It can be seen that all these codes have nearly the same error performance. At the BER of $10^{ -6}$, these codes perform within  $0.98$dB from their corresponding Shannon limits.
In \figurename~\ref{figsim5}, the error performance of the  codes obtained from $\mathbf{H}(4,32)$ and $\mathbf{H}(4,8)$ are compared  with the  codes constructed in \cite[Example 2]{12} and \cite[Example 2]{6}, respectively. The group ring based code with length $8192$ has the same error performance as the one based on finite fields in \cite[Example 2]{12}, but the constructed code in \cite[Example 2]{6} outperforms the group ring based code with length $2048$ about $0.15$dB  at the BER of $10^{-6}$.
The error performance of the other group ring based LDPC codes can be found in \cite{50}. It is shown that the performance of these codes can be compared with random MacKay LDPC codes.

%\begin{figure}[!t]
%\centering
%\includegraphics[width=3.4in]{D4results.png}
%\caption{BER performance of constructed codes based on $D_8$}
%\label{figsim1}
%\end{figure}
%
%\begin{figure}[!t]
%\centering
%\includegraphics[width=3.4in]{Allgroups.png}
%\caption{Simulation Results of constructed Codes based on $Z_8$, $D_8$ and $Q_8$}
%\label{figsim2}
%\end{figure}
\section{Conclusions}\label{conclusion}
In this paper, a new method has been proposed for constructing QC-LDPC codes from group rings. Simulation results show that the error performance of the group ring based codes outperforms the error performance of the random-like QC and non-QC LDPC codes. It has been shown that the error performance of group ring based QC-LDPC codes is as good as recently designed QC-LDPC codes based of finite fields. In addition, an algebraic framework has been proposed that describes the group ring based QC-LDPC codes as specific submodules in group rings. The relations between the parameters of the underlying group ring and the error performance of the obtained code have been illustrated using the simulation results. Applying the proposed algebraic framework, that authorizes the application of fast Fourier transform in computations, a new encoding method with faster implementation capability, compared to the available encoding methods, has been  proposed. The complexity of the proposed encoding method for group ring based QC-LDPC codes has been analyzed mathematically.  The proposed QC-LDPC codes in this paper together with the proposed encoding method, enable the exploiting of the benefits of algebraic codes (simple encoding) and modern codes (acceptable error performance in the AWGN channel) at the same time.
\section*{Acknowledgments}
%The authors would like to thank to the Institute for Research
%in Fundamental Sciences (IPM) for financial support. The
%research of the second author was in part supported by a grant
%from IPM (No. 93050220).
%========================================================================
The authors would also like to thank the anonymous referees for their helpful comments on the earlier versions of this paper. The authors also acknowledge the financial support of the Institute for Research in Fundamental Sciences (IPM). The research of the second author was in part supported by a grant from IPM (No. $95050116$). We would like to thank  Prof. Daniel Panario and Dr. Sara Saeedi Madani for proofreading the paper.

\end{document}